%% file: main.tex
\documentclass{jfm}

\usepackage{preamble}

\shorttitle{Generative AI in Turbulence ...}
\shortauthor{L. Biferale, M. Buzzicotti}

\title{Generative AI in Turbulence: Foundations, Applications, Validations and Benchmarks}

\author{
  Luca Biferale\aff{1} \corresp{\email{biferale@roma2.infn.it}}
 \and Michele Buzzicotti\aff{1}}

\affiliation{\aff{1}Department of Physics and INFN, University of Rome Tor Vergata, Rome, Italy}

\begin{document}

\maketitle

\begin{abstract}

Generative artificial intelligence offers new approaches to turbulence problems in which incomplete observations or unresolved degrees of freedom admit multiple physically plausible solutions. Potential applications include stochastic closures, super-resolution, data assimilation, uncertainty quantification, and the generation of Eulerian and Lagrangian flow ensembles. However, turbulence is not a conventional image-generation problem. Its multiscale dynamics, intermittency, extreme events, rough velocity fields and intrinsically stochastic transport make visual realism, pointwise error, and low-order spectral agreement insufficient measures of physical fidelity.

Here, we examines where generative models provide  novel capabilities  with emphasis on diffusion models, score-based methods, and related probability-transport formulations. We discuss their use in unconditional generation, probabilistic reconstruction, reduced-order modeling, temporal evolution, and mesh-free representations. Across these applications, we argue that validation must address the properties that distinguish turbulent flows: scale-by-scale energy transfer, structure functions, anomalous scaling, probability-density-function tails, coherent structures, conservation laws, temporal consistency, ensemble diversity and generalization across Reynolds numbers, forcing mechanisms, geometries, and observation conditions.
Finally, and importantly, we discuss potential limitations in physical validation, baseline selection, data availability and reproducibility, and the need for a federated benchmark ecosystem combining open datasets, standardized challenge problems, hidden test sets, physics-aware metrics, and documented failure cases. 

This Perspective is intended to serve as a bridge between two communities that have developed largely distinct languages, tools, and viewpoints. For practitioners from the AI community, our aim is to provide a fair and accessible introduction to those aspects of turbulence that are essential for appreciating the challenges posed by the field and for assessing the physical relevance of generative models. Conversely, for practitioners coming from the turbulence community, we aim to provide a fair introduction to the key technological and conceptual ingredients underlying modern Generative AI, clarifying what these tools actually do, what assumptions they rely upon, and where their potential and current limitations lie.

\end{abstract}

\tableofcontents


\include{chapters/Introduction}

\include{chapters/GenerativeAI}

\include{chapters/Applications}
\include{chapters/Bibliometric}
\include{chapters/Opendata}
\include{chapters/Outlook}
\include{chapters/Acknowledgements}


\bibliographystyle{jfm}
\bibliography{biblio,transitions,transitions2}

\end{document}

%% file: chapters/Introduction.tex
\section{Introduction}

\begin{figure*}
      \centering
\includegraphics[width=1.\linewidth]{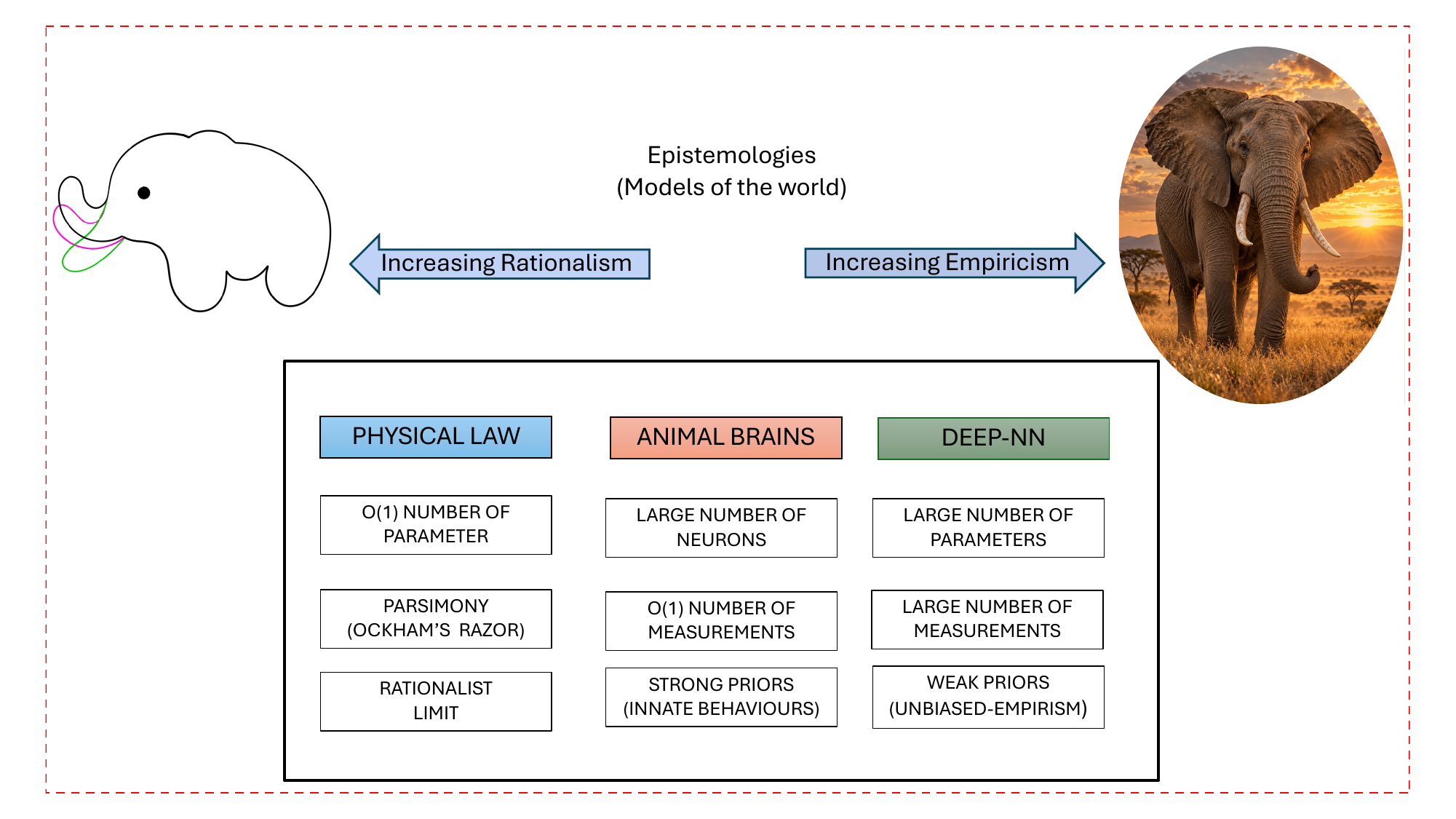}
      \caption{Top left: 
      Artistic reinterpretation of the results presented in \citep{mayer2010drawing}  where  an elephant contour is coded in a set of four complex parameters, the  real part of the 5th parameter is the “wiggle" parameter, a demonstration of von Neumann's statement cited in the main text. Top right: picture obtained from a Generative AI algorithm. Bottom table:  comparison of three epistemologies: physical laws (rationalist, theory-driven), animal brains (strong innate priors and efficient learning), and deep neural networks (data-driven, highly empirical). While physics relies on compact theories and biological brains on structured priors, modern ML learns primarily from large datasets with weak priors and high computational costs, representing a distinct “third epistemology” of knowledge generation \citep{mitra2021fitting} }
      \label{fig:elephant}
  \end{figure*}

  {\it “How many arbitrary parameters did you use for your calculations?” I thought for a moment about our cut-off procedures and said, “Four.” He said, “I remember my friend Johnny von Neumann used to say, with four parameters I can fit an elephant, and with five I can make him wiggle his trunk.” With that, the conversation was over (see also Fig. \ref{fig:elephant}).
  } \\ \\
  Before returning to this interesting  quote taken from a discussion between Freeman Dyson and Enrico Fermi \citep{dyson2004meeting}, let us present our disclaimer: {\it what this perspective paper is not about}.   The field of AI applications to basic sciences and engineering is booming, the number of publications, researchers, and practitioners involved, from academic institutions and commercial companies, as well as announcements of {\it achieved supremacy} are all increasing with an { \it inflationary supra-luminal} rate (see Sec. \ref{sec:biblio}).  The quantity of information 
appears  with a speed that does not probably allow the century-old system based on {\it checks and balances}, such as peer review  and reproducibility of results, to be appropriate any more. 

In the past four to five years, numerous excellent reviews have appeared on the applications of AI to fluid mechanics \citep{choi2025perspective,vinuesa2022enhancing,lino2023current,taira2025machine,bauerheim2026routes,wang2024recent,fukami2026practical}, collectively attracting thousands of citations. Remarkably, even one of the most comprehensive and widely cited \citep{brunton2020machine}, published as recently as 2020, preceded by just a few months the emergence of score-based diffusion models \citep{song2020score}, which are now regarded by many practitioners as the method of choice for generative AI.
This is just an example to say that we do not intend to write a review of recent results using AI tools for turbulence because  they are too many and not consolidated enough. Moreover,  with very high probability, considering the old-fashioned publication time-schedule, we will certainly miss the equivalent of another {\it Diffusion Model "revolution"} and our perspective would risk to be outdated before it is published. We also do not intend to write a review of reviews and will limit ourselves in the following to citing the papers that have been instrumental for us in getting a partial orientation in the field, without claiming to cover the whole relevant literature. \\
Similarly, we will not address the important developments associated with Agentic AI. 
The potential impact of AI on turbulence may ultimately extend well beyond data-driven modelling: remarkably, an advanced AI system has recently produced a proof of finite-time singularity formation for the three-dimensional Navier–Stokes equations with smooth forcing, addressing the long-standing Millennium Prize regularity problem. The result is accompanied by a machine-checked formalization in Lean, providing an independently verifiable formal proof of the key mathematical arguments \citep{OpenAI2026NavierStokes}. The implications of such a result for the communities working on the mathematics foundations, physical properties and engineering applications of Navier-Stokes equations are difficult to predict at the moment.  
Neither we open  the broader Pandora’s box of scientific discoveries and moral implications arising from the interaction with, and supervision of, AI tools. These issues are also closely connected to the increasing quantity, and perhaps quality, of scientific publications produced with the support of AI. Such developments may have a potentially disruptive impact on how science is organized, communicated, cited, and developed \citep{golan2023artificial,pope2026encyclical}. See, for example, the recent statement from the mathematics community \citep{LeidenDeclarationWebsite2026}.
A further set of questions that we will not touch concerns the epistemological implications of AI, the nature of the knowledge generated with its support, and the shift in paradigm,  from {\it Parsimony} to {\it Complexity} also known as the Ockham's razor dilemma  \citep{dubova2025ockham} and  from 'proof-scarcity' to 'proof-abundancy' \citep{tao2026math20}.
Related issues include the pursuit of superintelligence, recursive self-improvement, and the broader question of the kinds of reasoning AI systems may ultimately achieve. AI has already demonstrated substantial capabilities in induction, or statistical consistent interpolation \citep{mitra2021fitting}, and is rapidly advancing in deduction (formal proof). An open question is whether it will also be able to perform abduction, i.e. the generation of novel explanatory hypotheses \citep{zahavyposition}, see also \citep{bauerheim2026routes} for a JFM perspective on whether AI can rival, or potentially surpass, the effectiveness of traditional mathematical methods in addressing the challenges of Computational Fluid Mechanics (CFD). \\
On the other hand, we have chosen to structure this perspective as an attempt to capture an instantaneous snapshot of the rapidly evolving AI tools and ideas for turbulence, with particular emphasis on generative algorithms and their applications. We hope that the resulting picture will not be excessively blurred, without the ambition of predicting where the field will be a few years from now. In this respect, it is worth recalling Dante’s warning to those who presume to see too far into the future. He consigned diviners to Hell, condemned to walk with their faces turned backwards precisely because they had sought to look too far ahead: {\it "Perch\'e volse veder troppo davante,di retro guarda e fa retroso calle"}\footnote{{\it "Because she/he wanted to see too far ahead, she/he looks behind and turns back"}. Dante Alighieri. Divina Commedia (Inferno, XX, vv. 38-39)}. 
\subsection{What is Turbulence?} 
\begin{figure*}
      \centering      \includegraphics[width=1.\linewidth]{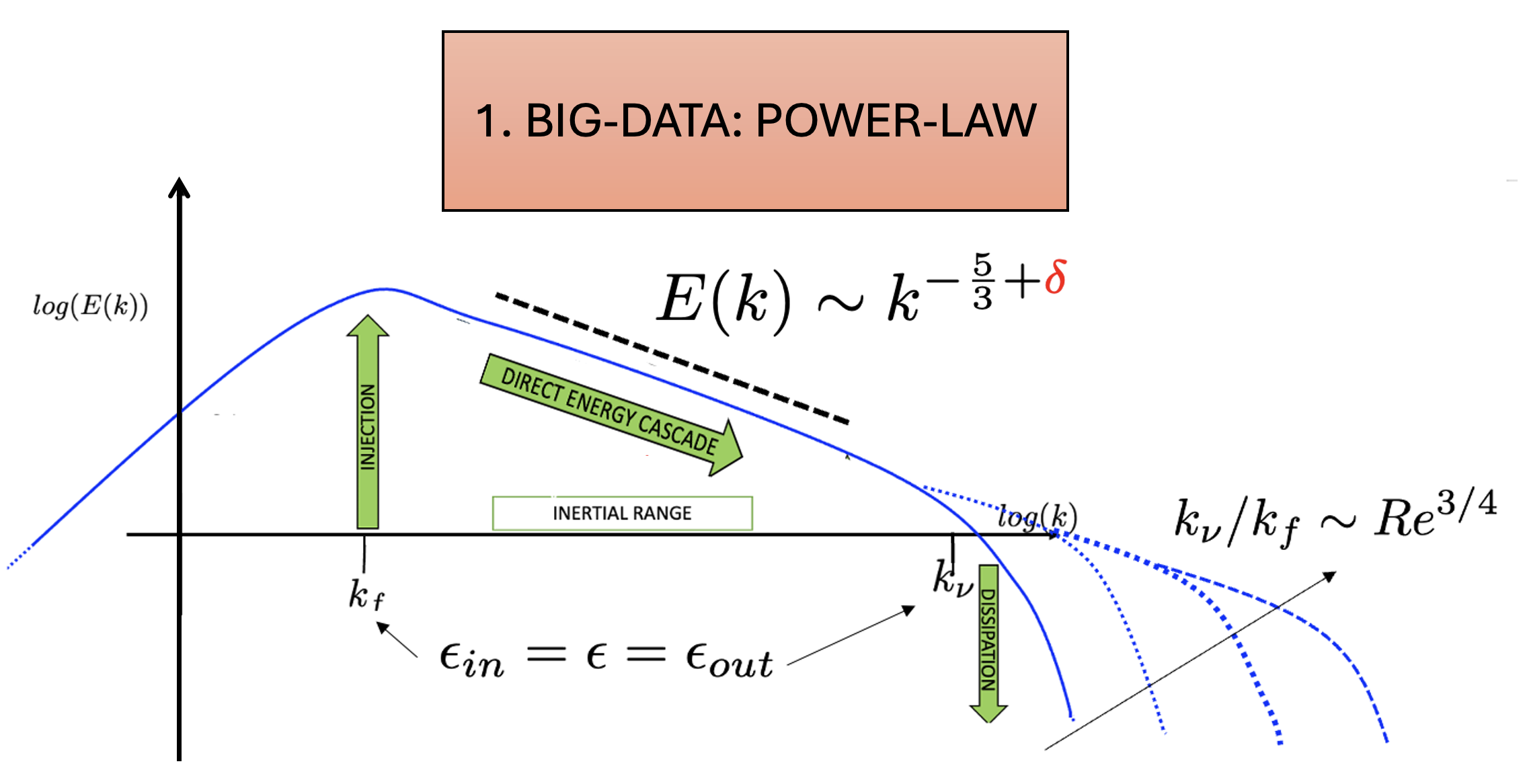}
      \caption{The Big-4. $\#1$: Scale separation and effective dimensionality. 3D Homogeneous and Isotropic Turbulence (HIT), is characterized by two   properties. First, the development of power-law spectra, $E(k) \sim k^{-5/3+\delta}$, within two cutoffs wavenumber: $k_f$, the infrared scale  where energy is injected, and $k_\nu$, the ultraviolet scale where energy is dissipated \citep{Frisch,pope2001turbulent}. In between,  the inertial terms, given by self advection $v_i \partial_i v_j$ and pressure stress, $\partial_i p$, are leading and  a power law distribution of energy per wavenumber is observed, with a  slope close to the Kolmogorov prediction $5/3$. Statistical stationary imposes a balance between the energy injection, $\epsilon_{in}$  and the energy dissipation, $\epsilon_{out}$. Second,  the    ratio between the IR an UV cutoffs grows with the Reynolds number $k_\nu/k_f \sim Re^{3/4}$, where the Reynolds number is given by the dimensionless ratio among non-linear and linear terms in the Navier-Stokes equations, $Re \sim U_0L_0/\nu$. Here,  $U_0$ and $L_0$ are the characteristic velocity and length scales of the flow and $\nu $ is the kinematic viscosity. The presence of a power-law inertial range with an exponent $<3$ is the indication of the development of a highly intricate set of hierarchical fluctuations (see Fig. \ref{fig:bigfour1.2})  which result in a quasi-singular velocity profile, never differentiable (see also Big-4 \#3 in Fig. \ref{fig:bigfour3}). Replacing typical Reynolds number  observed in real-life turbulent flows (from $Re\sim 10^6$ for turbo-machinery up to $Re\sim 10^{10}$ in the atmosphere, and more in the sky) one easy realizes that 3D HIT turbulence is to all possible meaning an {\it "infinite"} $\#_{dof}$ problem, requiring a number of $\sim Re^{9/4}$ to fully resolve all scales in between $k_\nu$ and $k_f$. This is a {\it discontinuous} difference from 2D flows, where the Energy spectrum goes in the UV  as $k^{-y}$, with $y \ge 3$ \citep{boffetta2012two}  leading to a smooth and differentiable small-scale flow, with much less energy content in the high wavenumber range (see also Fig. \ref{fig:turbulence2d} for more info about 2D set-ups). Other 3D configurations, e.g. wall bounded flows, are even more complicated than HIT, because of anisotropic and non-homogeneous dissipative, multi-scale statistics, which depends on the distance from the wall, as well as boundary layer physics dominated by the simultaneous  presence of production and dissipative terms \citep{pope2001turbulent}.  }
      \label{fig:bigfour1}
  \end{figure*}

\begin{figure*}
      \centering     
       \includegraphics[width=1\linewidth]{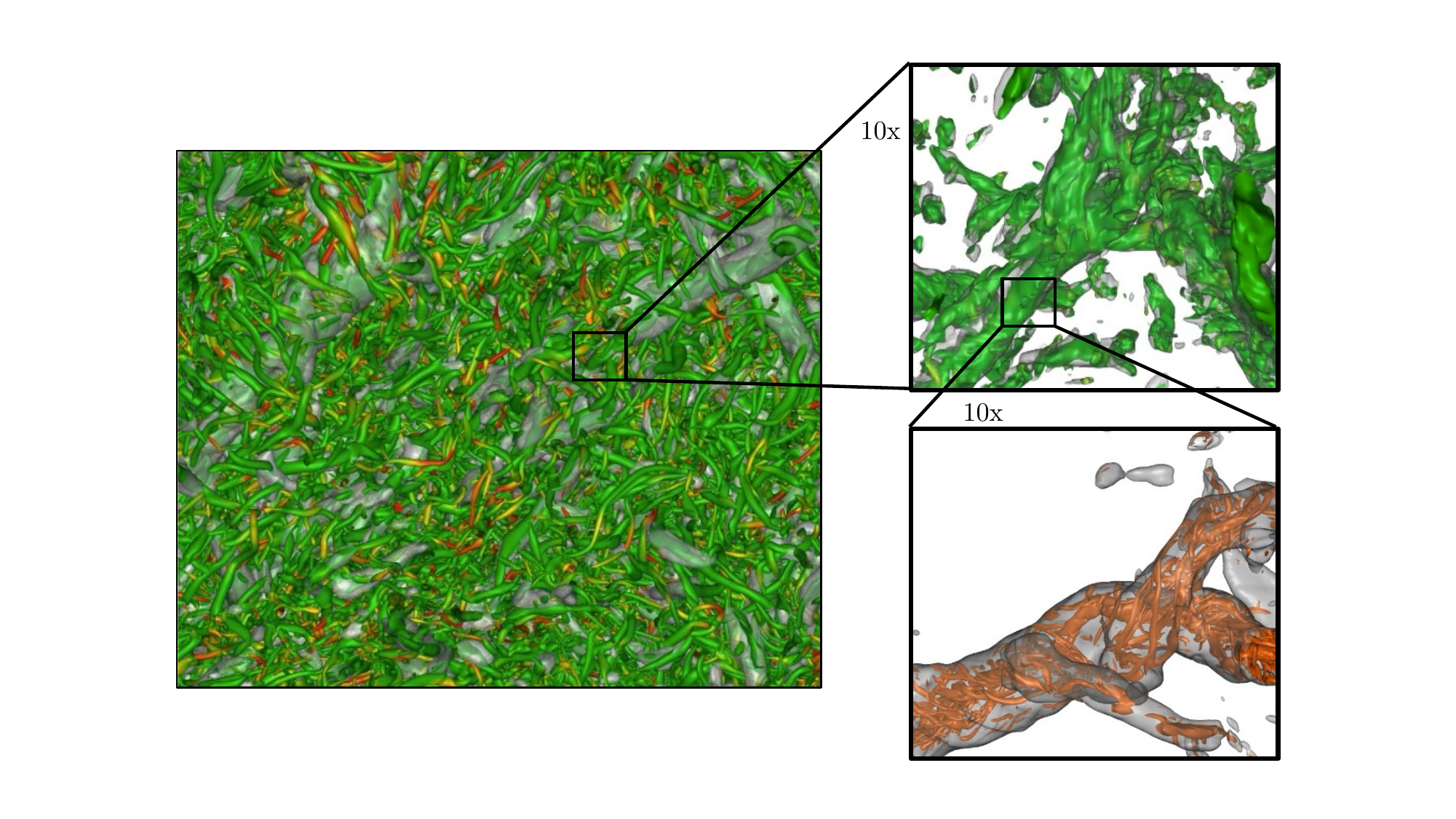}
      \caption{The Big-4 $\#1$ (continued): Three different zoomed ($\times 10$) rendering of the vortex tube structure observed at scales within the inertial range of 3D HIT turbulence (Fig. \ref{fig:bigfour1}), representing the nested hierarchy of fluctuations characterising 3D turbulence. Frames extracted from \citep{JHTDB_Vortex}, {permission kindly provided by C. Meneveau,}  see also \cite{burger3325vortices}. It is unknown how to fully characterize the distribution of these huge number of {\it dof}, and  it is an unresolved problem  to develop reliable {\it Reduce Order Models} (ROM) that fully describe the intricate set of sub-grid-scale fluctuations, see also the other 3 Big-4 properties described in Figs. \ref{fig:bigfour2}-\ref{fig:bigfour4}. A huge body of applied literature is devoted to the development of Large-Eddy-Simulations (LES) approaches, modeling small-scales fluctuations \citep{meneveau2000scale}. Other attempts to compress the $\#_{dof}$  in a smaller support have shown that blind decimation might fully destroy important flow features already by removing fraction of percent of data \citep{alexakis2018cascades}, a bad news concerning possibility to {\it compress} the information in turbulence. One of the main obstacles for applications of ML is given by the huge size of any input data (a minimum of billions of  data point are needed to be representative of the minimal flow complexity at moderate Reynolds) and the lack of data (no one has thousands, or hundred-thousands, uncorrelated samples for training). Moreover, the huge set of hierarchically nested scales and frequencies involved, implies a non trivial importance of different multi-scale features for different observable, questioning the possibility to identify a unique set of optimal metrics for training and validation.  As a result, none is able to satisfactory generate, using data-driven or other stochastic phenomenological model, a 3D configuration of a turbulent flow with billions or trillions of grid points, as required for moderate Reynolds number flows. For the latter task the only fully reliable way is still to use the Navier-Stokes equations. }
      \label{fig:bigfour1.2}
  \end{figure*}

\begin{figure*}
      \centering
\includegraphics[width=1.\linewidth]{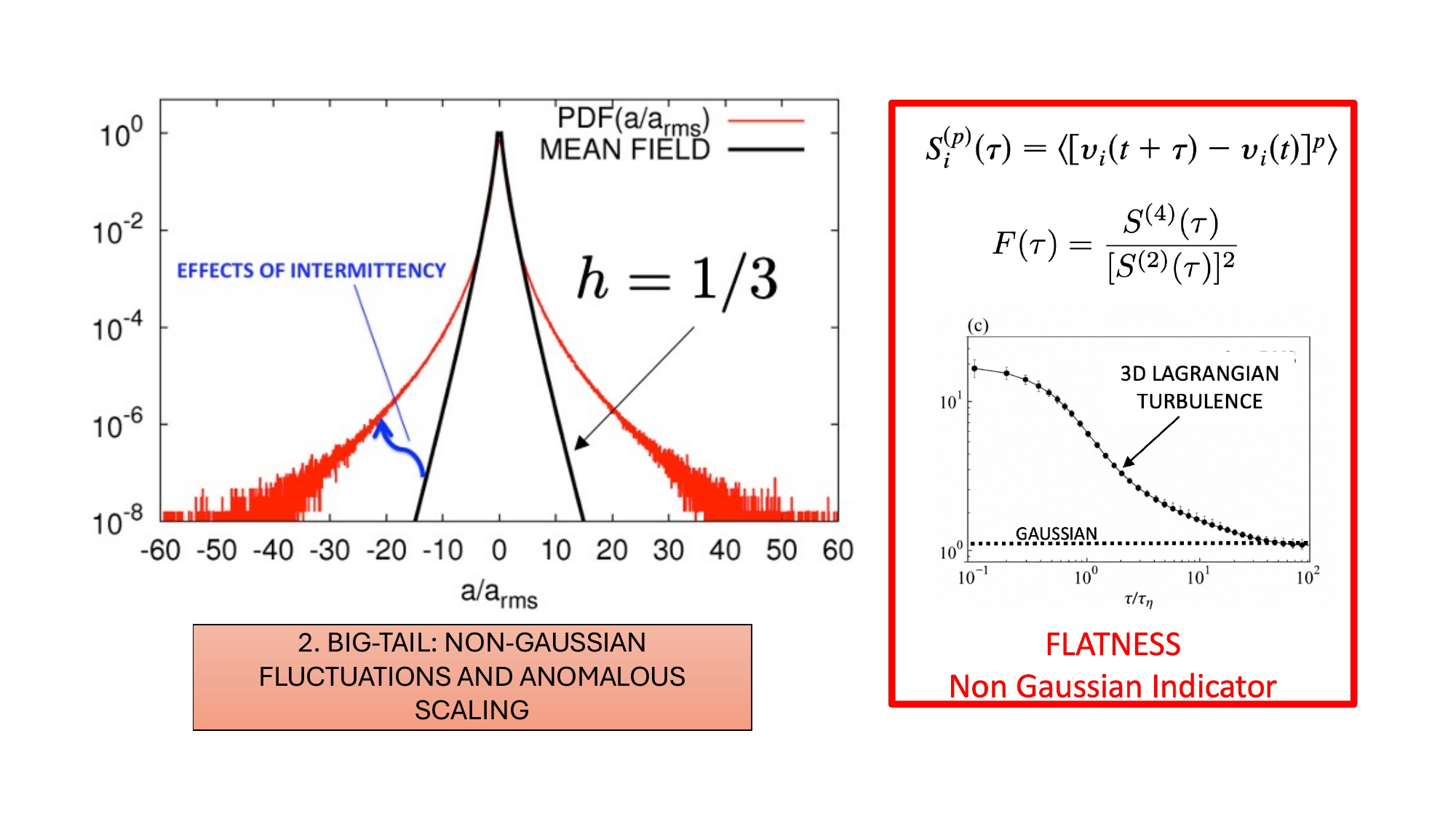}
      \caption{The Big-4. $\#2$: {\it Big-tails, intermittency, extreme and rare events,  non-perturbative and non-Gaussian fluctuations}. Left: 3D Eulerian and Lagrangian turbulence is characterized by strong deviations from Gaussianity and the presence of anomalous scaling \cite{alexakis2018cascades}. The most striking properties being the development of fat tails (fater and fater by increasing Reynolds)  as shown by the standardized PDF of the flow acceleration on the left panel, where extreme intense fluctuations up to 50-60 standard deviations can be easily detected. In the figure, red data comes from a DNS of HIT at $1024^3$ resolution points. The black curve corresponds to the prediction obtained by a global homogeneous $h=1/3$ scaling exponent for velocity increments as predicted by the {\it mean field} Kolmogorov 1941 theory (K41).   The strong discrepancy between K41 theory and the ground truth data  is known to be connected to multifractal small-scales properties \citep{Frisch,biferale2004multifractal}, see also Fig. \ref{fig:bigfour3}. Intermittency and fat tails are a key fingerprints of all multi-scale turbulent flows with and without boundaries at  small enough scales (in 3D) and must be considered as a primary feature to be reproduced. They emerge only going {\it beyond spectral properties}, observing third-order energy or helicity fluxes and high-order moments, as, e.g., Flatness and generalised Flatness, see \citep{Frisch} and right panel. Due to these strong fluctuations, it is not uncommon to observe extreme events at the gradient scale as large as the global large-scale rms signal, as it happens for cold/warm  fronts in atmosphere physics. As a results, small-scales turbulent fluctuations cannot be put under the carpet of the {\it low-energy-content},  even if one is interested only to large-scale smooth evolutions, because it might become locally $O(1)$  and disruptive for global control/predictability. Any ROM or data-driven approach which limits its benchmarks to spectral properties and/or low order moments might therefore  be blatantly wrong concerning these important effects.  Right: An example of 4th order Flatness,$F_4(\tau)$, defined in terms of the structure functions of single velocity components, $S_p(\tau)$  measured along Lagrangian particles evolution in 3D turbulence at changing the time lag $\tau$. Showing the relative increase of non-Gaussian (intermittent) fluctuations by decreasing the time lag in the inertial range, and the tendency to go from a Gaussian value $\sim 3$, for low frequency fluctuations, till very high intensities for $\tau \to 0$, close to the  inverse of the viscous Kolmogorov time, $\tau_\eta$.  }
      \label{fig:bigfour2}
  \end{figure*}

\begin{figure*}
      \centering
\includegraphics[width=0.8\linewidth]{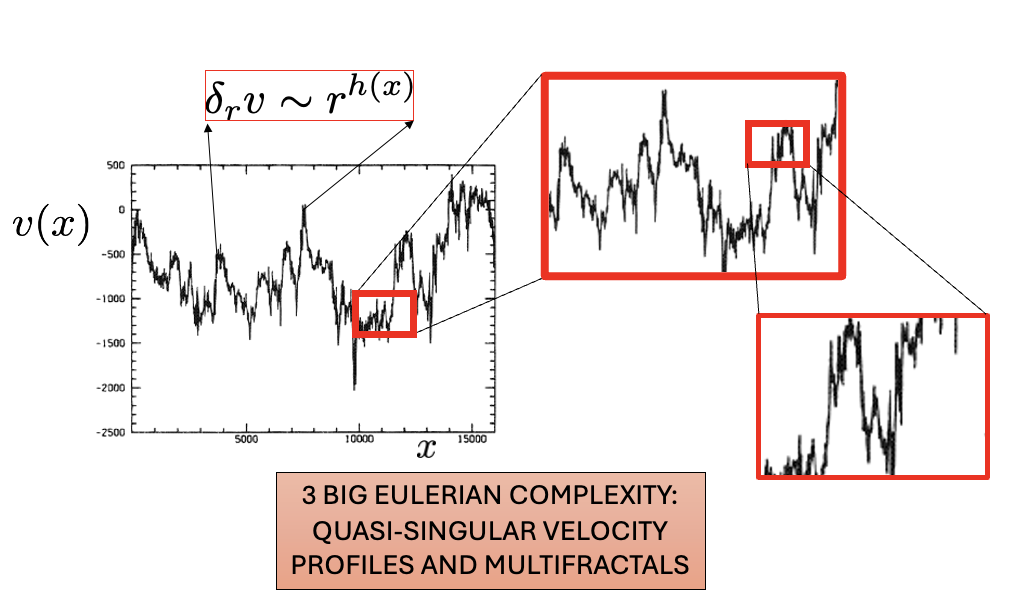}
      \caption{The Big-4. $\#3$: {\it Rough velocity profiles, loss of regularity and  Multifractal fluctuations}. Eulerian velocity profile (one component):  pictorical description of  one-dimensional cut of the 3D HIT configuration shown in Fig (\ref{fig:bigfour1.2}), with two consecutive zooms. The presence of power-law Fourier spectra is connected to fractional Brownian-like profiles in the real space. Velocity increments in the {\it inertial range}  of scales, between the IR and UV cutoffs, $1/k_\nu \ll r ll 1/k_f$,  do show anomalous power-law behavior in fully developed 3D turbulence \citep{Frisch}. Actual understanding supports a multifractal picture, i.e. the existence of a set of quasi-singularities characterizing the local scaling properties (Holder exponent) of generic velocity increments, $\delta_r v = v(x+r) -v(x) \sim r^{h(x)}$, with  $ h_{min} \le h \le h_{max}$ and the existence of different fractal dimensions characterizing the support of each exponent $D(h)$ \citep{Frisch}. Multifractals are connected  to anomalous scaling for the velocity structure functions: $ S_p(r) = \langle |\delta_ r v|^p \rangle  \sim r^{\zeta(p)}$, with a non-linear dependency of  $\zeta(p)$ on $p$. As a result, also the  Eulerian generalised Flatness of order $p$, $F_p(r) = S_p(r)/S_2(r)^{p/2} \sim r^{\zeta(p)-p/2 \zeta(2)}$ developed a scale-dependency. On the contrary, Kolmogorov theory predicts  $h=1/3$ everywhere, mono-fractal and self-similar scaling properties,  $\zeta(p)=p/3$,  a constant generalised Flatness, and the spectral slope $5/3$. Nowadays, we have more than 40 year of observations from experiments and direct numerical simulations supporting the more complex MF scenario, which is strictly needed to explain the growth of fat tails shown in Figure \ref{fig:bigfour2}. 
      Notice, that the limit $Re \to 
    \infty$ corresponds to send the UV cutoffs to zero $r_\nu = 1/k_\nu \to 0$, i.e. leading to a quasi-singular multifractal velocity profile, never differentiable almost everywhere. This is another {\it qualitative} discontinuity from 2D turbulence, where one can prove the regularity (smoothness) of solutions for all $Re$.  }
      \label{fig:bigfour3}
  \end{figure*}

\begin{figure*}
      \centering
\includegraphics[width=1.\linewidth]{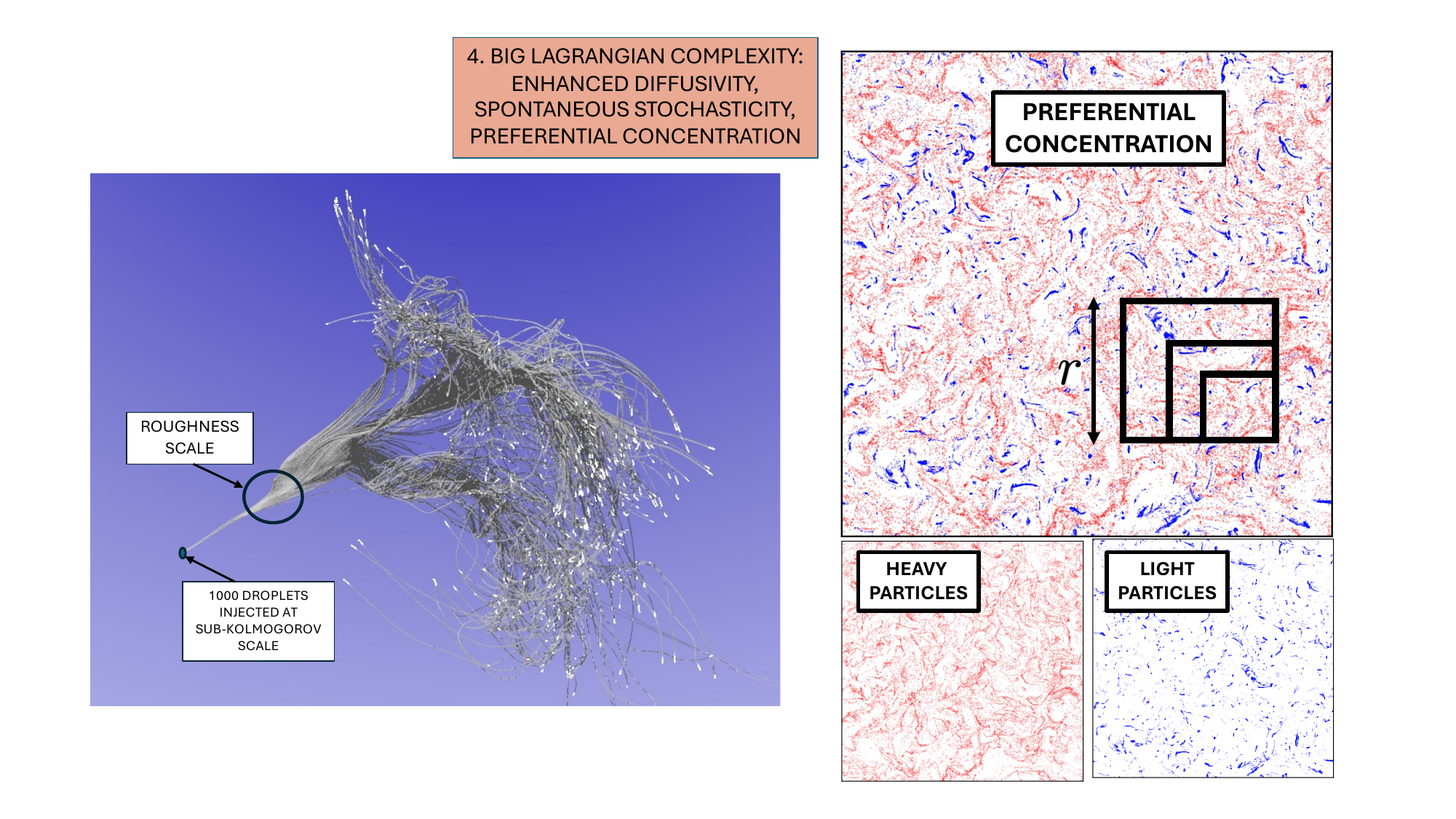}
      \caption{The Big-4. $\#4$: {\it Lagrangian (multiphase) turbulence. Long-time enhanced diffusivity, small-scales preferential concentration and intrinsic  stochastic transport}. The quasi-singular velocity profile present in  fully developed turbulence (Fig. \ref{fig:bigfour3}) leads to an extremely complex dispersion and spatial distributions of (small) particles  \citep{toschi2009lagrangian}. Including the presence of  Richardson like diffusivity (left panel), where clouds of particles can {\it explosively} disperse at large scales even if initially almost superposing, as soon as they reach the smallest distance where the velocity field exhibit rough-scaling properties (see Fig. \ref{fig:bigfour3}) A phenomenon known since the time of Lorenz, and recently baptized {\it spontaneous stochasticity} \citep{lorenz1969predictability,falkovich2001particles,thalabard2020butterfly,bandak2024spontaneous}. Small-scales mass distribution  exhibits clustering and inertia-dependent, multi-scale  multi-fractal properties (right panel). As a result, generating and/or inferring particles distributions in 3D turbulence is a challenge even more formidable then the one for the underlying Eulerian fields, not to mention the requirements to work on a mesh-less domain (see section \ref{sec:applications_eul_lag_mesh}) for more details on AI interesting perspectives. }
      \label{fig:bigfour4}
  \end{figure*}

\begin{figure*}
      \centering      
      \includegraphics[width=1.\linewidth]{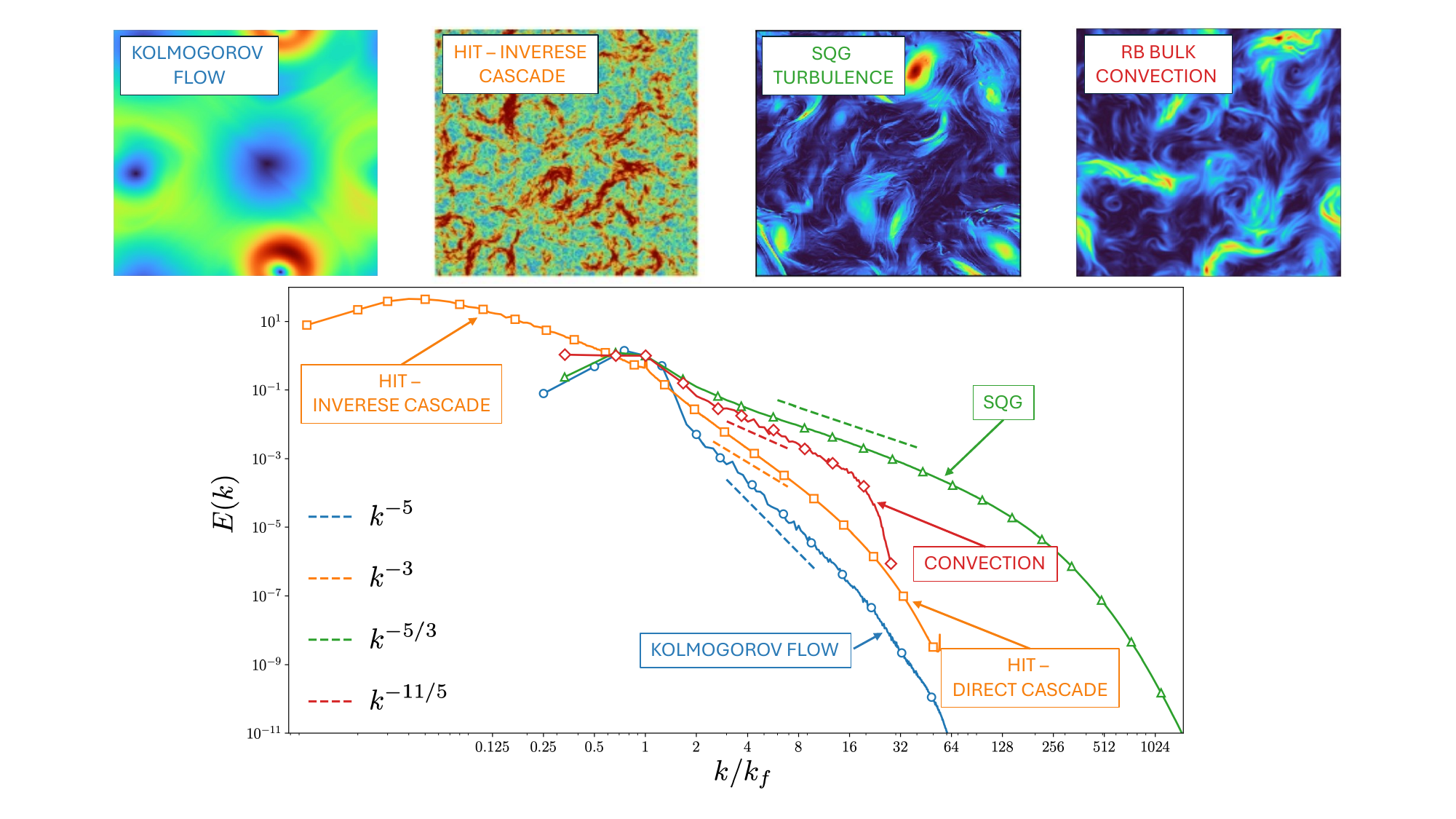}
      \caption{Not all 2D turbulent flows are equally informative as ML benchmarks. Top row: Representative velocity-field snapshots for four canonical two-dimensional regimes: Kolmogorov flow in a direct-cascade regime, homogeneous isotropic turbulence in the inverse-energy-cascade regime,  quasi-geostrophic turbulence, and  bulk fluctuations in two-dimensional Rayleigh–Benard convection. The snapshots highlight substantial differences in spatial regularity: Kolmogorov flow is comparatively smooth, whereas the inverse-cascade, convective, and quasi-geostrophic configurations exhibit rougher, multiscale structures. Bottom row: Energy spectra for the different flow realizations, horizontally aligned with respect to their forcing scale to facilitate a direct comparison of their ultraviolet and infrared  behaviour. The spectral slopes quantify the qualitative differences visible in the snapshots. Kolmogorov flow displays the steepest high-wavenumber decay very fast,  $\sim k^{-5}$,
      implying strongly suppressed small-scale fluctuations and a smooth, differentiable velocity field. By contrast, inverse-cascade turbulence and quasi-geostrophic turbulence exhibit rough-velocity configurations and  spectra close to the 3D K41 scaling  $\sim k^{-5/3}$,
       while Rayleigh–Benard convection shows an approximate Bolgiano scaling  $\sim k^{-11/5}$, also leading to quasi-singular velocity profiles with $\delta_r v \sim r^{6/5}$ in the range with constant energy flux.
      The spectrum of two-dimensional HIT spanning both inverse-energy  and direct-enstrophy cascade ranges is also shown, with an ultraviolet decay close to $k^{-3}$ as predicted by Kraichnan \citep{alexakis2018cascades}.
      The superposition of all spectra makes clear that although two-dimensional turbulence can provide physically meaningful and challenging test cases, the commonly used (by AI practitioners) Kolmogorov-flow is the smoothest and least demanding end of these paradigmatic examples and does not provide a strong benchmark for typical challenges of fully developed turbulence. Data were kindly provided by  G. Boffetta (2D double cascade), F. Fossella (Convection) and V. Valadao (SQG).}
    \label{fig:turbulence2d}
\end{figure*}

\begin{figure*}
      \centering      \includegraphics[width=1.0\linewidth]{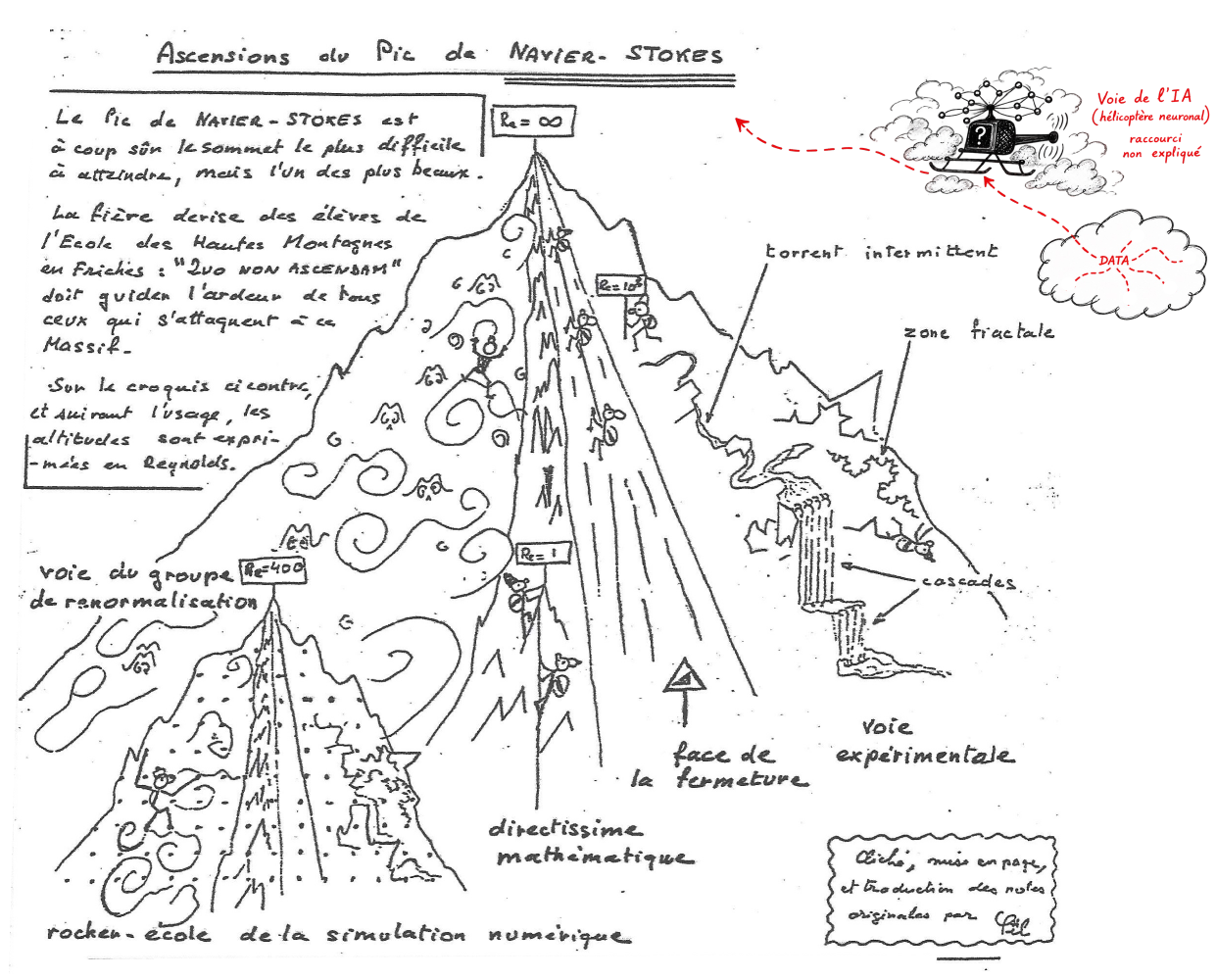}
      \caption{Adapted {with permission} from \citep{Frisch}; original image courtesy of Hélène Frisch. The original illustration, the celebrated “Guide de l’alpiniste turbulent” (Route to the Navier-Stokes peak) was drawn by P. Delache, where the challenge of reaching the infinite-Reynolds-number limit is represented as the ascent of a difficult mountain. The original sketch depicts several classical routes used to study turbulence: the direct mathematical ascent, the renormalization-group path, the experimental route through cascades and intermittent torrents, and the numerical-simulation rock-climbing school, each with its own obstacles, approximations and partial successes. In the present version, we add a new route: the AI helicopter, represented as a black box, collecting turbulent data in the 'cloud' with a question mark. This “neural helicopter” symbolizes the promise of modern data-driven and generative methods to bypass some traditional bottlenecks. At the same time, its black-box nature and unexplained shortcut emphasize the central caution of this Perspective: AI may offer a powerful new way to approach high-Reynolds-number turbulence, but its ability to “land” on physically meaningful statistics must be assessed through turbulence-specific validation. 
      }
      \label{fig:picturbulence}
  \end{figure*}

\begin{figure*}
      \centering \includegraphics[width=1.0\linewidth]{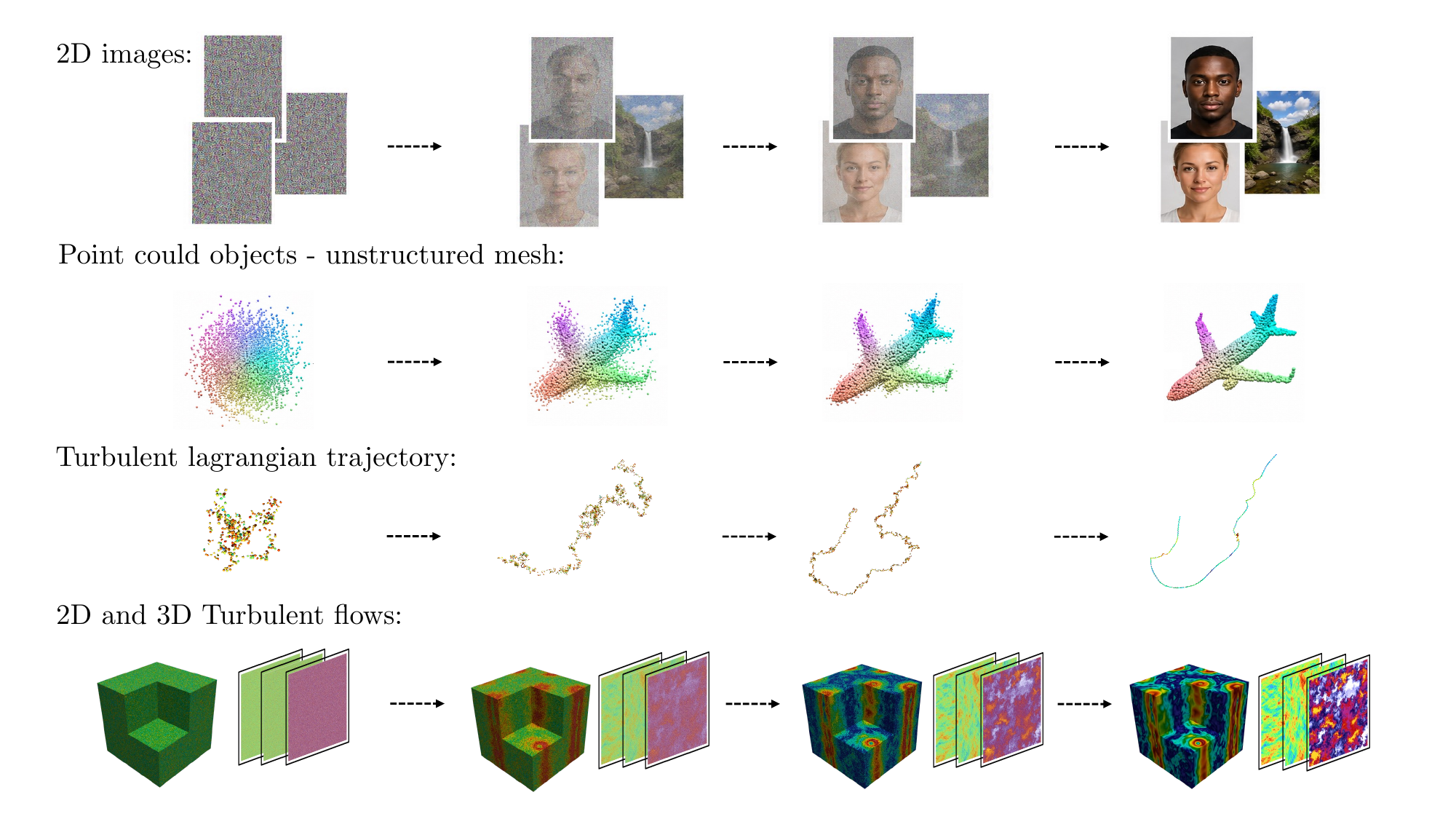}
      \caption{\textbf{Generative modelling across heterogeneous data representations.} Schematic examples illustrating the flexibility of modern generative models across markedly different classes of high-dimensional data. From top to bottom: two-dimensional images on regular pixel grids; point-cloud objects represented by irregular sets of spatial coordinates; turbulent Lagrangian trajectories described as time-dependent particle paths; and two- and three-dimensional turbulent flow fields defined on structured spatial domains. In each case, the generative process starts from a simple noise distribution and progressively transforms it into samples with the geometry, correlations, and statistical structure characteristic of the target dataset. The comparison highlights that the underlying generative principle is largely independent of the particular data representation. At the same time, this flexibility does not imply methodological uniformity: both the neural architecture and the benchmarks used to assess generative quality must be adapted to the geometry of the data and to the physically relevant observables of each problem.}
      \label{fig:DM_catsandflows}
  \end{figure*}

\begin{figure*}
      \centering      \includegraphics[width=1\linewidth]{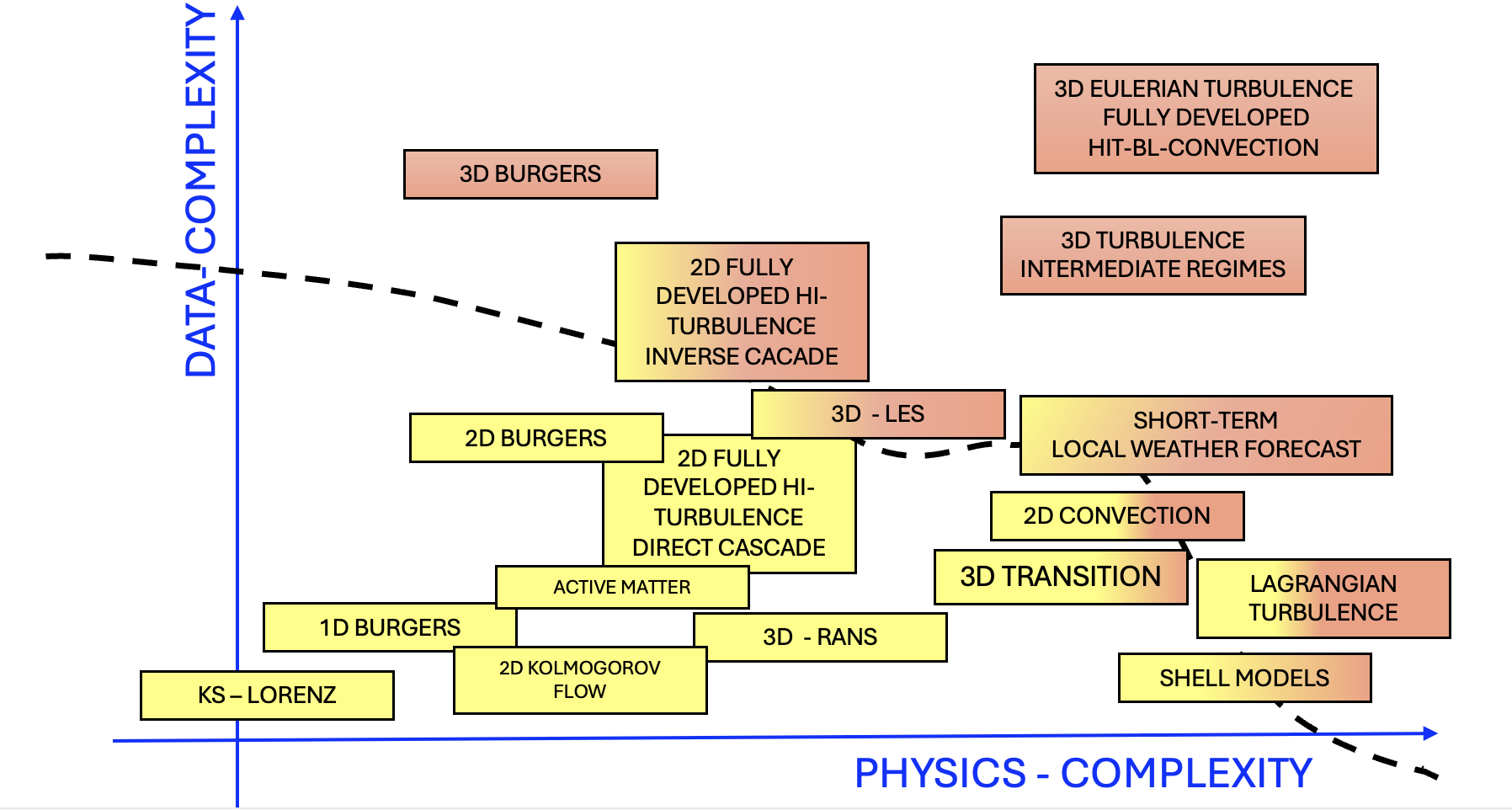}
      \includegraphics[width=1\linewidth]{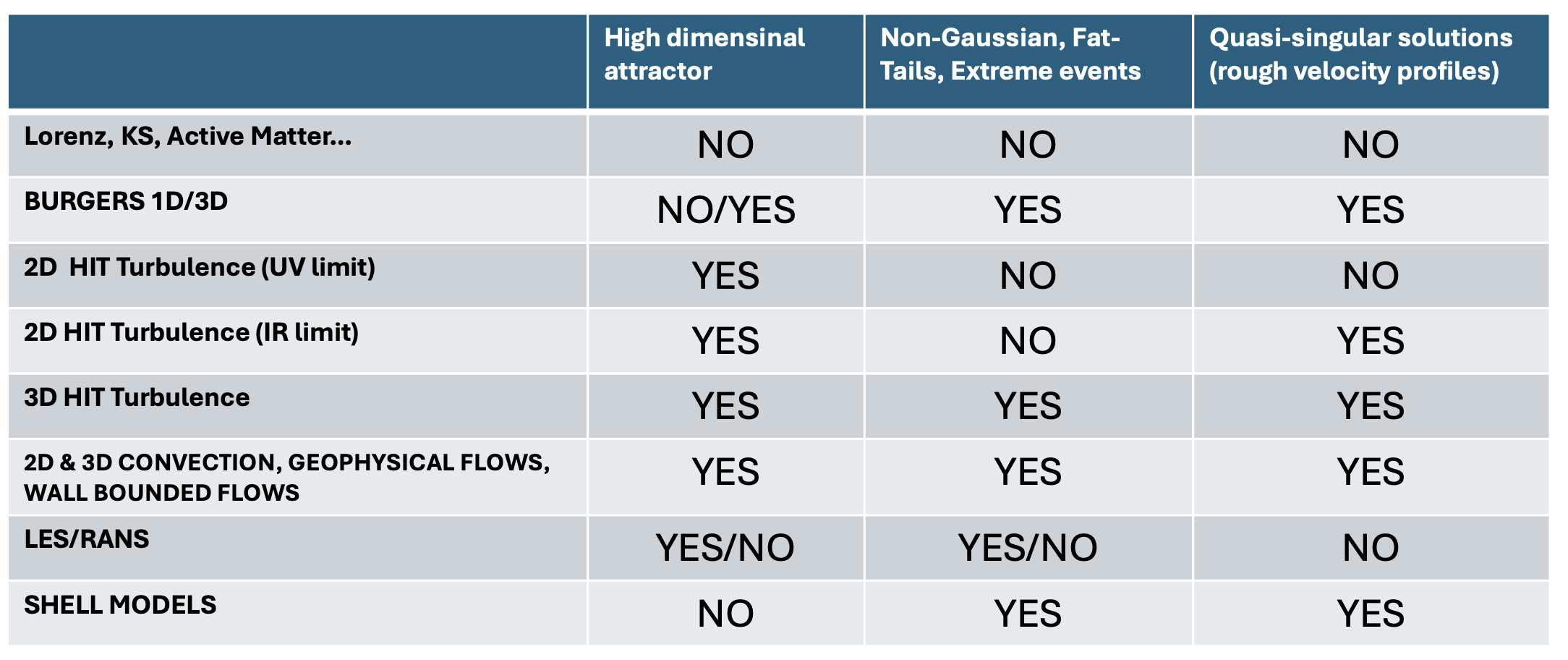}
      \caption{Top. Schematic phase-space defined by increasing physical and data complexity. The dashed curve denotes a heuristic frontier of ML applicability at the time of writing. Its non-convex shape reflects the uneven, problem-specific optimization efforts of the ML community, which have advanced selected applications beyond what their nominal complexity alone would suggest. Problems shown in yellow are those for which ML methods have demonstrated established utility and, in some cases, advantages over conventional approaches. Orange-colored problems correspond to regimes in which ML remains comparatively less explored or has not yet been validated and bench-marked sufficiently for quantitative advancements.  Canonical low-dimensional systems and simplified 1D and 2D flows, including Lorenz and KS dynamics, 1D and 2D Burgers equations, Kolmogorov flow, and 2D HIT in the direct cascade regime form the principal proof-of-concept region. Fully developed 3D HIT, wall-bounded and convective turbulence, and geophysical flows involving rotation, stratification, or multiphase effects occupy the high-complexity region. 2D HIT in the inverse cascade regime is somehow in between due  its reduced complexity wrt 3D case \citep{boffetta2012two}. 3D problems may nevertheless approach the current ML frontier when restricted to reduced-order descriptions, LES or RANS closures, local super-resolution, data assimilation, or small spatial subdomains. Lagrangian turbulence \citep{toschi2009lagrangian} occupies a distinct position because, for a moderate number of tracked particles, its data complexity can remain relatively low even when the physical complexity is comparable to that of 3D HIT. ROMs span a broad range, from low-dimensional Lorenz-type systems to physically complex Shell Models \citep{Frisch,bohr1998dynamical,biferale2003shell}. Bottom: qualitative comparison of selected problem classes with respect to the sources of complexity discussed in Figs. \ref{fig:bigfour1}-\ref{fig:bigfour4}. YES/NO ranking are subjected to specific conditions and do not apply necessarily to all possible flow configurations for each row.}
\label{fig:phasespace}
  \end{figure*}

A better question is: what do we mean by turbulence in this Perspective? Our central message is that machine-learning tools and algorithms developed in other disciplines cannot be transferred uncritically to systems that are multiscale, far from equilibrium, dominated by extreme events, characterized by quasi-singular field realizations, endowed with spontaneous stochasticity, and subject to strongly non-Gaussian fluctuations. These features are central to three-dimensional turbulence \citep{alexakis2018cascades,benzi2023lectures,buaria2026turbulence,dubrulle2019beyond} and also arise in two-dimensional turbulence under specific physical conditions \citep{boffetta2012two}. Indeed,  In this perspective we identify {\bf four big challenges} for data-driven modeling of turbulence: (i)   the huge number of active degrees of freedom, (ii) non-Gaussian intermittent fluctuations and extreme events, (iii) quasi-singular velocity fluctuations, (iv)  and spontaneous stochasticity (see Figures~\ref{fig:bigfour1}--\ref{fig:bigfour4}). They also support a second central message, often overlooked by practitioners entering turbulence from other fields: fully developed three-dimensional turbulence is not merely a quantitatively more difficult version of a low-dimensional chaotic system. It is qualitatively and discontinuously different, belonging to a distinct class of physical and computational complexity. This distinction is important because systems such as the Lorenz model, the Kuramoto--Sivashinsky (KS) equation, Burgers equations, active-matter models, and many other ordinary or partial differential equations are frequently used to validate new algorithms. The success of these systems is then too easily interpreted as evidence that the same methods will remain effective for fully developed turbulent flows.
A particularly careful distinction is required for solutions of the two-dimensional Navier--Stokes equations. In the forward enstrophy-cascade regime, including commonly studied Kolmogorov-flow configurations, solutions are globally regular and display spatially smooth, comparatively simple chaotic dynamics. This regime is structurally different from the strongly intermittent small-scale dynamics of three-dimensional turbulence. In particular, it lacks the multifractal, quasi-singular rough velocity fields characteristic of three-dimensional flows \citep{Frisch}, as well as the associated Eulerian and Lagrangian manifestations of spontaneous stochasticity \citep{lorenz1969predictability,falkovich2001particles,thalabard2020butterfly,bandak2024spontaneous}.
Nevertheless, two-dimensional turbulence can still provide a paradigmatic and demanding test bed for AI methods, provided that the physical regime is selected with care. Particularly relevant examples include the inverse energy cascade, in which rough velocity fields with quasi-singular features emerge \citep{boffetta2012two}, and flows driven or coupled across multiple scales, where a non-trivial ultraviolet, far-from-equilibrium balance develops. Examples include Rayleigh--Benard convection and quasi-geostrophic flows \citep{lohse2010small,valade2025surface}. Thus, the usefulness of two-dimensional turbulence as an AI benchmark depends not simply on dimensionality, but on whether the chosen configuration reproduces the physical complexities  that make fully developed turbulence fundamentally challenging (see a summary if  Fig. \ref{fig:turbulence2d}).
This clarification pushes us to introduce a key question: {\it What is the metric of turbulence? which kind of Loss/Cost must be used to train our models? } The answer is obviously not unique, and in most of the cases the typical $L_2$ loss, at the basis of many ML tools developed for other fields,  based on point-wise displacement between the ground truth and the data-driven guess is the less efficient and the less interesting. This is because the $L_2$ norm is mainly sensitive only to the pick of the energy spectrum, i.e. the energy containing scales, and blind to the intricate multi-scale development of extreme events, and small-scales unpredictable fluctuations. As a result, many claims of {\it AI supremacy} for applications to turbulence risk to be based on uninteresting properties, or -to the very least- to the simplest actor in the game, the slow and smooth evolution of large scale properties. A paradigmatic example for the latter is the claimed supremacy by data-driven tools for weather forecasts, proven to be  well funded for medium scale ad global properties, and less robust for extreme events and rare local fluctuations \citep{zhang2026physics,soliman2024deepmind}. In the sequel, we discuss a few suggestions about how  to identify the  correct (if any) metrics to be used, both for training and validations and for different applications to Eulerian and Lagrangian turbulence. \\
The general questions that we will 
try to address are the following.   Why can generative models be useful in turbulent dynamics? Their primary value being  in learning distributions rather than deterministic correlations, especially when unresolved degrees of freedom remain statistically correlated under the same observations. What does it mean to generate a physically faithful turbulent flow? Here, we will go deep in discussing 
how validation should extend beyond visual quality, mean-squared error, and energy spectra to high order correlations functions, probability-density-function tails, structure functions, intermittency, extreme events, physical  (equations-based) constraints, temporal consistency, and uncertainty quantification.  How (if) physical knowledge should be incorporated? Where possible, strategies include physics-informed loss functions, conditioning, sampling guidance, projection operators, latent-space constraints, and structure-preserving model design.   How to move from the (too) many proof-of-concept studies to trustworthy tools for novel discoveries? Implying the need to develop community-based open-data challenges and benchmarks. \\
A useful metaphor is provided in Fig. \ref{fig:picturbulence}, adapted from the beautiful “Guide de l’alpiniste turbulent” drawn by P. Delache and reproduced in \cite{Frisch}. In the original sketch, the ascent to the Navier–Stokes peak, corresponding to the infinite-Reynolds number limit, could be attempted through several classical routes: the direct mathematical ascent, the renormalization-group path, the experimental route, and the rock-climbing school of numerical simulation. Each route has its own strengths, dangers, and unresolved obstacles. Here, we add a new possible route: an AI helicopter, represented as a black box with a question mark. The metaphor is intentionally ambivalent. On the one hand, modern ML and generative-AI tools may offer shortcuts to regions that are difficult to access with traditional methods, for instance, by reconstructing missing information, super-resolving partial observations, learning stochastic subgrid effects, generating conditional ensembles, or assisting data assimilation in regimes where direct numerical or experimental access is incomplete. On the other hand, the helicopter is not a proof that the summit has been reached. Indeed, as already discussed, it is difficult for AI tools to make {\it jumps} and extrapolate to unchartered heights. Moreover, its flight path is not necessarily transparent, its landing target is uncertain, its success cannot be judged solely by speed or visual plausibility and, as recently said by a prominent Field Medalist "Airdropping someone on the summit of Mount Everest is rather different from climbing it" \citep{Economist2026OpenAIMathematics}.\\
The Perspective is structured as follows. First,   we briefly discuss  a few general issues related to best practices and open problems in using AI for scientific research, related to reproducibility, baselines and biases, including a recent critical assessment  concerning weak practises used by {\it seminal} papers for the case of accelerating PDEs (Sec. \ref{sec:pdebaseline}).
Sections \ref{sec:generative}  and \ref{sec:applications} constitute the two main quantitative components of this Perspective. In the former, we provide a pedagogical introduction to diffusion models, among the most powerful stochastic generative-AI frameworks currently available. We discuss both their continuous- and discrete-time formulations, together with their connections to related approaches, including flow matching, stochastic interpolants, and moment-guided generation. Particular attention is devoted to posterior-guidance techniques for {\it conditional} generation.
These methods are then exploited in the second technical section, devoted to applications, where we provide a bird’s-eye view of some of the most promising directions for their use in Eulerian, Lagrangian, and mesh-free formulations of turbulent flows. 
{Figure~\ref{fig:DM_catsandflows} illustrates the flexibility of generative models across heterogeneous data representations, while the appropriate architecture and validation metrics remain strongly application dependent. For turbulent flow,}  key challenges include reproducing and inferring inertial-range intermittency in Eulerian and Lagrangian settings, for both single- and multiphase flows; generating Lagrangian distributions within mesh-free frameworks; developing subgrid-scale local/non-local and deterministic/stochastic closures; achieving generalization across Reynolds numbers and other control parameters; enforcing physical constraints and enhancing interpretability; and uncovering novel reduced stochastic descriptions. These developments ultimately help delineate the frontier between AI- and equation-based tools in the complex-physics–complex-data phase space (see Fig. \ref{fig:phasespace}).  
Finally, in Sec. \ref{sec:biblio} we will present an updated summary of the bibliometric impact of AI in our field (fluid mechanics and turbulence) and in Sec. \ref{sec:open} a survey of the few attempts to offer  solid open-data infrastructures to advance the deployment of AI tools for our community  and the definition of {\it community Benchmarks} and community {\it  Grand Challenges}. 
\subsection{AI for turbulence: From demonstration to scientific evidence}
The rapid adoption of artificial intelligence in fluid mechanics has been accompanied at the same time by excessive optimism and excessive scepticism. Neither position is particularly useful. Machine-learning models should be regarded as scientific and computational tools whose value depends on the questions they address, the evidence supporting their predictions, and the conditions under which they fail, a sort of novel epistemological way to study Nature, next to physical law and animal brains, as
summarized in the right panel of Fig. \ref{fig:elephant}, and discussed by \cite{mitra2021fitting}. For curiosity, the specific problem of fitting an elephant (see Fig. \ref{fig:elephant}) has recently been solved using 4 parameters \citep{mayer2010drawing}  and even only one \citep{piantadosi2018one}.
Physics seeks parsimonious models with minimal parameters, focusing on strong priors and theoretical laws, following Ockham's razor logic \citep{dubova2025ockham}. Statistics avoid data interpolation, favoring regularization to balance signal and noise. Biology encodes learning through species-specific priors, resulting in efficient training with limited data.  ML contrasts with these, employing over-parameterized models that fit data exhaustively but still generalize effectively. 
For applications to turbulence, the central issue is therefore not whether AI is intrinsically superior or inferior to equation-based approaches, but whether a particular model provides reliable information that cannot be obtained as effectively through established alternatives.
Concerns about the opacity of neural networks are legitimate, but they are not unique to machine learning. The introduction of numerical simulation was likewise accompanied by uncertainty about stability, convergence and the physical significance of computed solutions \citep{goldstine1990remembrance}. Confidence in numerical methods developed through mathematical analysis, controlled experiments, convergence studies and comparisons with theoretical and experimental results. Generative models require a comparable culture of verification. Their outputs should be treated neither as automatically trustworthy because they appear realistic nor as scientifically irrelevant because their internal representations are difficult to interpret.
A generated field may reproduce the large-scale appearance of a reference flow and agree with its energy spectrum while failing to capture scale-by-scale energy transfer, intermittency, velocity-gradient statistics or extreme events. Similarly, a reconstructed trajectory may have a small pointwise error while underestimating the probability of violent accelerations or losing the temporal correlations associated with turbulent transport. Scientific evaluation must therefore be formulated in terms of the physical questions posed by the application. A useful model should make falsifiable predictions for observables that were not explicitly optimised during training and should be tested against conditions that differ meaningfully from those represented in the training data.
Generative models introduce an additional concern because they can produce synthetic fields and trajectories that may be difficult to distinguish visually from simulations or observations. Reproducibility consequently requires clear tags: generated data should be identified as such, and the training data, preprocessing operations, model architecture, sampling procedure and physical constraints should be documented whenever possible. 
The probabilistic character of generative models also creates opportunities unavailable to deterministic predictors. When sparse observations, coarse-grained fields or unresolved degrees of freedom are compatible with multiple turbulent states, the 
scientifically relevant target is not necessarily a single reconstruction. It is the conditional distribution of physically admissible states and the uncertainty associated with quantities that cannot be inferred uniquely. For example, suppose you have partial information along the trajectory of a tracer particles, neither its future evolution, nor the state values in the intermediate gaps are uniquely defined in terms of the gappy past trajectories, depending on the whole past history of the Eulerian velocity field. This makes generative approaches promising for data assimilation, stochastic closure,  and super-resolution. The advantage must be demonstrated against independent strong deterministic, stochastic and physics-based baselines.
The appropriate standard is therefore methodological. Claims should be proportional to the range of flows, Reynolds numbers and observables that have been tested; negative results and failure regimes should be reported; and computational cost should include data generation, training and sampling. These principles motivate the discussion of reproducibility, baselines and biases in the following section and the turbulence-specific validation and benchmark framework developed later in this Perspective.
There is no doubt that there is a  growing trend of over-reliance on artificial intelligence in scientific research, driven by hype and funding incentives \citep{buchanan2024don}. 
A dramatic increase in the use of AI in diverse fields, such as biology, engineering, and medicine, follows an exponential like growth pattern, and the same happened in our field (see Sec. \ref{sec:biblio}).
This rapid adoption appears to be driven more by trends and funding opportunities than by  scientific needs. It is the so called {\it fear-of-missing-out} FOMO phenomenon, where scientists adopt the tool even if with mixed feelings about its efficiency \citep{basu2026scientists}. The high number of citations that many of these papers received is often more the reflection of a trendy phenomenon than a real indicator of short/long term scientific impact, producing another critical aspect in the way bibliometrics  can be used to assess the career of (young) scientists. Furthermore, the natural multi-disciplinary nature of AI tools, "{\it data are data}", foster the blind application of the same specific algorithm to different disciplines, provide the data structure is similar, flooding the research environment with statements of cross-applicability to many different fields completely disconnected from  the specific topic-dependent benchmark.
\subsection{ Reproducibility, baselines and biases  in turbulent flows}
\label{sec:Reproducibility}
Before moving to the more technical sections devoted to Generative AI for turbulence, let us summarize, for the sake of clarifying the novel challenges posed by data-driven research, a set of general questions concerning  reproducibility, baselines and biases  for general research environments and for the specific target of turbulent research \citep{mehrabi2021survey}. Readers eager to explore applications of generative AI to turbulence can skip ahead to Sect.~\ref{sec:generative}.\\
Obviously, reproducibility in turbulence does not require two simulations, experiments or generated samples to produce identical instantaneous fields. Due to spontaneous stochastifcity, turbulent dynamics can amplify even microscopic thermal fluctuations up to the largest scales within a few eddy-turnover times \citep{boffetta2017chaos,bandak2024spontaneous}, while different realizations may nevertheless remain statistically equivalent.
The relevant question is therefore whether independent implementations lead to compatible physical conclusions and reproduce the same statistical laws over the resolved range of scales. As said, for generative models, agreement in visual appearance, pointwise error or low-order moments is insufficient: a model may reproduce the energy spectrum while failing on scale-by-scale fluxes (including back-scatter), anomalous scaling, Reynolds dependency of velocity-gradient statistics or PDF's fat-tail and extreme  events.
General sources of irreproducibility in machine-learning research have been classified by \cite{gundersen2022sources}, including choices related to: {\bf   experimental design},  avoiding P-hacking and  P-fishing; {\bf  algorithms}, specifying  weights initialisation, data shuffling, batch ordering;  {\bf  implementation}, with details about the software used and parallel/serial implementaion, including single/double precision, high-level libraries auto-tuning;  {\bf   observations}, including data set biases and pre-processing ; {\bf   evaluation}, as selective reporting, lack of baselines; and {\bf   documentation}, concerning readability of results, access to data and access to code. These principles remain relevant to fluid mechanics, but turbulence requires additional controls because the data are strongly correlated, multiscale and non-Gaussian. The field-agnostic standards proposed by the REFORMS project \citep{kapoor2023reforms} provide a useful minimum set of best practices. Authors should motivate the use of ML; document data provenance, quality, preprocessing and exclusions; specify training, validation and test splits while guarding against data leakage and hidden correlations; describe models, hyperparameters, training procedures, model-selection criteria and computing infrastructure; provide data, code and reproduction scripts whenever possible; justify the adopted metrics and compare against strong, relevant baselines; report variability, uncertainty and statistical significance; and clearly discuss external validity, limitations and failure cases.  The purpose of the REFORMS Checklist is multifold. For Researchers, to ensure transparency and completeness in reporting, aiding reproducibility and credibility. For reviewers, serve as a guideline for evaluating the rigor of ML-based scientific studies.  For Journals, to help set consistent standards for publishing ML research across different scientific fields. The REFORMS checklist aims to address common errors that occur when applying ML methods in scientific contexts, promoting better practices for robust and reliable ML-based research  which should be supplemented by the following turbulence-specific checklist.
\subsection{ A turbulence-specific checklist for generative modelling}
\noindent {\sc Define the physical question and justify generation.} State why a deterministic/probabilistic generative model is required.  Generative modelling is most strongly justified when unresolved variables or incomplete observations admit multiple physically plausible solutions.\\
{\sc Document the turbulent ensemble.} and report the governing equations, geometry, forcing, boundary conditions, Reynolds number, and other relevant dimensionless parameters. Specify the numerical scheme or experimental procedure, spatial and temporal resolution, filtering, sampling frequency, statistical stationarity, and resolution criteria. Training data should be accompanied by sufficient provenance to determine which physical ensemble they represent. Always give physical hyperparameters defining the numerical or experimental  set-up used to generate the training data in terms of dimensionless quantities, normalized with proper physical time and length scales, typical of the specific turbulent problems.\\ 
{\sc Prevent data leakage and hidden correlations;} Turbulent snapshots, spatial patches, and Lagrangian trajectory segments may remain strongly correlated even when treated as separate samples. Training, validation, and test sets should therefore be separated by independent realizations, sufficiently distant time intervals, spatial regions, or physical control parameters, according to the application. Random splitting of neighboring snapshots or overlapping patches is generally inadequate. \\
{\sc Report preprocessing as part of the physical model.} Normalisation, interpolation, coarse-graining, spectral truncation, filtering, patch extraction, temporal subsampling and treatment of outliers can alter the physical content of the learning problem. These operations must be defined explicitly and applied consistently. For super-resolution and subgrid reconstruction, the operator producing the coarse field must be specified because different filters define different inverse problems. \\
{\sc Use strong and relevant baselines.} Comparisons should include both established physics-based methods and simpler statistical controls. Depending on the task, suitable baselines may include, e.g.,  interpolation, Gaussian random fields with matched spectra, POD or Gappy-POD, wavelet representations, classical stochastic models, established LES closures, data-assimilation methods and deterministic neural networks.  \\
{\sc Evaluate a hierarchy of turbulence metrics.} 
Image-quality metrics borrowed from computer
vision such as the Mean Squared Error (MSE), Peak-signal-to-noise-ratio (PSNR), Structural Similarity
Index Measure (SSIM) or Learned Perceptual Image Patch Similarity (LPIPS) can
be useful diagnostic tools but are generally insufficient for turbulence.
Validation should progress from low-order to increasingly discriminating quantities. These may include mean fields, variances and spectra; scale-by-scale transfers and fluxes; structure functions and anomalous exponents; probability-density-function tails of increments, gradients, accelerations, dissipation and enstrophy; coherent-structure statistics; temporal correlations; conservation laws; and, where appropriate, irreversibility or Lagrangian dispersion. The selected metrics should be tied to the scientific claim and not reported as an undifferentiated collection (Fig. \ref{fig:metrics}). \\
{\sc Test memorisation vs generation. } Report sample diversity, convergence of ensemble statistics, sensitivity to sampling parameters and uncertainty calibration. Generated ensembles should reproduce both the variability of the reference data and the conditional uncertainty left unresolved by the observations. Memorisation tests should compare generated samples with their nearest training examples at multiple scales and for physically meaningful observables (Sect. \ref{sec:mem_gen}).\\
{\sc Test interpolation vs extrapolation.} Distinguish interpolation within the training ensemble from generalization across Reynolds numbers, forcing mechanisms, geometries, boundary conditions, observation operators, and numerical-to-experimental domains. Out-of-distribution tests should be specified before model selection whenever possible, and failure outside the training range should be reported explicitly. \\
{\sc Report uncertainty, variability and computational cost.} Results should include variability across training runs, data subsets and generated ensembles, together with confidence intervals where meaningful. Computational comparisons should account for data generation, preprocessing, training, and sampling, not inference time alone, and should report hardware, memory use, parameter count, and numerical accuracy. \\
{\sc Make the workflow inspectable and preserve failure cases.} Release data, code, trained parameters, preprocessing routines and evaluation scripts whenever possible, with persistent identifiers and version information. Since software links may become unavailable over time, their stability should be considered when archiving research outputs. When full release is impossible, the unavailable components and resulting limitations should be stated. Every study should identify at least one regime, observable or condition for which the proposed model fails or remains unvalidated.
\subsection{A paradigmatic test case: accelerating PDE solvers}
\label{sec:pdebaseline}
The use of machine learning to accelerate partial-differential-equation solvers provides a useful example of how weak baselines and incomplete cost accounting can produce exaggerated claims. In a critical assessment of this literature, \cite{mcgreivy2024weak} found that approximately 80\% of the reviewed studies employed weak numerical baselines (including articles with thousands of citations as the application to 2D turbulence by \cite{kochkov2021machine}), while 95\% emphasized positive results and comparatively few reported negative findings. Although this analysis was not restricted to generative models or only to fully developed turbulence, its conclusions are directly relevant to the evaluation of data-driven fluid-mechanics methods.
Claims of computational advantage are meaningful only when accuracy, resolution and physical fidelity are controlled. Comparisons should use state-of-the-art solvers configured for the problem under consideration, not general-purpose or outdated implementations. A specialised neural model trained for one equation, geometry and parameter range should not be compared uncritically with a numerical library designed for robustness and generality. Hardware, software optimisation and numerical tolerance must also be reported because apparent speed-ups may depend strongly on implementation.
Cost accounting should include the generation and storage of training data, model optimisation and inference. For generative models, sampling cost and the number of samples required to converge ensemble statistics must also be considered. Conversely, comparisons with equation-based approaches should distinguish the cost of developing a solver from the marginal cost of producing an additional solution. No single efficiency measure is therefore sufficient across all applications.
More importantly, computational speed is not itself evidence of scientific progress. In turbulence, an accelerated surrogate that reproduces low-order quantities but fails on the most important finger-prints may be unsuitable even when it is substantially faster. Performance must be evaluated at fixed physical accuracy using observables appropriate to the intended application.
The broader lesson is that machine-learning methods should be assessed against strong numerical, statistical and phenomenological alternatives, with both successes and failures reported. 
The current literature is probably oriented  toward an overstatement of the success of ML tools due to biased comparisons and reporting (see also the bibliometric analysis shown in Sec.\ref{sec:biblio}).  As discussed above, this could be induced by the use of weak baselines (inefficient or outdated comparison methods), leading to inflated claims about ML efficiency, often caused by the fact that there is a {\it lack of domain expertise}, i.e. many papers on ML for fluids-and also for turbulence- are published by  researchers outside the field which may not always have the background to select the most efficient numerical methods as baselines. Furthermore,
the problem of {\it Limited Benchmarking} exists,   the community has not always focused on establishing benchmarks, leading to inconsistent standards for evaluating ML methods (see Sec. \ref{sec:bestpractise}). \\
 Among the possible remedies to this very important problem, we mention:   ML journals and conferences could adopt registered reports (where researchers write their research question, hypotheses, methods, and analysis plan before running the experiments), and the  study designs are peer-reviewed before results are known, reducing the bias toward positive findings.
	Challenge Problems: Funding agencies could sponsor challenge problems that are well-defined and have clear evaluation criteria. This could direct ML research toward meaningful scientific problems and ensure fairer benchmarking.
	Awareness Campaigns: Promote cultural change and awareness of reproducibility issues and the importance of strong baselines among ML researchers.
	Internal Controls: research groups should develop practices to ensure unbiased reporting, such as discussing baseline selection openly in publications.
 	Education: departments and universities should include discussions on reproducibility and fair comparison practices in their ML and numerical methods courses.

%% file: chapters/GenerativeAI.tex
\section{Generative AI}
\label{sec:generative}
Sections 2 and 3 constitute the technical core of this Perspective. In Sec. 2, we introduce the main concepts underlying modern generative AI, with particular emphasis on score-based and diffusion models, their continuous- and discrete-time formulations, and related approaches such as flow matching, stochastic interpolants, and posterior guidance. We also discuss how physical constraints can be incorporated and address the central distinction between memorization and genuine generalization. Section 3 then turns to turbulence applications, illustrating how these tools can be used for Eulerian, Lagrangian, and mesh-free generation, data assimilation and reduced-order modelling, anomaly detection, and reinforcement learning. Throughout, the emphasis is not on specific architectures, which evolve rapidly, but on the physical questions, observables, and validation criteria that determine whether a generative model is genuinely useful for turbulence.
The ultimate objective of a generative model is to learn the statistical structure of a complex system and subsequently generate new realizations that are statistically indistinguishable from independent observations of the same system. In the context of fluid mechanics, such realizations may correspond to velocity fields, pressure distributions, coherent structures, reduced-order representations, or entire spatio-temporal flow evolutions.
This objective is particularly challenging because turbulent flows are intrinsically high-dimensional and with complex multi-scale physics (see Figs. \ref{fig:bigfour1}:\ref{fig:bigfour4}).  As a consequence, directly characterizing the underlying data distribution is hindered by the well-known curse of dimensionality: the number of samples required to accurately resolve a probability density generally grows exponentially with the dimension of the state space~\citep{scott2015multivariate}.
\noindent
However, advances in experimental diagnostics, high-performance computing, and data acquisition technologies have led to an unprecedented increase in the quantity and quality of available fluid-dynamics data, as will be discussed in Sec. \ref{sec:open}. High-fidelity direct numerical simulations, large experimental campaigns, remote sensing observations, and geophysical databases now routinely produce datasets of a size that was unimaginable only a decade ago. Such developments have motivated the exploration of black-box data-driven approaches capable of learning and the statistical information contained in these large datasets and provide for novel sampling in those  applications where  the governing equations may be prohibitively expensive to solve, and experimental measurements may be sparse, incomplete, or altogether impossible to obtain. This situation arises, for example, in atmospheric and oceanic flows, environmental monitoring, astrophysical systems, and other complex multiscale systems. In such settings, generative models offer the possibility of constructing statistically consistent flow realizations directly from the observed data, without any need of physical prior.
Examples include reconstructing missing measurements, inferring unresolved flow structures, generating physically plausible ensemble realizations, super-resolving coarse observations, quantifying uncertainty, and exploring rare or poorly sampled events. \\
 Before discussing specific algorithms, it is useful to formulate the generative modeling problem in a precise mathematical framework and to review the principal ideas that have led to modern generative models. \\

\subsection{Generative Models - background formulation} Let us consider a dataset composed of $N$ independent samples drawn from an unknown probability distribution, $p_{\mathrm{data}}(x)$:
\begin{equation}
\{a^\mu\}_{\mu=1}^{N},
\qquad
a^\mu \in \mathbb{R}^{d},
\qquad
a^\mu \sim p_{\mathrm{data}}(x).
\label{eq:dataset}
\end{equation}

\noindent
The objective of generative models is to learn the statistical structure of the dataset and subsequently sampling  new realizations that are statistically consistent with the training data. Formally, one introduces a parameterized probability distribution
\begin{equation}
p_\theta(x),
\end{equation}
where $\theta$ denotes a set of trainable parameters, and seeks to approximate the unknown data distribution $p_{\mathrm{data}}(x)$.
Generative modeling can therefore be viewed as a problem of distribution matching. The target is to determine $\theta$ such that $p_\theta(x)$ reproduces the statistical properties of $p_{\mathrm{data}}(x)$ while remaining amenable to efficient sampling.

\noindent
A common measure of discrepancy between two probability distributions is the Kullback--Leibler (KL) divergence,
\begin{equation}
\label{eq:DKL}
D_{\mathrm{KL}}
\left(
p_{\mathrm{data}}
\,\Vert\,
p_\theta
\right)
=
\int
p_{\mathrm{data}}(x)
\log
\frac{p_{\mathrm{data}}(x)}
     {p_\theta(x)}
\,dx.
\end{equation}
The KL divergence is non-negative and vanishes only when the two distributions coincide. Consequently, many generative learning algorithms can be interpreted as seeking a parameterized distribution that minimizes the KL divergence, or a related statistical discrepancy, with respect to the data distribution.\\

\noindent
A general way to represent a probability distribution is through an energy-based model,
\begin{equation}
p_\theta(x)
=
\frac{\exp\!\left[-E_\theta(x)\right]}
     {Z_\theta},
\end{equation}
where $E_\theta(x)$ is an energy function that contains the dependence on the learning parameters $\theta$, and
\begin{equation}
Z_\theta
=
\int
\exp\!\left[-E_\theta(x)\right]
\,dx
\end{equation}
is the normalization constant, commonly referred to as the partition function.

\noindent
This formulation is closely related to equilibrium statistical mechanics and is extremely flexible: in principle, an appropriately chosen energy function can represent an arbitrarily complex probability~\citep{lecun2006tutorial}. However, the normalization constant $Z_\theta$ is generally intractable in high-dimensional spaces ($d \gg 1$). 
As a consequence, evaluating the log-likelihood of a generic sample $x$,
\begin{equation}
\label{eq:likelihood}
\log p_\theta(x)
=
-E_\theta(x)-\log Z_\theta,
\end{equation}
and the computation of its gradients with respect to the model parameters, as required for likelihood-based training, are generally computationally prohibitive.\\

\noindent
The construction of generative models is complicated by another major obstacle: the probability distribution associated with the underlying process is typically unknown and accessible only through a finite number of realizations. In turbulence, in particular, the stationary data distribution is not known in closed form, nor is there a reduced description of the corresponding effective energy, $E(x)=-\log p_{\mathrm{data}}(x)+\mathrm{const}$, in terms of a small set of physically meaningful observables. The historical evolution of modern generative modeling can be interpreted largely as a sequence of strategies designed to overcome these two challenges while retaining the ability to generate realistic samples from high-dimensional distributions~\citep{bond2021deep}.\\

\noindent
The first major class of approaches addresses the normalization problem by imposing architectural constraints that make probability densities explicitly tractable (see Fig. \ref{fig:DM_flowpdf}). 
Representative examples include {\bf Variational Autoencoders} (VAEs)~\citep{kingma2013auto, salimans2015markov} and {\bf Normalizing Flows}~\citep{rezende2015variational}.
Variational autoencoders introduce latent variables and optimize a variational lower bound on the data likelihood, using a MSE for the reconstruction term and a KL divergence for the latent space distribution. 
Their ability to represent turbulent fluctuations therefore depends on the latent representation, decoder expressiveness (modeled as a Gaussian) and training objective.
{Discrete} Normalizing Flows adopt a fundamentally different strategy. They represent the target probability density through a sequence of deterministic changes of variables that transform a simple reference distribution, typically a Gaussian, into the data distribution. This formulation provides exact likelihood evaluation through the change-of-variables formula and is, in principle, capable of representing arbitrarily complex probability distributions. However, every transformation must remain analytically invertible with a tractable Jacobian determinant, imposing strong architectural constraints that substantially restrict the class of admissible models.
\noindent
A second class of approaches circumvents density estimation altogether. {\bf Generative Adversarial Networks} (GANs)~\citep{goodfellow2014generative} formulate generation as a competition between two neural networks: a generator and a discriminator. The generator maps samples drawn from a simple latent distribution, standard Gaussian noise, directly into the data space, with the aim of producing realizations that are indistinguishable from the training data, and hence distributed according to the data distribution. The discriminator instead receives both real and generated samples and is trained to classify whether a given realization originates from the data distribution or from the generator. The two networks are optimized adversarially: as the discriminator becomes better at detecting synthetic samples, the generator is simultaneously trained to confuse it. In this way, GANs do not require the explicit evaluation of a probability density or likelihood; the target data distribution is learned indirectly through the discriminator's ability to distinguish generated samples from real observations.
GANs have achieved remarkable success for complex data generation, including some attempts for turbulence~\citep{xie2018tempogan, li2023jfm, kim2024prediction}, although training can be unstable and quantitative likelihood evaluation is generally unavailable.

\noindent
More recently, a third class of methods has emerged that combines ideas from stochastic processes, non-equilibrium statistical mechanics, and machine learning~\citep{sohl2015deep, ho2020denoising}. These approaches, known as {\bf score-based generative models} or {\bf diffusion models} (DM), do not attempt to learn the probability density of the data itself. Instead, they learn its logarithmic gradient, commonly referred to as the \emph{score-function}:
$$
s(x) = \nabla_x \log p_{\mathrm{data}}(x) 
$$

\noindent
The key observation is that the score is independent of the normalization constant in (\ref{eq:likelihood}), consequently, it can be learned without ever evaluating the partition function $Z_\theta$. Once the score has been estimated, it can be used to construct a stochastic model that transforms simple Gaussian random noise into samples drawn from the target distribution (see Fig. \ref{fig:DM_flowpdf}), as will be detailed in Section \ref{sec:score-based}. 
\noindent
Recent developments have further extended the score-based paradigm toward more general formulations based on probability transport. In particular, approaches such as {\bf Flow Matching} (Sec. \ref{sec:flowmatching})~\citep{lipman2022flow} can be viewed as a natural evolution of diffusion- and score-based generative models. In Flow Matching one seeks to directly learn the {\it velocity} governing the transport of probability mass  (not to be confused with the physical velocity of the turbulent flow) in the high-dimensional data space from a simple reference distribution to the target distribution~\citep{lipman2022flow}. As a consequence, generation is formulated in terms of deterministic transport dynamics. These emerging approaches highlight an increasingly unified perspective in which generative modeling is interpreted as the problem of learning and controlling probability flows in high-dimensional spaces.
Without dwelling on the mathematical details, we can summarize similarities and differences as follows (see also Fig. \ref{fig:DM_flowpdf}). VAEs, GANs, and Normalizing flows construct mappings between a simple latent distribution, from which sampling is performed, and the target data distribution. In VAEs and GANs, this transformation is represented by a deep nonlinear neural network mapping the latent variables to data space. Normalizing flows instead construct the transformation as a sequence of explicitly invertible mappings, whose functional form is prescribed while their parameters are learned from the data.
On the other hand, score based, DM and Flow matching can be recast as special cases of stochastic interpolants, where the map connecting initial and final distributions is defined via  a continuum  process. 
The connection between this last class, which represents the state-of-the-art in generative modeling, and non-equilibrium statistical mechanics will be the focus of the remainder of this chapter. Three fundamental questions must be addressed: (1) How can the score function be learned directly from data? (Sec.~\ref{sec:score-based}). (2) Once learned, how can it be used to generate new samples? (Sec.s~\ref{sec:backward_proc} and~\ref{sec:diffusion_models}). (3) How can we distinguish memorization (overfitting) from poor and/or good generalization? (Sec.~\ref{sec:mem_gen}). To answer these questions, we first review the stochastic processes that form the foundation of modern score-based generative models.

\subsection{Continuous-time score-based generative modelling}
\label{sec:score-based}
\begin{figure*}
      \centering      \includegraphics[width=1.\linewidth]{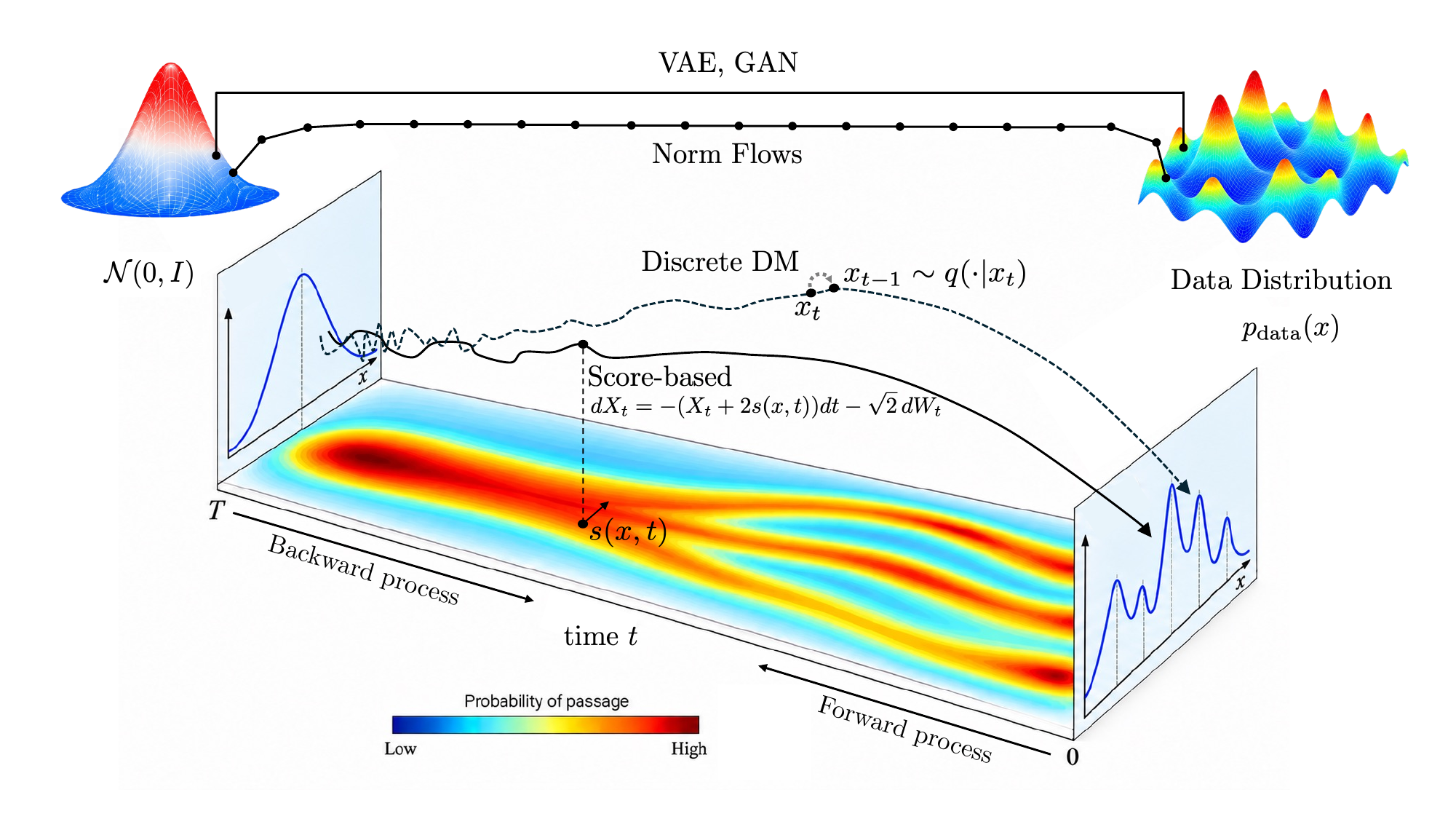}
      \caption{Unified view of generative modelling as probability transport in high-dimensional data space. The target data distribution, $p_{\mathrm{data}}(x)$, shown on the right as a highly structured and multimodal probability landscape, is connected to a simple reference distribution, typically an isotropic Gaussian, shown on the left. Different generative architectures realize this connection in distinct ways. {\bf Variational Autoencoders} (VAE) and {\bf Generative Adversarial Networks} (GAN) construct an implicit mapping between latent and data spaces, whereas {\bf Normalizing Flows} use a sequence of deterministic and invertible transformations, allowing the probability density to be tracked explicitly along the transport path. {\bf Diffusion models}  instead introduce a progressive stochastic deformation of the data distribution: in the forward direction (from right to left), probability mass is gradually dispersed until the structured data distribution approaches Gaussian noise; in the generative, reverse -backward-  direction (form left to right), samples initialized from the Gaussian reference are transported back toward the data manifold. In discrete-time diffusion models, this evolution is represented by a finite sequence of noisy states, $ x_t$, while continuous-time {\bf Score-based models} describe it through a stochastic differential equation. The coloured surface at the bottom provides a schematic representation of the probability density encountered along the transport process. At each intermediate time, the score field, $s(x,t)$ points toward regions of increasing probability and supplies the additional drift required to reverse the diffusive spreading. Learning this time-dependent vector field therefore makes it possible to steer -during the backward process- random trajectories from the simple reference measure toward the multiple high-probability modes of the target distribution without explicitly evaluating the generally intractable normalization constant of the underlying density. {\bf Flow-matching} and related probability-flow formulations offer a deterministic counterpart to this construction by learning directly the velocity field that transports probability mass between the same endpoints. The trajectories displayed in the figure should thus be interpreted as alternative stochastic or deterministic paths in probability space.}
      \label{fig:DM_flowpdf}
  \end{figure*}

The final aim of score-based models is to generate samples from $p_{\mathrm{data}}(x)$ by starting from pure noise and evolving it through a continuous sequence of infinitesimal denoising steps. This is not possible without a suitable drift that progressively guides the dynamics toward the target data distribution. To determine this guidance, one first defines a stochastic evolution that performs the opposite transformation, the so-called {\bf forward process}: starting from the training samples $\{a^\mu\}$, progressively increasing amounts of noise are added until the resulting distribution becomes approximately indistinguishable from a simple Gaussian distribution (see Fig.~\ref{fig:DM_flowpdf}). 
This direction is straightforward to construct and corresponds to moving from right to left in Fig.~\ref{fig:DM_flowpdf}.

The second step consists in learning how to reverse this evolution during the {\bf backward process}. The reverse dynamics is guided by the score, $\nabla_x \log p(x,t)$, of the time-dependent probability distributions generated along the forward process. Once this score is learned, it provides the additional drift required to progressively transform Gaussian noise back into samples distributed according to $p_{\mathrm{data}}(x)$, corresponding to moving from left to right in Fig.~\ref{fig:DM_flowpdf}.
More precisely, the data distribution $p_{\mathrm{data}}(x)$ is embedded as one endpoint of a continuous trajectory of probability densities, ${p(x,t)}_{t\in[0,T]}$, whose other endpoint is a simple reference Gaussian distribution:
\begin{equation}
p(x,0)=p_{\mathrm{data}}(x),
\qquad
p(x,T)\approx \mathcal{N}(0,I),
\end{equation}
{where time, $t$, here has nothing to do with the  physical time of the fluid evolution: it is just a label used to parametrise the evolution of $p(x,t)$  in the space of admissible data distributions}, 
see also Fig.~\ref{fig:DM_forward_backward} for a specific example of forward and backward dynamics applied to 2d passive scalar turbulent realization.

\subsubsection{Forward process}
The forward diffusion process is described by the stochastic differential equation
\begin{equation}
\begin{cases}
dX_t=f(X_t)\,dt+g(t)\,dW_t,\\[0.3em]
X_0\sim p_{\mathrm{data}}.
\end{cases}
\label{eq:forward_sde}
\end{equation}
where the initial state $X_0$ is taken as one of the data distribution samples $p_{\mathrm{data}}$, and  $W_t$ denotes a standard Wiener process, and $g(t)$ is the diffusion coefficient controlling the amplitude of the stochastic term. Its time dependence can be chosen, or optimized, to define different noise schedules and control how rapidly the data distribution is transformed into the noise distribution.
The corresponding marginal probability density of $X_t$, denoted by $p(x,t)$, evolves according to the Fokker-Planck equation
\begin{equation}
\begin{cases}
\displaystyle
\partial_t p(x,t)
=
-\nabla_x\cdot
\bigl(
f(x)p(x,t)
\bigr)
+
\frac{1}{2}
\nabla_x\cdot
\bigl(
g^2(t)\nabla_x p(x,t)
\bigr),\\[0.3em]
p(x,0)=p_{\mathrm{data}}(x).
\end{cases}
\label{eq:fokker_planck}
\end{equation}
\noindent
Equivalently, the evolution may be characterized by the transition density $p(x,t\,|\,x_0,0)$, namely the probability density that $X_t=x$ given that $X_0=x_0$. The two quantities are related through
\begin{equation}
p(x,t)
=
\int
p(x,t\,|\,x_0,0)\,
p_{\mathrm{data}}(x_0)\,
dx_0.
\label{eq:marginal_transition}
\end{equation}

\noindent
A common choice is the Ornstein-Uhlenbeck (OU) process,
\begin{equation}
f(x)=-x,
\qquad
g(t)=\sqrt{2},
\end{equation}
which converges toward a Gaussian equilibrium independently of the initial condition. The forward dynamics therefore progressively destroys the detailed structure of the data distribution while preserving a tractable stochastic evolution (see top row of Fig. \ref{fig:DM_forward_backward}).

\noindent
An important feature of the OU process is that its transition density can be obtained analytically. Conditioned on the initial state $X_0=x_0$, the random variable $X_t$ is Gaussian distributed,
\begin{equation}
p(x,t\,|\,x_0,0)
=
\mathcal{N}
\Bigl(
e^{-t}x_0,
\,(1-e^{-2t})I
\Bigr),
\label{eq:ou_transition}
\end{equation}
or, equivalently,
\begin{equation}
X_t
=
e^{-t}x_0
+
\sqrt{1-e^{-2t}}\,\varepsilon,
\qquad
\varepsilon\sim\mathcal{N}(0,I).
\label{eq:ou_sampling}
\end{equation}
As $t\rightarrow\infty$, the dependence on the initial condition is progressively lost and the distribution converges toward the stationary Gaussian distribution $\mathcal{N}(0,I)$. Obviously, even if the transition density $p(x,t\,|\,x_0,0)$ can be computed analytically, the marginal density in Eq.~(\ref{eq:marginal_transition}) remains unknown because it depends on the unknown data distribution $p_{\mathrm{data}}(x)$. Nevertheless, knowledge of the explicit form of the forward transition density in Eq.~(\ref{eq:ou_transition}) plays a key role in the training procedure, as it allows us to evaluate analytically the conditional score
\begin{equation}
\label{eq:score_cond}
\nabla_x \log p(x,t\,|\,x_0,0).
\end{equation}

\subsubsection{Backward process}
\label{sec:backward_proc}
As said, the generative problem consists of reversing the forward deformation of the probability density.
  \begin{figure*}
      \centering      \includegraphics[width=1\linewidth]{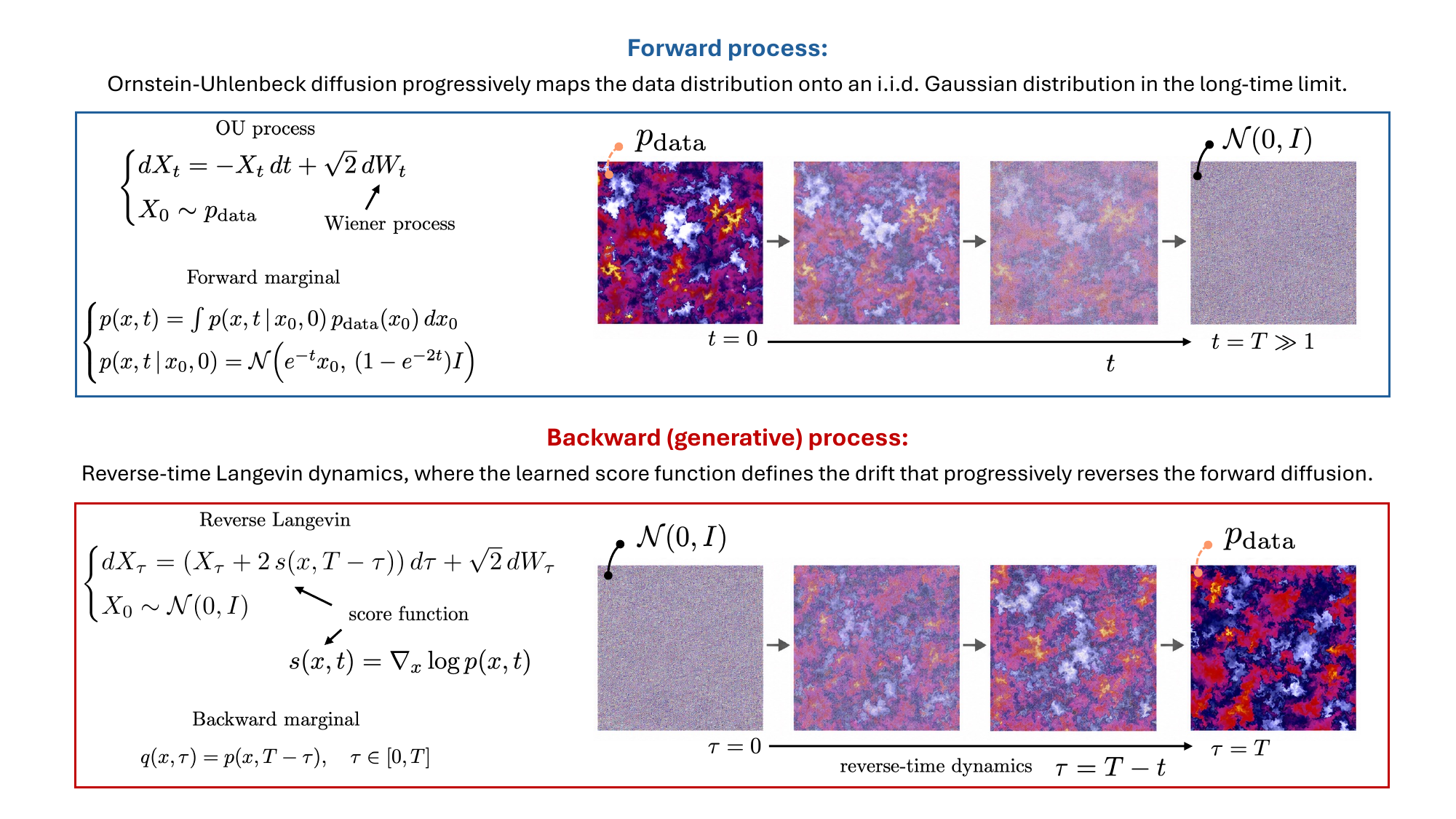}
      \caption{Schematic representation of the forward and backward processes underlying score-based generative models.
      \textit{Top:} The forward Ornstein-Uhlenbeck process progressively perturbs samples drawn from the data distribution (here taken from a passive scalar advected by a 2d inverse cascade turbulent flow, courtesy of C. Calascibetta), $p_{\mathrm{data}}(x)$, driving the corresponding marginal probability density $p(x,t)$ toward the attracting Gaussian distribution $\mathcal{N}(0,I)$ in the long-time limit. The forward transition probability, $p(x,t|x_0,0)$,  is analytically known, while the marginal distribution is obtained by averaging over the unknown data distribution.
      \textit{Bottom:} The generative process follows the corresponding reverse-time Langevin dynamics, initialized from the Gaussian prior. If learning succeed, the learned score function, $s_\theta(x,t)\approx \nabla_x\log p(x,t)$, enters the drift and drives the stochastic dynamics through the sequence of forward marginals in reverse order, ultimately recovering the target data distribution.}
      \label{fig:DM_forward_backward}
  \end{figure*}
The objective is not to reverse individual stochastic trajectories, but  to construct a new stochastic process whose probability distributions retrace the same path in probability space in the opposite direction (see bottom row of Fig. \ref{fig:DM_forward_backward}). To this end, we introduce the backward time variable $\tau$,
\begin{equation}
\tau=T-t,
\qquad
\partial_\tau=-\partial_t.
\end{equation}
The time-evolving probability density associated with the backward process, $\{q(x,\tau)\}_{\tau\in[0,T]}$, is defined so that its marginal distributions coincide with those of the forward process in reverse temporal order. In other words,
\begin{equation}
q(x,\tau)=p(x,T-\tau),
\qquad
\tau\in[0,T],
\label{eq:time_reversed_density}
\end{equation}
so that the matching of probability densities is enforced at every instant of time and not only at the endpoints. 

\begin{figure*}
      \centering      \includegraphics[width=1.0\linewidth]{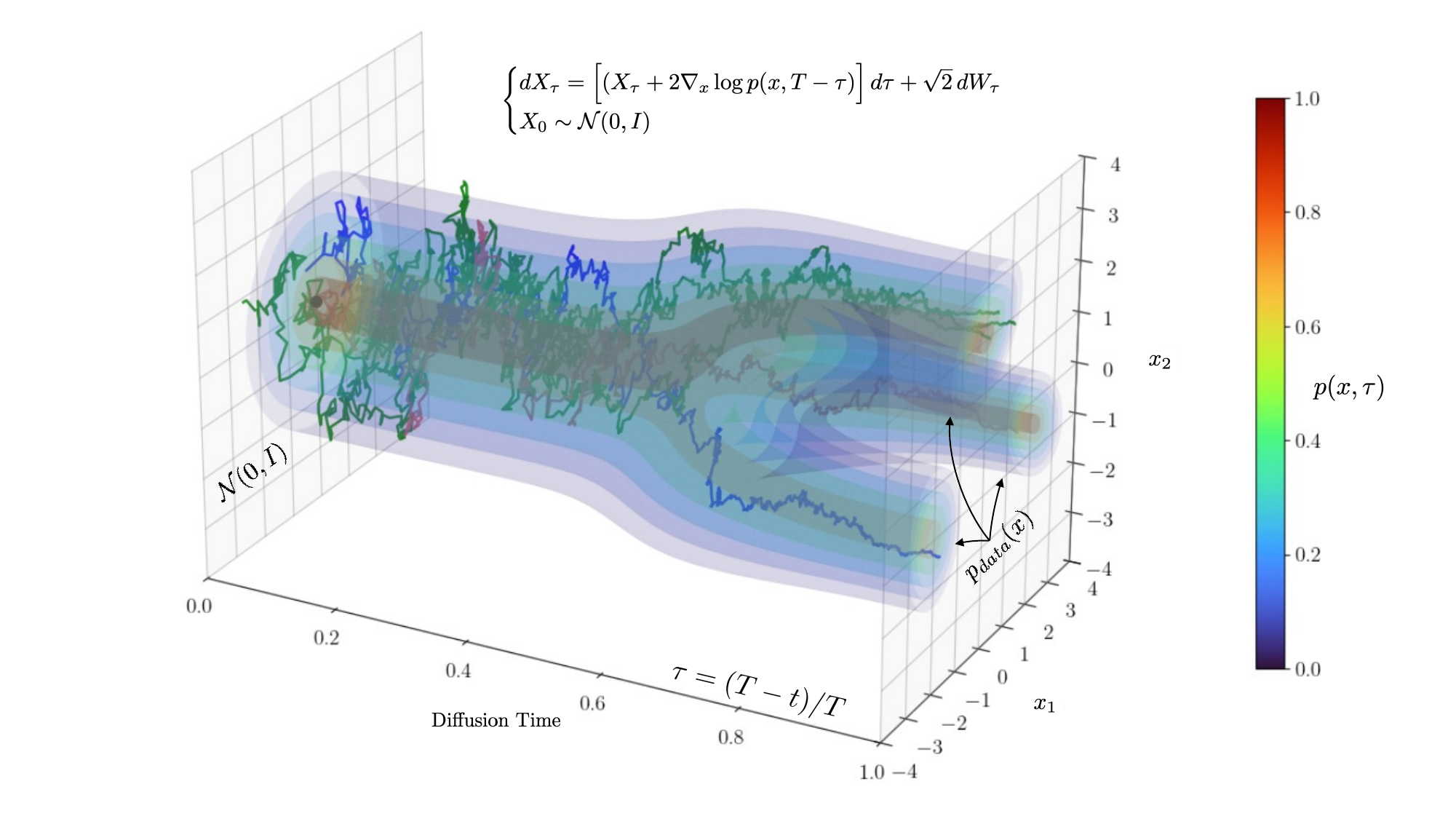}
      \caption{Schematic representation of the reverse diffusion process in probability space. The colored surfaces represent the time-dependent marginal probability density $p(x,t)$ along the diffusion path, interpolating between the Gaussian prior $\mathcal{N}(0,I)$ and the target data distribution $p_{\mathrm{data}}(x)$. Starting from independent realizations sampled from the Gaussian prior, the reverse-time Langevin dynamics generates stochastic trajectories that progressively follow the evolving probability landscape toward regions of high probability of the data distribution.}
      \label{fig:DM_trajectories}
  \end{figure*}

\noindent
A remarkable result due to \cite{anderson1982reverse} states that such a reverse process is itself a diffusion process, provided that its drift is appropriately modified. The resulting reverse-time dynamics is governed by
\begin{equation}
\begin{cases}
dX_\tau
=
\Bigl[
-f(X_\tau)
+
g^2(T-\tau)
\nabla_x\log p(x,T-\tau)
\Bigr]\,d\tau
+
g(T-\tau)\,dW_\tau, \\[0.3em]
X_0\sim\mathcal{N}(0,I),
\end{cases}
\label{eq:reverse_sde}
\end{equation}
where, to simplify the notation, the dependence of the score fields on the current process realization is left implicit, i.e., $\nabla_x\log p(x,T-\tau) = \nabla_x\log p(x,T-\tau)|_{x=X_\tau}$,  and $g(T - \tau)$ takes the same value of the noise schedule used in the forward process at time $t$. Exact recovery of the forward marginals requires the exact score and initialization from the forward terminal distribution, which is reached only $T \to \infty$; using  a standard Gaussian as input for the backward process in (\ref{eq:reverse_sde}) with finite $T$ introduces an additional approximation. Equation~(\ref{eq:reverse_sde}) constitutes the fundamental result underlying score-based generative modeling. Relative to the forward dynamics, the reverse process contains an additional drift term proportional to the score function,
\begin{equation}
\label{eq:scoreDM}
    s(x,t)=\nabla_x\log p(x,t),
\end{equation}
where $p(x,t)$ is the marginal probability density generated by the forward diffusion process as defined in Eq.~(\ref{eq:marginal_transition}). This is precisely the additional drift required to ensure that the marginal distributions of the reverse process satisfy the matching condition~(\ref{eq:time_reversed_density}) throughout the evolution.
\noindent
Whereas the forward diffusion progressively disperses the probability mass and smooths the data distribution, the score term steers the reverse dynamics toward regions of higher probability under the forward marginal distribution. In doing so, it counteracts the effects of diffusion and reconstructs the prescribed sequence of marginal distributions in reverse order.
For the OU process, the forward drift corresponds to a relaxation toward the minimum of a quadratic potential, while the score contribution biases trajectories toward the regions of state space where the probability density is largest. See Fig.~\ref{fig:DM_trajectories} for a schematic example in which two-dimensional trajectories evolve under the backward process, progressively mapping samples from an initial Gaussian distribution onto a multivariate target distribution, while the shaded regions denote the time-dependent marginal probability density $p(x,t)$ throughout the process.
The combined action of these two terms generates a stochastic dynamics whose marginals evolve backward along the path defined by the forward process.

\noindent
How to learn the score (\ref{eq:scoreDM}) is addressed in the sections below through {\it score matching}, which provides a practical strategy for estimating the score directly from data without requiring explicit knowledge of the probability density. Once the score field has been learned, sample generation proceeds by numerically integrating the reverse-time stochastic differential equation~(\ref{eq:reverse_sde}) using the Euler-Maruyama scheme.

\subsubsection{Learning the score function}
The objective is to introduce a neural network, denoted $s_\theta(x,t)$, and determine its parameters $\theta$ such that it accurately reproduces the score field (\ref{eq:scoreDM}). A natural choice for the corresponding loss function is the mean-squared error
\begin{equation}
\int
p(x,t)
\left|
s_\theta(x,t)
-
\nabla_x \log p(x,t)
\right|^2
\,dx.
\label{eq:score_matching}
\end{equation}
In the above expression, the expectation with respect to the marginal distribution $p(x,t)$ can be estimated empirically by Monte Carlo sampling obtained by drawing data points from the training set, $\{a^\mu \}$, 
    and evolving them according to the analytically known forward process. Direct minimization of Eq.~(\ref{eq:score_matching}) is neverthelss impossible because {the gradient defining the target score function, $\nabla_x \log p(x,t)$, cannot be computed analytically and  cannot be evaluated from the training data alone}. 
However, Eq.~(\ref{eq:score_matching}) can be reformulated into a tractable objective that can be estimated by Monte Carlo sampling. The key observation is that, after conditioning the forward process on a specific initial state $x_0$, the unknown score of the marginal distribution can be replaced by the score of the conditional forward distribution,
$\nabla_x\log p(x,t\,|\,x_0,0)$, which can be computed exactly because the forward transition density is analytically known, Eq.~(\ref{eq:ou_transition}).
To derive this formulation, we first expand the square in Eq.~\eqref{eq:score_matching} and isolate the additive constant independent of the network parameters $\theta$,
\begin{align}
\int p(x,t)\,|s_\theta(x,t)|^2\,dx
- 2 \int p(x,t)\,s_\theta(x,t)\cdot \nabla_x \log p(x,t)\,dx
+ \text{const}.
\end{align}
Using the identity, $p(x,t)\,\nabla_x \log p(x,t) = \nabla_x p(x,t)$, the cross term becomes
$$
\int s_\theta(x,t)\cdot \nabla_x p(x,t)\,dx.
$$
Integrating by parts (assuming suitable decay at infinity) yields
$$
\int s_\theta(x,t)\cdot \nabla_x p(x,t)\,dx
=
- \int p(x,t)\,\nabla_x \cdot s_\theta(x,t)\,dx.
$$
The advantage of this manipulation is that the unknown score function disappears from the objective. Eq.~(\ref{eq:score_matching}) can therefore be rewritten as~\citep{hyvarinen2005estimation},
\begin{equation}
\int p(x,t)\left(
|s_\theta(x,t)|^2
+ 2\,\nabla_x \cdot s_\theta(x,t)
\right)\,dx
+ \text{const},
\label{eq:Fisher_form}
\end{equation}
where now $\nabla_x \cdot s_\theta(x,t)$ can be computed   by backward differentiation of the neural network. Eq.~\eqref{eq:Fisher_form} can be rewritten by substituting the marginal density with its representation in Eq.~\eqref{eq:marginal_transition}, discarding the additive constant independent of the network parameters, we get:
\begin{align}
\int \int p(x,t\,|\,x_0,0) \, p_{\mathrm{data}}(x_0) &\left(
|s_\theta(x,t)|^2
+ 2\,\nabla_x \cdot s_\theta(x,t)
\right)\,dx_0 \,dx = \\[0.3em]
\mathbb E_{x,x_0}
&\left[
|s_\theta(x,t)|^2+2\nabla_{x}\cdot s_\theta(x,t)
\right],
\label{eq:Fisher_expectation}
\end{align}
where, in the second line, the notation 
$\mathbb E_{x,x_0}=\mathbb E_{x\sim p(\cdot,t\,|\,x_0,0),\;x_0\sim p_{\mathrm{data}}}$
indicates that one first samples $x_0$ from the training dataset and subsequently samples $x$ from the conditional distribution $p(x,t\,|\,x_0,0)$. The above formulation is known as the {\it Fisher score matching}.
This definition of the loss function is still computationally demanding because of the need to perform the back differentiation, and a more efficient formulation is obtained integrating by parts the second term on the RHS, assuming boundary terms vanish:
\begin{equation}
\int p(x,t\,|\,x_0,0)\,\nabla_x\cdot s_\theta(x,t)\,dx
=
-\int p(x,t\,|\,x_0,0)\,s_\theta(x,t)\cdot \nabla_x \log p(x,t\,|\,x_0,0)\,dx,
\label{eq:ibp_conditional}
\end{equation}
leading to: 
\begin{equation}
\mathbb E_{x,x_0}
\left[
|s_\theta(x,t)|^2
-2\,s_\theta(x,t)\cdot \nabla_{x}\log p(x,t\,|\,x_0,0)
\right].
\label{eq:cross_term_conditional_score}
\end{equation}
Finally, completing the square one obtains \citep{vincent2011connection,song2020score} the \emph{denoising score matching} objective, $\mathcal L_t(\theta)$, equal to (\ref{eq:Fisher_expectation}) up to an additive constant independent of $\theta$:
\begin{equation}
\mathcal L_t(\theta)
\equiv
\mathbb E_{x,x_0}
\left[
\left|
s_\theta(x,t)
-
\nabla_{x}\log p(x,t\,|\,x_0,0)
\right|^2
\right].
\label{eq:dsm_loss}
\end{equation}

\noindent
For the Ornstein-Uhlenbeck forward transition density Eq.~\eqref{eq:ou_transition}, the conditional score is available in closed form:
\begin{equation}
\nabla_{x}\log p(x,t\,|\,x_0,0)
=
-
\frac{x-e^{-t}x_0}
     {(1-e^{-2t})}.
\end{equation}
Substituting into (\ref{eq:dsm_loss}) yields
\begin{equation}
\mathcal{L}_t(\theta)
=
\mathbb E_{x,x_0}\;
\left[
\left|
s_\theta(x,t)
+
\frac{x-e^{-t}x_0}
     {(1-e^{-2t})}
\right|^2
\right].
\label{eq:practical_loss}
\end{equation}

\noindent
The final training objective is obtained by averaging over the entire diffusion history:
\begin{equation}
\mathcal{L}(\theta)
=
\int_0^{T}
\lambda(t)\,
\mathcal{L}_t(\theta)\,
dt,
\end{equation}
where $\lambda(t)$ is a weighting hyperparameter that can be further optimized empirically.\\  
\noindent
Finally, the expectation over the unknown data distribution in Eq.~\eqref{eq:practical_loss}  is replaced by an empirical (Monte Carlo) average over the training dataset $\{a^\mu\}_{\mu=1}^{N}$. The corresponding empirical loss is
\begin{equation}
\mathcal{L}^{\mathrm{emp}}_t(\theta)
=
\frac{1}{N}
\sum_{\mu=1}^{N}
\left|
s_\theta(x_t^\mu,t)
+
\frac{x_t^\mu-e^{-t}a^\mu}
     {(1-e^{-2t})}
\right|^2 =
\frac{1}{N}
\sum_{\mu=1}^{N}
\left|
s_\theta(x_t^\mu,t)
+
\frac{\varepsilon_t^\mu}
     {\sqrt{1-e^{-2t}}}
\right|^2,
\label{eq:empirical_dsm}
\end{equation}
where the second equality follows directly from the sampling representation of the Ornstein-Uhlenbeck process (Eq.~\eqref{eq:ou_sampling}),
\begin{equation}
x_t^\mu
=
e^{-t}a^\mu
+
\sqrt{1-e^{-2t}}\,
\varepsilon_t^\mu,
\qquad
\varepsilon_t^\mu\sim\mathcal N(0,I).
\label{eq:ou_sampling_emp}
\end{equation}
The summation in Eq.~\eqref{eq:empirical_dsm} approximates the expectation over the data distribution,
$\frac{1}{N}\sum_{\mu=1}^{N} [\cdot] \approx \mathbb{E}_{x_0\sim p_{\mathrm{data}}}[\cdot]$, since the samples $\{a^\mu\}_{\mu=1}^N$ are drawn from the training dataset. The second expectation, over the conditional Gaussian distribution $\mathbb{E}_{x\sim p(\cdot,t\,|\,x_0,0)}$, is not computed explicitly. Instead, at each training iteration (epoch) a new Gaussian perturbation is independently sampled for every training example. Consequently, since the training is repeated over many training epochs, the stochastic optimization progressively approximates the expectation over the conditional distribution.
\noindent
The second identity in Eq.~\eqref{eq:empirical_dsm}, reveals the denoising interpretation of score matching: the neural network is trained to predict the Gaussian perturbation added during the forward diffusion process. Since, for the OU process, the conditional score is proportional to the injected noise, learning this perturbation is equivalent to learning the score. Repeating this denoising operation along the reverse diffusion trajectory enables the generation of new samples from the target data distribution.
One of the most fundamental theoretical questions surrounding generative models remains largely unanswered: why do these highly over-parameterized models generate novel samples instead of simply memorizing the training data?
The question can already be appreciated by considering the empirical score-matching objective, Eq.~(\ref{eq:empirical_dsm}), which is minimized during training. If this loss were optimized exactly, the learned score would coincide with the one of the empirical marginal data distribution, given by  a sum of Dirac delta centered on all samples pertaining to the training set, smoothed by the forward transition density:
\begin{equation}
    {p}^{\mathrm{emp}}(x,t) =
\frac{1}{N}
\sum_{\mu=1}^{N} \int p(x,t|x_0,0) \delta(x_0-a^\mu) dx_0
\end{equation}
 In this limit,the reverse stochastic dynamics would  reproduce  only the training samples, without  generating genuinely new realizations. Perfect optimization of the empirical loss would therefore correspond to perfect memorization, i.e. complete overfitting.
\noindent
From the perspective of statistical learning theory, this reflects the fundamental distinction between the empirical distribution defined by a finite dataset and the underlying probability distribution from which the data have been sampled, $p_{\mathrm{data}}(x)$. The objective of generative modelling is to infer the underlying  probability distribution compatible with the finite number of  observations~\citep{biroli2024dynamical}. Generalization is achieved only to the extent that the learned score captures the underlying data distribution without overfitting the empirical distribution associated with the finite training set. The success of modern score-based models therefore relies on this subtle balance: the score approximation must be sufficiently accurate to reconstruct the correct probability bridge between the target distribution and the initial Gaussian, while avoiding {\it perfect} convergence toward the empirical score associated with the finite training set. Section~\ref{sec:mem_gen} focuses on recent developments in this direction.
More recently, the diffusion trajectory has itself become an object of study: by analysing different temporal regimes of the forward and backward processes, recent works seek to identify signatures of memorization, information loss, and possible redundancies, with the broader goal of understanding more deeply how the full generative process operates~\citep{behjoo2025uturn}.

\subsection{Discrete-time diffusion models}
\label{sec:diffusion_models}
\begin{figure*}
      \centering  \includegraphics[width=1.0\linewidth]{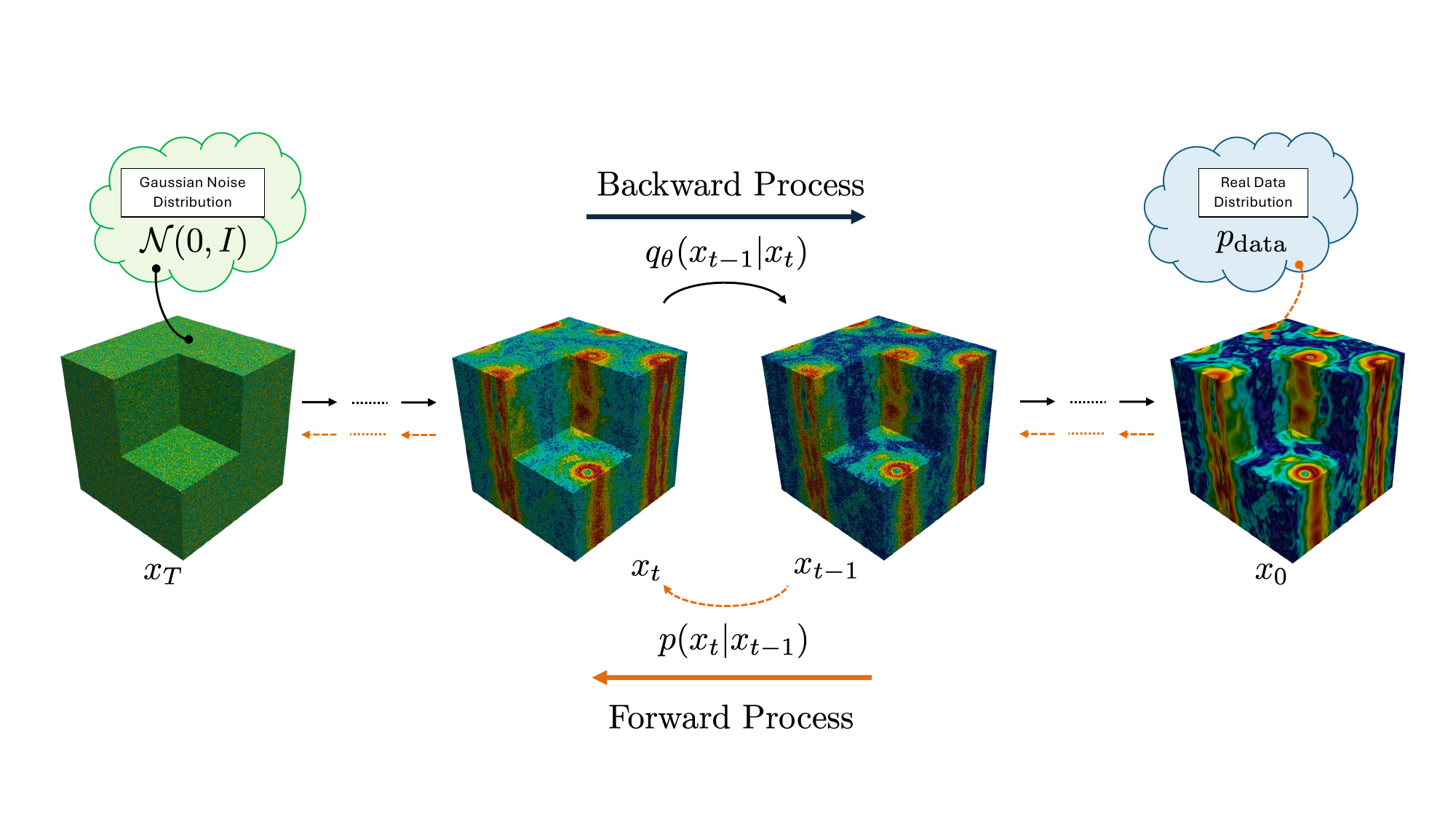}
      \caption{Schematic representation of the Forward noising 
      and learned Backward denoising processes in a discrete-time diffusion model. The forward process, from right to left, is prescribed (not  learned): at every step, a small amount of Gaussian noise is added according to a chosen variance schedule, gradually destroying the fine structure and statistical correlations of the original data.  Intermediate states $x_t$ therefore contain different levels of signal and noise, while the full sequence $x_0, \dots, x_T$  provides a discrete approximation to the continuous deformation of probability densities introduced in the score-based formulation (see Fig. \ref{fig:DM_forward_backward}). Generation proceeds through the backward process, from left to right. Starting from an independently sampled Gaussian realization $x_T \sim N(0,I)$, a neural network parameterizes the reverse conditional distributions, $q_\theta (x_{t-1}|x_t)$ and iteratively removes the estimated noise to reconstruct increasingly structured states. The reverse transition is not obtained by simply undoing the particular noise realization used in the forward process. Instead, the model learns, from the training data, the statistical rule required to map a noisy state $x_t$  onto a distribution of plausible less-noisy states $x_{t-1}$.  Repetition of this stochastic denoising step eventually produces a new realization $x_0$  sampled from the learned approximation to $p_{\mathrm{data}}$ (see Sec. \ref{sec:mem_gen} for a discussion about the memorization/generalization transitions). The central pair of states illustrates the local learning problem at a generic diffusion step.  As discussed in the text, for Gaussian forward kernels, noise prediction and conditional-score estimation are directly related; hence, learning to denoise at all noise levels provides the information needed to approximate the full reverse Markov chain. The intermediate arrows emphasize that practical models employ many intermediate steps, so that the complex transport between the data and Gaussian distributions is decomposed into a sequence of comparatively simple local transformations.} 
      \label{fig:DM_example}
  \end{figure*}

The score-based formulation described above constructs a diffusion process that progressively transforms the data distribution into a Gaussian equilibrium state. Practically speaking, this convergence is achieved only in the asymptotic limit $t\rightarrow\infty$ with a continuous stochastic evolution. Alternatively, it is also possible to define an equivalent probabilistic transport from a Gaussian to $p_{\mathrm{data}}(x)$  over a finite temporal window, say $t \in [t_0,t_f] = [0,T]$, and with a finite sequence  of Markov transitions, as sketched in Fig.~\ref{fig:DM_example}. However, the quality of the learned generation will be good enough only for a high enough number of intermediate steps (see below). 
\noindent
In this discrete approach, we do not consider a continuous family of probability densities, the diffusion process is discretized into a sequence of latent variables
\begin{equation}
x_0,x_1,\ldots,x_T,
\end{equation}
where $x_0$ denotes a sample drawn from the training dataset
$\{a^\mu\}_{\mu=1}^{N}$, providing an empirical realization of
$p_{\mathrm{data}}(x)$, while the terminal state satisfies
\begin{equation}
x_T\sim\mathcal N(0,I).
\end{equation}
Let us stress that the finite set of intermediate times  $t=0,1,\ldots,T$, does not necessarily require to be equally spaced. 
The forward diffusion process is defined as the Markov chain
\begin{equation}
p(x_{1:T}|x_0)
=
\prod_{t=1}^{T}
p(x_t|x_{t-1}),
\label{eq:forward_chain}
\end{equation}
with Gaussian transition probabilities
\begin{equation}
p(x_t|x_{t-1})
=
\mathcal N
\Bigl(
\sqrt{1-\beta_t}\,x_{t-1},
\,
\beta_t I
\Bigr),
\label{eq:forward_transition}
\end{equation}
where $\{\beta_t\}_{t=1}^{T}$ is a prescribed variance schedule satisfying
$0<\beta_t<1$. Introducing
\begin{equation}
\alpha_t=1-\beta_t,
\qquad
\bar{\alpha}_t
=
\prod_{s=1}^{t}\alpha_s,
\end{equation}
the schedule is chosen such that
$\bar{\alpha}_0=1$ and
$\bar{\alpha}_T\approx0$.
Consequently, the signal contribution is progressively attenuated while the noise level increases, driving the forward process toward an approximately Gaussian terminal distribution.

\noindent
As in the continuous-time Ornstein-Uhlenbeck process, the forward dynamics is completely specified and requires no learning. Moreover, because the composition of Gaussian transitions remains Gaussian, the conditional probability from the initial data sample to an arbitrary diffusion step can be computed analytically as
\begin{equation}
p(x_t|x_0)
=
\mathcal N
\Bigl(
\sqrt{\bar{\alpha}_t}\,x_0,
\,
(1-\bar{\alpha}_t)I
\Bigr),
\label{eq:forward_conditional}
\end{equation}
which immediately leads to the sampling representation;
\begin{equation}
x_t
=
\sqrt{\bar{\alpha}_t}\,x_0
+
\sqrt{1-\bar{\alpha}_t}\,\epsilon,
\qquad
\epsilon\sim\mathcal N(0,I).
\label{eq:diffusion_reparam}
\end{equation}
Equation~\eqref{eq:forward_conditional} is the discrete counterpart of the OU forward transition density in Eq.~\eqref{eq:ou_transition}. In both formulations, the analytical conditional distribution allows direct sampling at arbitrary diffusion times without explicitly integrating the underlying stochastic dynamics. This property will again play a central role in the determination of the training objective.

\medskip

As in continuous-time formulation, the reverse dynamics responsible for sample generation is unknown and must be learned from data. The reverse process is also represented as a Markov chain
\begin{equation}
q(x_{0:T})
=
p(x_T)
\prod_{t=1}^{T}
q(x_{t-1}|x_t),
\label{eq:reverse_chain}
\end{equation}
which transports probability mass from the Gaussian reference distribution back toward the data distribution. As in the continuous-time setting, the endpoint distributions are fixed,
\begin{equation}
q(x_T)=p(x_T) \approx \mathcal N(0,I),
\qquad
q(x_0)=p(x_0)=p_{\mathrm{data}},
\end{equation}
while only the intermediate transition probabilities remain unknown.
\noindent
The exact reverse conditional distribution can be expressed through Bayes' theorem as
\begin{equation}
q(x_{t-1}|x_t)
=
\frac{
p(x_t|x_{t-1})
p(x_{t-1})
}{
p(x_t)
},
\label{eq:forward_posterior}
\end{equation}
where the forward transition
$p(x_t|x_{t-1})$
is given by Eq.~\eqref{eq:forward_transition}. However,  the two marginal distributions of this type
\begin{equation}
p(x_t)
=
\int
p(x_t|x_0)\,
p_{\mathrm{data}}(x_0)\,
dx_0,
\label{eq:forward_marginal}
\end{equation}
depend on the unknown data distribution and therefore cannot be computed analytically. Consequently, the reverse conditional distribution, $q(x_{t-1}|x_t)$ is also unknown. This situation is entirely analogous to the continuous-time formulation, where the score function depends on the unknown marginal distribution generated by the forward diffusion process. 
\noindent
The goal of discrete diffusion models is therefore to approximate the reverse conditional distribution using a neural network,
\begin{equation}
\label{eq:back_appr}
q_\theta(x_{t-1}|x_t),
\end{equation}
whose parameters $\theta$ are learned from data.
\noindent
A key result is that when the discretization is sufficiently fine, corresponding to a large number of diffusion steps and therefore small perturbations between successive states, the reverse conditional is itself well approximated by a Gaussian distribution~\citep{sohl2015deep},
\begin{equation}
q_\theta(x_{t-1}|x_t)
=
\mathcal N
\Bigl(
\mu_\theta(x_t,t),
\Sigma_\theta(x_t,t)
\Bigr).
\label{eq:reverse_gaussian}
\end{equation}
The learning problem is thus reduced to estimating the parameters of the Gaussian reverse transition kernel. The resulting generative model defines the probability distribution
\begin{equation}
\label{eq:generative_chain}
q_\theta(x_0)
=
\int\cdots\int
p(x_T)
\prod_{t=1}^{T}
q_\theta(x_{t-1}|x_t)
\,dx_1\cdots dx_T,
\end{equation}
which approximates the unknown data distribution.
\noindent
In practice,~\cite{ho2020denoising} observed that approximating the variance of the backward step in Eq.~\eqref{eq:reverse_gaussian} with the choice $\Sigma_\theta=\beta_t I$ has little effect on the quality of the generated samples, so the the neural network is only needed to predict the reverse mean, $\mu_\theta(x_t,t)$. The discrete single steps embedded in both the forward and backward diffusion processes are schematized in Fig.~\ref{fig:DM_example}. As shown below, this prediction can be reformulated into an equivalent denoising objective as in the score approach, in which the neural network learns to estimate the Gaussian perturbation introduced during the forward diffusion process.

\medskip

The training objective of discrete diffusion models can formally be formulated in terms of the cross-entropy between $p_{\mathrm{data}}(x_0)$ and $q_\theta(x_0)$. Equivalently, one seeks to maximize the expected log-likelihood assigned by the generative model to samples drawn from the data distribution,
\begin{equation}
\mathcal{J}(\theta)
=
\int
p_{\mathrm{data}}(x_0)
\log q_\theta(x_0)\,
dx_0.
\label{eq:maximum_likelihood_diffusion}
\end{equation}
This objective is directly related to the Kullback-Leibler divergence between the data distribution and the distribution generated by the model, see Eq.~\eqref{eq:DKL}, up to a term that depends only on $p_{\mathrm{data}}(x_0)$ and is therefore independent of the model parameters $\theta$. Consequently, maximizing $\mathcal{J}(\theta)$ with respect to $\theta$ is equivalent to minimizing $D_{\mathrm{KL}}(p_{\mathrm{data}}\Vert q_\theta)$.\\

Although conceptually straightforward, this objective cannot be directly evaluated. As discussed in connection with Eq.~\eqref{eq:empirical_dsm}, the expectation on the data distribution can be replaced by a Monte Carlo estimate on the training dataset. However, this would still require evaluating, for each training realization $x_0=a^\mu$, the probability density $q_\theta(a^\mu)$ assigned to it by the generative model. This quantity is not explicitly available, since its evaluation would require integrating Eq.~\eqref{eq:generative_chain} over all possible intermediate states of the generative process, which is computationally intractable in the high-dimensional spaces characterizing turbulent flows.
Diffusion models bypass this difficulty by constructing a tractable variational bound of the log-likelihood ($\log q_\theta (x_0)$). In particular, multiplying and dividing the argument of the logarithm by the forward trajectory distribution $p(x_{1:T}|x_0)$ and using the definition of $q_\theta(x_0)$ in Eq.~\eqref{eq:generative_chain}, Jensen's inequality gives
\begin{equation}
\begin{aligned}
\log q_\theta(x_0)
&=
\log \mathbb E_{p(x_{1:T}|x_0)}
\left[
\frac{q_\theta(x_{0:T})}{p(x_{1:T}|x_0)}
\right]
\\
&\geq
\mathbb E_{p(x_{1:T}|x_0)}
\left[
\log q_\theta(x_{0:T})
-
\log p(x_{1:T}|x_0)
\right]
\\
&\equiv
\mathcal L_{\mathrm{ELBO}}(x_0;\theta).
\end{aligned}
\end{equation}
The $\mathcal L_{\mathrm{ELBO}}$ thus provides a lower bound on the log-likelihood for each data realization. Equivalently, its negative provides an upper bound on the negative log-likelihood.
Averaging over the data distribution, we define the variational loss
\begin{equation}
\mathcal L_{\mathrm{VLB}}(\theta)
\equiv
-\mathbb E_{p_{\mathrm{data}}(x_0)}
\left[
\mathcal L_{\mathrm{ELBO}}(x_0;\theta)
\right],
\end{equation}
leading to an upper bound for  (\ref{eq:maximum_likelihood_diffusion}):
\begin{equation}
-\mathcal J(\theta)
\leq
\mathcal L_{\mathrm{VLB}}(\theta).
\end{equation}
Minimizing this upper bound provides a tractable surrogate for maximum-likelihood training. Note that the bound becomes tight when the posterior distribution of the generated trajectories, $q_\theta(x_{1:T}|x_0)=q_\theta(x_{0:T})/q_\theta(x_0)$, coincides with the forward-process distribution conditioned on the same data realization, $p(x_{1:T}|x_0)$, for almost every $x_0$ under the data distribution~\citep{blei2017variational}. 
The complete derivation of the variational bound can be found in the original works ~\citep{sohl2015deep, ho2020denoising} and will not be repeated here. Instead, we focus on the key observation connecting the discrete formulation with the score-based framework discussed in the previous section.
The key idea is to replace the direct evaluation of the log-likelihood ($\log q_\theta (x_0)$) by a sequence of local comparisons along the diffusion trajectory. Exploiting the Markov structure of the forward and reverse processes, the variational loss $\mathcal L_{\mathrm{VLB}}(\theta)$ can be decomposed into a sum of Kullback-Leibler divergences associated with the individual diffusion steps, together with two boundary contributions at the first and last transition steps (see below Eq.~\eqref{eq:elbo_diffusion}).
In this way, the learning problem is reduced to matching local reverse transitions along trajectories generated from the training data, without requiring evaluation of the probability density assigned by the complete generative process to each training realization.
Still the ground-truth local reverse conditional distribution $q(x_{t-1}|x_t)$ of the forward process is generally not accessible, since it depends on the unknown data distribution, see Eqs.~\eqref{eq:forward_posterior} and~\eqref{eq:forward_marginal}.
The crucial simplification closely parallels that encountered in the continuous-time score-based formulation, where replacing the intractable marginal score by a conditional score made the training objective analytically accessible. Here, further conditioning the reverse distribution on the initial data $x_0$, $q(x_{t-1}|x_t,x_0)$, applying Bayes' theorem and exploiting the Markov property of the forward process, one gets
\begin{equation}
q(x_{t-1}|x_t,x_0)
=
\frac{
p(x_t|x_{t-1})
p(x_{t-1}|x_0)
}{
p(x_t|x_0)
}.
\label{eq:posterior_conditioned}
\end{equation}
Both $p(x_{t-1}|x_0)$ and $p(x_t|x_0)$ are analytically known from the prescribed forward diffusion, Eq.~\eqref{eq:forward_conditional}, as is the transition probability $p(x_t|x_{t-1})$. Consequently, $q(x_{t-1}|x_t,x_0)$ can be analytically evaluated for any trajectory obtained by perturbing a training realization $x_0=a^\mu$ through the forward process.
The variational loss is constructed such that it can be expressed as 
\begin{equation}
\begin{aligned}
\mathcal L_{\mathrm{VLB}}(\theta)
={}&
\underbrace{
\mathbb E_{x_0}
\left[
D_{\mathrm{KL}}
\left(
p(x_T|x_0)
\,\Vert\,
p(x_T)
\right)
\right]
}_{\mathcal L_{\mathrm{prior}}}
\\[2mm]
&+
\underbrace{
\sum_{t=2}^{T}
\mathbb E_{x_t,x_0}
\left[
D_{\mathrm{KL}}
\left(
q(x_{t-1}|x_t,x_0)
\,\Vert\,
q_\theta(x_{t-1}|x_t)
\right)
\right]
}_{\displaystyle \sum_{t=2}^{T}\mathcal L_t}
\\[2mm]
&+
\underbrace{
\mathbb E_{x_1,x_0}
\left[
-\log q_\theta(x_0|x_1)
\right]
}_{\mathcal L_{\mathrm{rec}}}.
\end{aligned}
\label{eq:elbo_diffusion}
\end{equation}
Here, the expectations are taken over $x_0\sim p_{\mathrm{data}}(x_0)$, sampled from the training data, and over the noisy states generated by the prescribed forward process, $x_t\sim p(x_t|x_0)$.  
The three contributions have a direct interpretation. The first term, $\mathcal L_{\mathrm{prior}}$, does not depend on the model parameters and acts as the terminal boundary of the diffusion chain measuring the discrepancy between the endpoint distribution of the forward process $p(x_T|x_0)$ and the prescribed Gaussian prior $p(x_T)$. The last term, $\mathcal L_{\mathrm{rec}}$, is a reconstruction contribution associated with the final reverse transition from $x_1$ to the data sample $x_0$.
Between these two boundaries, the sum over $t=2,\ldots,T$ compares, at every intermediate diffusion step, the learned reverse transition $q_\theta(x_{t-1}|x_t)$ with the analytically accessible posterior $q(x_{t-1}|x_t,x_0)$, conditioned on the initial realization $x_0=a^\mu$ taken from the training data. In this way, at every diffusion step, $q(x_{t-1}|x_t,x_0)$, Eq.~\eqref{eq:posterior_conditioned}, provides a training target for the learned reverse process. The neural network can be trained to learn a reverse transition that depends only on the current noisy state $x_t$ and the diffusion time $t$, without requiring the clean realization $x_0$ as input.
Since $q(x_{t-1}|x_t,x_0)$ is the product of Gaussian distributions, see Eq.~\eqref{eq:posterior_conditioned}, it remains Gaussian,
\begin{equation}
q(x_{t-1}|x_t,x_0)
=
\mathcal N
\left(
\mu(x_t,x_0),
\tilde{\beta}_t I
\right),
\end{equation}
whit mean,
\begin{equation}
\mu(x_t,x_0)
=
\frac{\sqrt{\bar{\alpha}_{t-1}}\beta_t}
{1-\bar{\alpha}_t}
x_0
+
\frac{\sqrt{\alpha_t}(1-\bar{\alpha}_{t-1})}
{1-\bar{\alpha}_t}
x_t =
\frac{1}{\sqrt{\alpha_t}}
\left(
x_t
-
\frac{\beta_t}
{\sqrt{1-\bar{\alpha}_t}}
\epsilon
\right),
\label{eq:posterior_mean}
\end{equation}
where the second equality follows from Eq.\eqref{eq:diffusion_reparam}, while its variance is
$\tilde{\beta}_t =
\frac{1-\bar{\alpha}_{t-1}}
{1-\bar{\alpha}_t}
\beta_t$. 

Knowing the exact mean of the conditioned posterior $q(x_{t-1}\!\mid x_t,x_0)$ and adopting the approximation of \citet{ho2020denoising} for the model covariance, $\Sigma_\theta=\beta_t I$, each intermediate Kullback-Leibler contribution $\mathcal L_t$ reduces, up to terms independent of the model parameters (due to the variance approximation), to a weighted mean-squared error between the exact posterior mean $\mu(x_t,x_0)$ and the mean $\mu_\theta(x_t,t)$ predicted by the neural network,
\begin{equation}
\label{eq:dm_loss_mean}
\mathcal L_t
=
\mathbb E_{x_0,x_t}
\left[
\frac{1}{2\beta_t}
\left|
\mu(x_t,x_0)
-
\mu_\theta(x_t,t)
\right|^2
\right]
+\mathrm{const}.
\end{equation}
The expectation over $x_0$ and $x_t$ can be estimated by sampling $x_0\sim p_{\mathrm{data}}$ from the training dataset and independently sampling a noise realization $\epsilon\sim\mathcal N(0,I)$, from which $x_t$ is obtained according to Eq.~\eqref{eq:diffusion_reparam}. 

Consequently, optimization reduces to learning the posterior mean $\mu(x_t,x_0)$ through neural-network prediction $\mu_\theta(x_t,t)$, without providing the clean sample $x_0$ as input.
This means that the network must infer, from noisy observation $x_t$, the underlying clean sample that is required to reconstruct the posterior mean, as shown in the first expression of Eq.~\eqref{eq:posterior_mean}.
Using instead the second expression in Eq.~\eqref{eq:posterior_mean}, this problem can equivalently be reformulated in terms of the Gaussian perturbation $\epsilon$ that transformed $x_0$ into $x_t$ during the forward diffusion process. A widely used parameterization of the reverse process therefore writes
\begin{equation}
\mu_\theta(x_t,t)
=
\frac{1}{\sqrt{\alpha_t}}
\left(
x_t
-
\frac{\beta_t}{\sqrt{1-\bar{\alpha}_t}}
\epsilon_\theta(x_t,t)
\right),
\end{equation}
so that the neural network directly predicts the noise term $\epsilon_\theta(x_t,t)$ rather than the posterior mean itself. With this parameterization, the loss in Eq.~\eqref{eq:dm_loss_mean} becomes,
\begin{equation}
\mathcal L_t
=
\frac{\beta_t}{2\alpha_t(1-\bar{\alpha}_t)}
\mathbb E_{x_0,x_t}
\left[
\left\|
\epsilon-\epsilon_\theta(x_t,t)
\right\|^2
\right] + \text{const}.
\label{eq:noise_prediction_loss}
\end{equation}
Thus the network can be trained to only infer the Gaussian perturbation that produced the noisy state $x_t$ from the clean sample $x_0$.

\noindent
This observation reveals the intimate connection between diffusion probabilistic models and continuous-time score matching. Indeed, using Tweedie's formula, which relates the mean of a clean random variable ($x_0$) observed through additive Gaussian noise ($x_t$) to the score of the corresponding noisy
distribution ($\nabla_{x_t}\log p(x_t)$)~\citep{efron2011tweedie}. Applying Tweedie's to the forward diffusion process it gives
\begin{equation}
\mathbb E_{x_0}
\left[
x_0
\,\middle|\,
x_t
\right]
=
\frac{1}{\sqrt{\bar{\alpha}_t}}
\left[
x_t
+
\left(1-\bar{\alpha}_t\right)
\nabla_{x_t}\log p(x_t)
\right],
\label{eq:tweedie}
\end{equation}
where the mean of $x_0$, (the clean data), is taken over its posterior distribution conditioned on the noisy observation, namely $p(x_0|x_t)$, while $1-\bar{\alpha}_t$ is the noise variance prescribed by the forward
diffusion process at time $t$. Using the definition $x_t = \sqrt{\bar{\alpha}_t}\,x_0 + \sqrt{1-\bar{\alpha}_t}\,\epsilon$ and rearranging Eq.~\eqref{eq:tweedie}, it is immediate to see that the score of $p(x_t)$ is proportional to the conditional expectation of the Gaussian noise injected into $x_0$ to generate $x_t$, namely
\begin{equation}
\nabla_{x_t}
\log p(x_t)
=
-
\frac{1}
{\sqrt{1-\bar{\alpha}_t}}
\,
\mathbb E_{x_0}
\left[
\epsilon
\,\middle|\,
x_t
\right].
\label{eq:tweedie_reshaped}
\end{equation}
Therefore, predicting the injected noise into $x_t$, estimating the score function, and learning the mean of the reverse Gaussian transition are mathematically equivalent parameterizations of the same denoising problem. Their practical training objectives may nevertheless differ in the sampling of diffusion times and in the time-dependent weighting assigned to different noise levels.

From this perspective, diffusion probabilistic models and continuous-time score-based generative models should not be regarded as distinct methodologies, but as two complementary formulations of the same probabilistic transport problem.
A less obvious observation is that the same family of evolving distributions can also be generated by purely deterministic dynamics. This insight has motivated the recent development of flow matching methods, which replace stochastic diffusion by deterministic probability transport while preserving the same endpoint distributions as will be discussed in Sect.~\ref{sec:flowmatching}.

\subsection{Why do generative models not memorize?}
\label{sec:mem_gen}
  \begin{figure*}
      \centering      \includegraphics[width=1.0\linewidth]{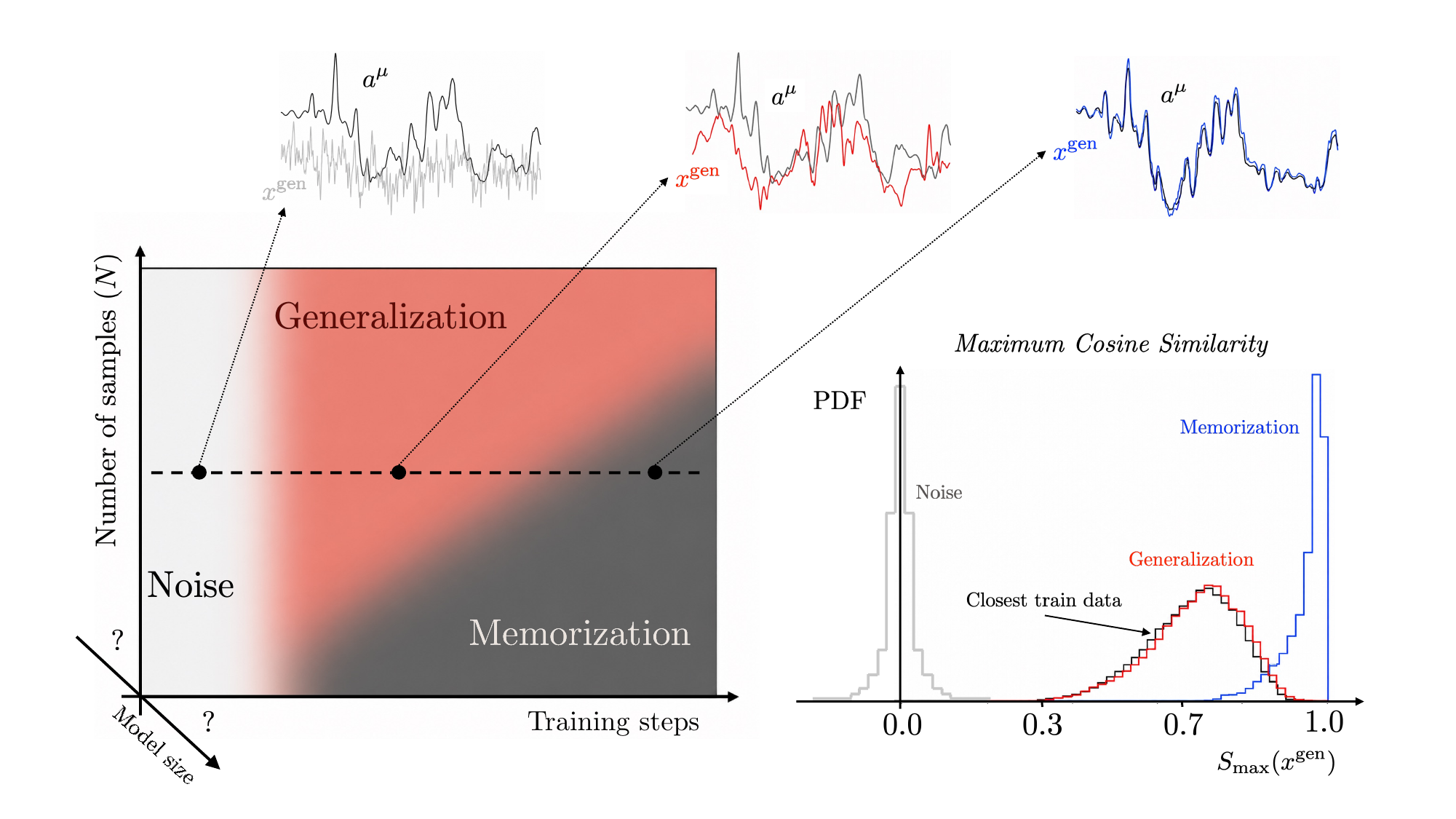}
      \caption{Schematic representation of the generalization-memorization transition in diffusion models and its characterization through maximum cosine similarity, introduced in Eq.~\eqref{eq:max_cosine}. \textit{Left:} Qualitative phase diagram of the learning dynamics as a function of the number of training backpropagation steps, the number of training samples $N$, for a fixed model size. At early training stages, generated samples are dominated by noise; at intermediate times, the model enters a generalization regime in which it reproduces the statistical features of the data while generating novel realizations; after long training time, assuming a model size large enough, optimization eventually drives the model toward memorization of individual training samples. Representative generated trajectories $x^{\mathrm{gen}}$ and their closest training samples $a^\mu$ illustrate the three regimes along a fixed-$N$ section of the diagram.
      \textit{Right:} Probability density of the maximum cosine similarity $S_{\max}(x^{\mathrm{gen}})$ between each generated sample and the training dataset. In the noise regime, generated and training samples have negligible similarity. In the generalization regime, the distribution of $S_{\max}$ approaches the intrinsic nearest-neighbor similarity between distinct samples of the training set (black curve), indicating that generated samples are statistically comparable to, but do not reproduce, individual training examples. In the memorization regime, the distribution instead concentrates close to $S_{\max}=1$, revealing the near-exact reproduction of training samples.}
      \label{fig:cosine}
  \end{figure*}
Concerning the problem of memorization {\it vs} generalization, already briefly discussed at the end of Sec. \ref{sec:score-based}, a  new perspective has recently emerged from the analysis of the learning dynamics itself~\citep{bonnaire2026diffusion}.  In sufficiently expressive models capable of resolving the empirical score, continued optimization can eventually produce memorization. Recent work identifies a separation between the timescales associated with learning distributional structure and memorizing individual samples, while also showing that the outcome depends on model capacity and dataset size. As a result, in principle, asymptotic convergence for a long enough number of epochs does not necessarily involve good  generalization.
The key observation is, instead, that learning proceeds through two well-separated dynamical regimes. During the early stages of optimization, the neural network rapidly learns the smooth large-scale components of the score associated with the underlying probability distribution and is already sufficient to generate realistic samples. Only after much longer optimization epochs the sample-specific high-frequency corrections required to exactly reconstruct the empirical distribution gradually emerge, eventually leading to memorization. Generalization is therefore interpreted as a transient dynamical phenomenon that occurs before the onset of memorization. Importantly, the number of optimization epochs required for memorization is predicted to increase with the size of the training dataset. As a consequence, larger datasets widen the temporal window over which the model captures the underlying distribution without yet reproducing individual training samples, making successful generalization increasingly likely in practice~\citep{bonnaire2026diffusion}.
The corresponding phase diagram is illustrated in Fig.~\ref{fig:cosine}, which summarizes the different learning regimes as a function of the number of training steps in the large-network limit. Although the figure emphasizes this asymptotic regime, the network size itself is an additional control parameter that shifts the boundary between generalization and memorization and therefore plays a central role in the learning dynamics.

\noindent
A practical way to quantify the degree of memorization is through the \emph{maximum cosine similarity} between generated samples and the training dataset, see \citep{lahitani2016cosine} and~\citep{li2026deterministic} for an application to generative AI for Lagrangian turbulence. Given a generated sample $x^{\mathrm{gen}}$ and the training set $\{a^\mu\}_{\mu=1}^{N}$, one computes
\begin{equation}
S_{\max}(x^{\mathrm{gen}})
=
\max_{\mu}
\frac{
\langle x^{\mathrm{gen}},a^\mu\rangle
}{
\|x^{\mathrm{gen}}\|\,
\|a^\mu\|
},
\label{eq:max_cosine}
\end{equation}
where $\langle x^{\mathrm{gen}},a^\mu\rangle$ denotes the scalar product between the generated sample and the $\mu$-th training example, while $\|x^{\mathrm{gen}}\|$ and $\|a^\mu\|$ are their Euclidean norms. The ratio therefore measures the cosine of the angle between the two high-dimensional vectors, taking the value $1$ for identical samples, $0$ for orthogonal samples, and $-1$ for opposite vectors. The maximum is taken over all samples in the training dataset, thereby identifying the training realization that is most similar to the generated one. The largest cosine similarity between the generated realization and
all samples in the training set indicates that the
generated sample is nearly identical to one of the training examples, whereas smaller values correspond to genuinely novel realizations.
To determine whether this similarity reflects memorization, and not the intrinsic redundancy of the dataset, it must be compared with a baseline computed from the training data themselves. This baseline is obtained by repeatedly selecting one sample from the training set, computing its maximum cosine similarity with all the remaining training samples, and averaging over many realizations. In the absence of memorization, the distribution of $S_{\max}$ measured for generated samples should be statistically indistinguishable from this intrinsic training-set baseline. Conversely, a systematic increase of the maximum cosine similarity for generated samples indicates that the model is reproducing individual training examples and  failing to generate new realizations. 
As discussed in other sections of this perspective, the multi-scale nature of turbulent realizations implies that also the cosine-similarity test is not fully objective: one may obtain a very good generalization score for some specific observables and poor results for others, in particular when small-scales and high-frequencies properties are tested.

\noindent
Memorization issues must also be taken into account in conditional generation, where the aim is to sample from a distribution constrained by additional observations or measurements. Guidance-based methods such as Diffusion Posterior Sampling (DPS)~\citep{chung2022diffusion} do not learn a new conditional generative model, but instead rely on a pretrained unconditional diffusion model to represent the prior distribution. Consequently, any tendency of the pretrained model to memorize the training dataset is directly inherited by the conditional posterior, irrespective of the guidance strategy employed. Recent work \citep{burns2026and} further shows that the approximation underlying DPS may become inaccurate when the posterior distribution is broad, multimodal, or strongly non-Gaussian. These observations indicate that understanding the memorization-generalization trade-off of the pretrained prior is essential for unconditional generation,  and for the reliability of conditional generation and uncertainty quantification. Developing a unified theoretical framework capable of simultaneously describing memorization, generalization, and posterior approximation therefore remains one of the major open challenges in modern diffusion-based generative modelling.

\noindent
More broadly, a complete theoretical understanding of why over-parameterized neural networks learn smooth probability distributions before memorizing finite datasets is still lacking. Establishing quantitative connections between the optimization dynamics, the geometry of the data manifold, the network architecture and the quality of the learned probability transport represents one of the most important open problems at the interface between statistical physics, machine learning and generative modelling.

\subsection{Deterministic probability transport: flow matching}
\label{sec:flowmatching}

Besides the stochastic formulation underlying diffusion and score-based models, probability distributions can also be connected through a deterministic transport process. This is the basic idea behind \textit{flow matching}~\citep{lipman2022flow}: rather than describing
the evolution of the samples through a stochastic differential equation, one
introduces a time-dependent velocity field $v(x,t)$ in the space of generated states $x$, and considers the ordinary differential equation
\begin{equation}
    \frac{dX_t}{dt}=v(X_t,t).
    \label{eq:flow_matching}
\end{equation}
Here, $v(x,t)$ describes how samples move through the state space, $x$, as the probability distribution is continuously deformed. It should therefore not be confused with the physical velocity field of the fluid, which may itself constitute part or all of the state $x$ being generated. The corresponding probability density $p(x,t)$ evolves according to the continuity equation
\begin{equation}
    \partial_t p(x,t)
    +
    \nabla_x\cdot\left[v(x,t)p(x,t)\right]
    =0,
    \label{eq:continuity}
\end{equation}
so that the velocity field directly describes the transport of probability mass. In the following, we retain the time convention used for the diffusion processes introduced above: $p(x,0)=p_{\rm data}(x)$ is the data distribution, whereas $p(x,T)$ is a simple reference distribution, typically a standard Gaussian. Notice that this convention is opposite to the one commonly adopted in the flow-matching literature, where the Gaussian distribution is usually taken as the initial distribution and the data as the final one. The two descriptions are equivalent up to a reversal of the time parametrization.

The central idea of flow matching is to prescribe a simple family of
interpolation trajectories connecting the two endpoint distributions and
to learn the velocity field that generates the resulting probability path.
Importantly, this can be done without explicitly evaluating the
intermediate marginal densities. Given a data sample $x_0\sim p_{\rm data}$ and a noise sample $x_T\sim p_{\rm noise}$, one introduces a simple interpolation
\begin{equation}
    X_t=I_t(x_0,x_T),
    \qquad
    I_0(x_0,x_T)=x_0,
    \qquad
    I_T(x_0,x_T)=x_T.
    \label{eq:flow_interp}
\end{equation}
For fixed endpoints $(x_0,x_T)$, the map defines a deterministic trajectory. We refer
to it as a \emph{conditional trajectory}, since it is obtained by conditioning on the particular choice of endpoints. Its instantaneous velocity is
\begin{equation}
    u_t(x_0,x_T)
    =
    \frac{d}{dt}I_t(x_0,x_T),
\end{equation}
which we call the \emph{conditional velocity}. Since the interpolation is prescribed, this velocity is known analytically and can therefore be used as a training target.
Since the endpoints $x_0$ and $x_T$ are themselves random variables, sampling many pairs $(x_0,x_T)$ generates an ensemble of such trajectories. At any fixed time $t$, the corresponding intermediate states $X_t$ are therefore also random and define a probability density $p(x,t)$.
As $t$ varies from $0$ to $T$, these densities form a probability path connecting
\begin{equation}
    p(x,0)=p_{\rm data}(x),
    \qquad
    p(x,T)=p_{\rm noise}(x).
\end{equation}
We refer to $p(x,t)$ as the \emph{marginal probability path}, because the dependence on the particular endpoints $(x_0,x_T)$ has been marginalized out.

For example, the simplest possible choice is the linear interpolation,
\begin{equation}
    I_t(x_0,x_T)
    =
    \left(1-\frac{t}{T}\right)x_0
    +
    \frac{t}{T}x_T,
\end{equation}
whose conditional velocity is $u_t(x_0,x_T) = (x_T-x_0)/T$. 

The objective is in general to determine a single velocity field $v(x,t)$ whose deterministic dynamics generates this same family of marginal distributions $p(x,t)$, see Eq.~\eqref{eq:flow_matching}. We call $v(x,t)$ the \emph{marginal velocity field} to distinguish it from the conditional velocity $u_t(x_0,x_T)$ associated with one particular interpolation trajectory.

In general, different endpoint pairs may generate trajectories that pass through the same state $x$ at the same time $t$, while having different conditional velocities. The marginal velocity must instead assign a single velocity to $(x,t)$. It is given by
\begin{equation}
    v(x,t)
    = \int
    u_t(x_0,x_T)\,
    p(x_0,x_T \mid X_t=x)\,
    dx_0\,dx_T,
    \qquad
    X_t=I_t(x_0,x_T).
    \label{eq:marginal_flow_velocity}
\end{equation}
Thus, $v(x,t)$ is the probability-weighted average of the conditional velocities associated with all endpoint pairs whose interpolating trajectories can reach $x$ at time $t$. This averaging preserves the total probability flux at $(x,t)$, so that the resulting field satisfies the continuity  equation~\eqref{eq:continuity} for the marginal density $p(x,t)$. Consequently, $v(x,t)$ generates exactly the marginal probability path induced by the prescribed interpolation~\citep{lipman2024guide}.

Furthermore, $v(x,t)$ is precisely the quantity learned by a neural network $v_\theta(x,t)$ when trained to minimize the mean-squared flow-matching loss,
\begin{equation}
    \mathcal{L}_{\rm FM}(\theta)
    =
    \mathbb{E}_{t,x_0,x_T}
    \left[
        \left\|
        v_\theta\!\left(I_t(x_0,x_T),t\right)
        -
        u_t(x_0,x_T)
        \right\|^2
    \right]
    \label{eq:flow_matching_loss}
\end{equation}
where the expectation is taken over sampled data/noise endpoints, and over interpolation times.
As shown in~\cite{lipman2024guide}, the minimizer of Eq.~\eqref{eq:flow_matching_loss} is precisely the conditional expectation defined in Eq.~\eqref{eq:marginal_flow_velocity}, and therefore recovers the marginal velocity field $v(x,t)$.

Once $v_\theta$ has been learned, generation proceeds in the reverse direction: one draws $X_T$ from the Gaussian reference distribution and integrates Eq.~(\ref{eq:flow_matching}) backward from $T$ to $0$, thereby transporting noise into samples from the data distribution. Thus, in contrast with score-based diffusion models, flow matching directly learns a deterministic transport field.

The deterministic and stochastic descriptions are nevertheless closely related. In particular, a diffusion process admits an associated \textit{probability-flow ODE} whose marginal probability densities coincide with those of the corresponding stochastic process \citep{song2020score}. Thus, diffusion models can also be represented within a deterministic probability-transport framework, while flow matching provides a direct way of learning such deterministic transport fields without first parameterizing the score.

\subsection{Stochastic interpolants: a unified perspective}
\label{sec:stoch_interp}
The probability path introduced in the previous section represents only one possible choice of interpolation. More generally, the endpoints of transport may be arbitrary probability
distributions. This observation has motivated the recent development of the stochastic interpolants framework~\citep{albergo2022building,albergo2025stochastic}, which provides a unified mathematical
description of score-based diffusion models, and flow matching.  The starting point is the construction of a probability bridge connecting two distributions over a finite time interval.
Let $\rho(x,0)$ and $\rho(x,T)$ denote two arbitrary probability densities. A stochastic interpolant is defined as
\begin{equation}
X_t = I_t(X_0,X_T)+\gamma_tZ,
\qquad t\in[0,T],
\label{eq:stochastic_interpolant}
\end{equation}
where $X_0\sim\rho(x,0)$, $X_T\sim\rho(x,T)$, and $Z\sim\mathcal{N}(0,I)$ is an independent Gaussian random variable. Here we consider stochastic interpolants defined over the finite time interval $t\in[0,T]$. Without loss of generality, the final time can be set to $T=1$ by a suitable rescaling of the time variable. The interpolation function satisfies
\begin{equation}
I_0(X_0,X_T)=X_0,
\qquad
I_T(X_0,X_T)=X_T,
\end{equation}
while the noise amplitude is chosen such that
\begin{equation}
\gamma_0=\gamma_T=0.
\end{equation}
Consequently,
\begin{equation}
X_{t=0}\sim\rho(x,0),
\qquad
X_{t=T}\sim\rho(x,T),
\end{equation}
that is, the random variable $X_t$ has marginal density $\rho(x,0)$ at the initial time and $\rho(x,T)$ at the final time. As in the previous generative frameworks discussed in this Perspective, these two endpoint distributions correspond to the prescribed initial and target distributions, typically the data distribution and a simple Gaussian reference distribution, respectively, although the stochastic-interpolant formulation itself does not privilege either temporal orientation. The stochastic process therefore defines, by construction, a finite-time bridge between these two distributions. The interpolation function $I_t(X_0,X_T)$ is the deterministic component of the transport, while $\gamma_t$ controls the stochastic fluctuations around it. Together, they generate the family of intermediate probability distributions connecting the two endpoints.
The conceptual advantage of this formulation is that the choice of the Gaussian reference distribution is no longer distinguished from any other probability measure. In contrast to conventional diffusion models, where the Gaussian prior is inherited from the OU process. The Gaussian prior therefore becomes only one particular instance of a much broader class of probability bridges, naturally encompassing conditional generation, domain translation, and entropy-regularized optimal transport problems such as Schr\"odinger bridges~\citep{de2021diffusion}.

The stochastic interpolant separates the construction of the probability path from the dynamics used to realize it. Score-based models, diffusion models, and flow matching therefore emerge as different realizations of the same underlying probability transport. They are not fundamentally distinct generative paradigms (see Sect. 3.1 in~\citep{albergo2025stochastic}). The probability bridge itself becomes the fundamental object of generative modeling, while the stochastic or deterministic dynamics used to sample it is a secondary algorithmic choice.

From the point of view of turbulence, this unified framework is particularly appealing. Many applications naturally require transporting probability between complex flow distributions rather than between turbulent states and Gaussian noise. Examples include conditional generation from sparse observations, data assimilation, super-resolution, translation across Reynolds numbers, and mappings between reduced-order and fully resolved representations. In these situations, finite-time probability bridges provide a flexible probabilistic framework while preserving the rigorous statistical interpretation that underlies score-based generative modeling.

\subsection{Moment Guided Diffusion. Can Generative Models Become Interpretable?}
\label{sec:mgd}

Modern generative models have demonstrated a remarkable ability to reproduce the statistics of complex natural data. Yet, despite this empirical success, the statistical representation learned by these models remains largely \emph{black boxes}. While considerable effort has been devoted to understanding why generative models generalize without  memorizing their training data, an equally fundamental question remains: \emph{can we identify, in physically or statistically meaningful terms, the information about the data distribution that is actually required for generation?}

This question lies at the heart of the rapidly growing field of Explainable Artificial
Intelligence~\citep{vinuesa2026explainable}, and is particularly relevant for scientific applications of generative modelling. In score-based diffusion models, for example, generation is achieved by estimating the high-dimensional score field, $\nabla_x \log p(x,t)$. The score contains sufficient local information about the probability density to construct a stochastic process transporting a simple prior distribution into the data distribution. However, the statistical information encoded in this learned field is not generally expressed in terms of a small set of interpretable observables. In physical applications, one would instead like to know whether the distribution can be characterized by quantities with a direct statistical or physical meaning, for example, correlations, spectra, structure functions, or multiscale observables, and to construct a generative model that reproduces these quantities by design.

The recently introduced \emph{Moment Guided Diffusion} (MGD) framework provides a possible route in this direction by combining the maximum-entropy principle with the non-equilibrium transport mechanism underlying diffusion models and stochastic interpolants \citep{lempereur2026mgd}. The central idea is to replace the problem of reproducing the complete data distribution by the weaker problem of reproducing a prescribed finite collection of statistical constraints. Given a set of $M$ deterministic functions (observables) $ \phi_i(x)$ where $i=1,\ldots,M$, the information extracted from the data is represented by their expectation values
\begin{equation}
    m_i
    =
    \mathbb{E}_{p_{\mathrm{data}}}
    \left[
    \phi_i(X)
    \right],
\end{equation}
where the notation uppercase $X$ denotes a random variable, ($X \sim p_{\mathrm{data}}$), whose possible realizations are the configurations $x$. It is useful here to distinguish between the \emph{observable} $\phi_i(x)$, which is a function evaluated on an individual configuration, $x$, and its \emph{moment} $m_i$, which is the statistical quantity constrained by the model. Thus, the term ``moment'' in MGD should be understood in this generalized sense: the functions $\phi_i$ do not need to be simple powers of the data, but can represent essentially arbitrary statistical descriptors.
{When the selected observables are invariant under an additive shift of the field (as for quantities constructed solely from data increments) the set of constraints must be complemented by an observable fixing the mean (zero) mode. Otherwise, the corresponding distribution remains invariant under global translations and is not normalizable over the full state space.
As a concrete example, consider a dataset consisting of Lagrangian velocity trajectories discretized at $N_x$ times, $ x=(v_1,v_2,\ldots,v_{N_x}).$ In this case, the observables $\phi_i(x)$ could be chosen as
\begin{equation}
    \phi_{j,p}(x)
    =
    \frac{1}{N_x-j}
    \sum_{i=1}^{N_x-j}
    \left|
    v_{i+j}-v_i
    \right|^p,
\end{equation}
such that imposing
\begin{equation}
    \mathbb{E}
    \left[
    \phi_{j,p}(X)
    \right]
    =
    S^{(p)}(\tau_j), \qquad
    \tau_j=j\,\Delta t_{\mathrm{phys}}
\end{equation}
corresponds to prescribing the $p$-th order Lagrangian structure function at the time lag $\tau_j$, where $\Delta t_{\mathrm{phys}}$ denotes the physical time increment between consecutive samples of the discretized Lagrangian trajectory (see also Fig.~\ref{fig:bigfour2}). Choosing, for instance, $p=2$ and $p=4$ at several time lags would define a finite set of constraints, $\phi(x) = \left\{ \phi_{1,2}(x),\ldots,\phi_{n,2}(x), \phi_{1,4}(x),\ldots,\phi_{n,4}(x) \right\},$ encoding second- and fourth-order statistics of the velocity increments and, consequently, their scale-dependent flatness. Other possible choices could include Fourier spectra, multiscale correlation functions, wavelet-scattering coefficients, and related statistical observables. In this sense, the generation procedure can be made \emph{physics-informed}: one explicitly specifies which physical or statistical properties of the data must be reproduced.}

Once these constraints have been specified, since infinitely many densities satisfy the moment constraints, entropy maximization acts as a concave regularization that selects a unique solution. So the maximum-entropy principle selects, among all probability distributions satisfying $\mathbb{E}_{p_\theta} \left[ \phi_i(X) \right] = m_i$, the distribution containing the least additional statistical information beyond these constraints. Formally, this distribution belongs to the exponential family
\begin{equation}
    p_{\theta^\ast}(x)
    =
    \frac{1}{Z_{\theta^\ast}}
    \exp
    \left[
    -\sum_{i=1}^{M}
    \theta_i^\ast \phi_i(x)
    \right],
    \qquad
    Z_{\theta^\ast}
    =
    \int dx\,
    \exp
    \left[
    -\sum_{i=1}^{M}
    \theta_i^\ast \phi_i(x)
    \right]
    \label{eq:max_entrop}
\end{equation}
where the superscript $*$ represent the maximum entropy solution and $Z_{\theta^\ast}$ is the partition function. The Lagrange multipliers $\theta_i^\ast$ are determined implicitly by the constraints
\begin{equation}
\label{eq:constraints_max_entr}
    \mathbb{E}_{p_{\theta^\ast}}
    \left[
    \phi_i(X)
    \right]
    =
    m_i,
\end{equation}
where now the notation $\mathbb{E}_{p_{\theta^\ast}}[\cdot]$ denotes that expectation over the maximum entropy distribution in Eq.~\eqref{eq:max_entrop}, ($X \sim p_{\theta^*}$). The choice of the observables therefore explicitly defines what statistical information the model is required to retain.

The difficulty is that writing down the formal Gibbs distribution does not
make either its parameters or its sampling immediately accessible. The
Lagrange multipliers $\theta_i^\ast$ must be chosen such that to satisfy the constraints in  Eq.~\eqref{eq:constraints_max_entr}. In a conventional maximum-entropy approach, this condition is typically
solved iteratively. For a trial value of the parameters $\theta$, one must
estimate the moments
$\mathbb{E}_{p_\theta}[\phi_i(X)]$, compare them with the prescribed values $m_i$, and update $\theta$ accordingly. Since these expectations are taken with respect to the Gibbs distribution $p_\theta(x) \propto \exp[-\theta_i\phi_i(x)]$, which is itself generally difficult to sample in high-dimensional systems, each update of the Lagrange multipliers may require an equilibrium sampling procedure such as Monte Carlo or Langevin dynamics. The estimation of the maximum-entropy parameters is therefore coupled to a potentially expensive sampling problem.
 
The main contribution of MGD~\citep{lempereur2026mgd} is to circumvent this equilibrium sampling problem by constructing instead a finite-time, non-equilibrium stochastic transport from a simple reference distribution toward the desired maximum-entropy distribution. The connection with modern diffusion models becomes particularly transparent through the stochastic-interpolant formulation~\citep{albergo2022building}. Let $Z$ denote a simple reference random variable, for example a Gaussian variable, and let $X$ denote a sample from the data distribution. One can define the deterministic term of the stochastic interpolant introduced in Eq.~\eqref{eq:stochastic_interpolant} as follow,
\begin{equation}
    I_t(Z,X)
    =
    \cos(\alpha_t) Z
    +
    \sin(\alpha_t) X,
    \qquad
    t\in[0,T],
\end{equation}
where $\alpha_t$ is a function that continuously interpolates between the reference and data distributions over the time interval $[0,T]$. Here, we adopt the time convention used in the original MGD formulation, which is opposite to that employed in the preceding sections: $t=0$ corresponds to the reference distribution, while $t=T$ corresponds to the data distribution. Accordingly, $\alpha_t$ satisfies the boundary conditions $\alpha_0=0$ and $\alpha_T=\frac{\pi}{2}$, so that $I_0=Z$ and $I_T=X$.
For example, a possible choice indicated in~\citep{lempereur2026mgd} is $\alpha_t = \frac{\pi}{2}\, \frac{t}{T}$.
A conventional stochastic interpolant model constructs a stochastic differential equation whose marginal distribution at each time $t$ reproduces the full probability distribution of $I_t$. 
MGD weakens this requirement in a crucial way, searching for the  equality of the selected moments only, without constraining, in principle, the entire distribution:
\begin{equation}
    \mathbb{E}_{p_{\theta}}
    \left[
    \phi_i(X_t)
    \right]
    =
    \mathbb{E}_{{\mathrm{emp}}}
    \left[
    \phi_i(I_t)
    \right]
    \equiv
    m_i(t).
\label{eq:constr_interpol}
\end{equation}
The first expectation is
computed over the distribution generated by the MGD dynamics at time $t$,
whereas the second defines the empirical expectation of the target moment $m_i(t)$ directly computed from sampling the training data and the noise along the prescribed interpolant $I_t$. Thus, whereas a conventional stochastic interpolant attempts to reproduce the complete time-dependent probability distribution, MGD follows only its projection onto a selected finite-dimensional set of observables.

\noindent
To enforce this prescribed moment path, MGD evolves an ensemble of realizations, $\{X_t\}$, according to the stochastic differential equation
\begin{equation}
    dX_t
    =
    \left[
    \eta_i(t)
    -
    \sigma^2\theta_i(t)
    \right]
    \partial_{x}\phi_i(X_t)\,dt
    +
    \sqrt{2}\sigma\,dW_t,
\end{equation}
where repeated indices are summed. The coefficients $\eta_i(t)$ and $\theta_i(t)$ must satisfy
\begin{equation}
    G_{ij}(t)\eta_j(t)
    =
    \frac{dm_i(t)}{dt},
\end{equation}
and
\begin{equation}
    G_{ij}(t)\theta_j(t)
    =
    \mathbb{E}_{p_{\theta}}
    \left[
    \Delta_x\phi_i(X_t)
    \right],
\end{equation}
where
\begin{equation}
    G_{ij}(t)
    =
    \mathbb{E}_{p_{\theta}}
    \left[
    \nabla_x\phi_i(X_t)
    \cdot
    \nabla_x\phi_j(X_t)
    \right],
\end{equation}
in order to make sure that moments are exactly preserved along the entire process: 
\begin{equation}
    \mathbb{E}_{p_{\theta}}
    \left[
    \phi_i(X_t)
    \right]
    =
    m_i(t).
\end{equation}
It is important to stress that MGD is expected to approach, at each time $t$, the maximum-entropy distribution, ($p_{\theta}\rightarrow p_{\theta^*}$), compatible with the prescribed moments $m_i(t)$ only in the large-$\sigma$ limit. See~\citet{lempereur2026mgd} for empirical arguments and diagnostics to estimate how far the generated distribution remains from the corresponding maximum-entropy solution. Consequently, only in this limit does the final-time distribution correspond to the maximum-entropy distribution associated with the statistical constraints extracted from the data; see
\citet{lempereur2026mgd} for further details.
This result also provides a useful connection with score-based generative modelling, Sec.~\ref{sec:score-based}. For a maximum-entropy distribution its score is known explicitly:
\begin{equation}
    \nabla_x\log p_{\theta^\ast}(x,t)
    =
    -
    \sum_{i=1}^{M}
    \theta_i^\ast(t)
    \nabla_x\phi_i(x).
\end{equation}
Thus, the score belongs to the finite-dimensional space spanned by the gradients of the selected observables,
\begin{equation}
    \left\{
    \nabla_x\phi_1(x),
    \ldots,
    \nabla_x\phi_M(x)
    \right\},
\end{equation}
providing perhaps the most direct connection between MGD and the problem of interpretability.
The discrepancy, $D_{\mathrm{KL}} \left( p_{\mathrm{data}} \Vert p_{\theta^\ast} \right)$, therefore represents a model-specification error associated with the choice of the observables $\phi_i$. This error is conceptually distinct from the numerical or sampling error made by MGD in approximating $p_{\theta^\ast}$. If the selected observables provide an incomplete description of the physical system, the resulting maximum-entropy distribution may be interpretable but nevertheless fail to reproduce important features of the data.

From this perspective, interpretability becomes a scientific question. Instead of asking how a neural network internally represents a dataset, one asks which statistical observables are necessary and sufficient to characterize the underlying physical system. In turbulence, for example, one may ask whether low-order correlations and structure functions are sufficient, whether intermittency requires higher-order constraints, multi-point and multi-time  correlations functions and fusion rules \citep{l1996fusion}  or whether multiscale quantities such as wavelet scattering coefficients \citep{cheng2024scattering} provide a more economical representation. More generally, the question {\it what has the generative model learned?} is replaced by the questions {\it which statistical constraints/metrics are necessary to reproduce the physics of interest? what is the minimal set of  bricks to describe  turbulence?} \\
This is also advantageous when learning from limited datasets. Once the observables have been chosen, the information extracted from the training data enters through estimates of quantities such as Eq.~\eqref{eq:constr_interpol}. The maximum-entropy principle then selects the most entropic distribution compatible with these estimated constraints. This does not eliminate statistical error: if the available data are insufficient to estimate the selected moments accurately, the resulting probability distribution may still be biased. Nevertheless, it makes explicit which statistical information from the training set is retained by the model and which information is deliberately left unconstrained.

\subsection{Conditional generation via posterior guidance}
\label{sec:cond_gen}
So far, the discussion has focused on unconditional generative models whose objective is to reproduce the probability distribution of the training data. However, many scientific applications require the generation of samples that are consistent with partial observations or additional physical constraints. Examples include the reconstruction of missing measurements, super-resolution, data assimilation, uncertainty quantification, and, more generally, the solution of inverse problems.
  \begin{figure*}
      \centering  \includegraphics[width=1.0\linewidth]{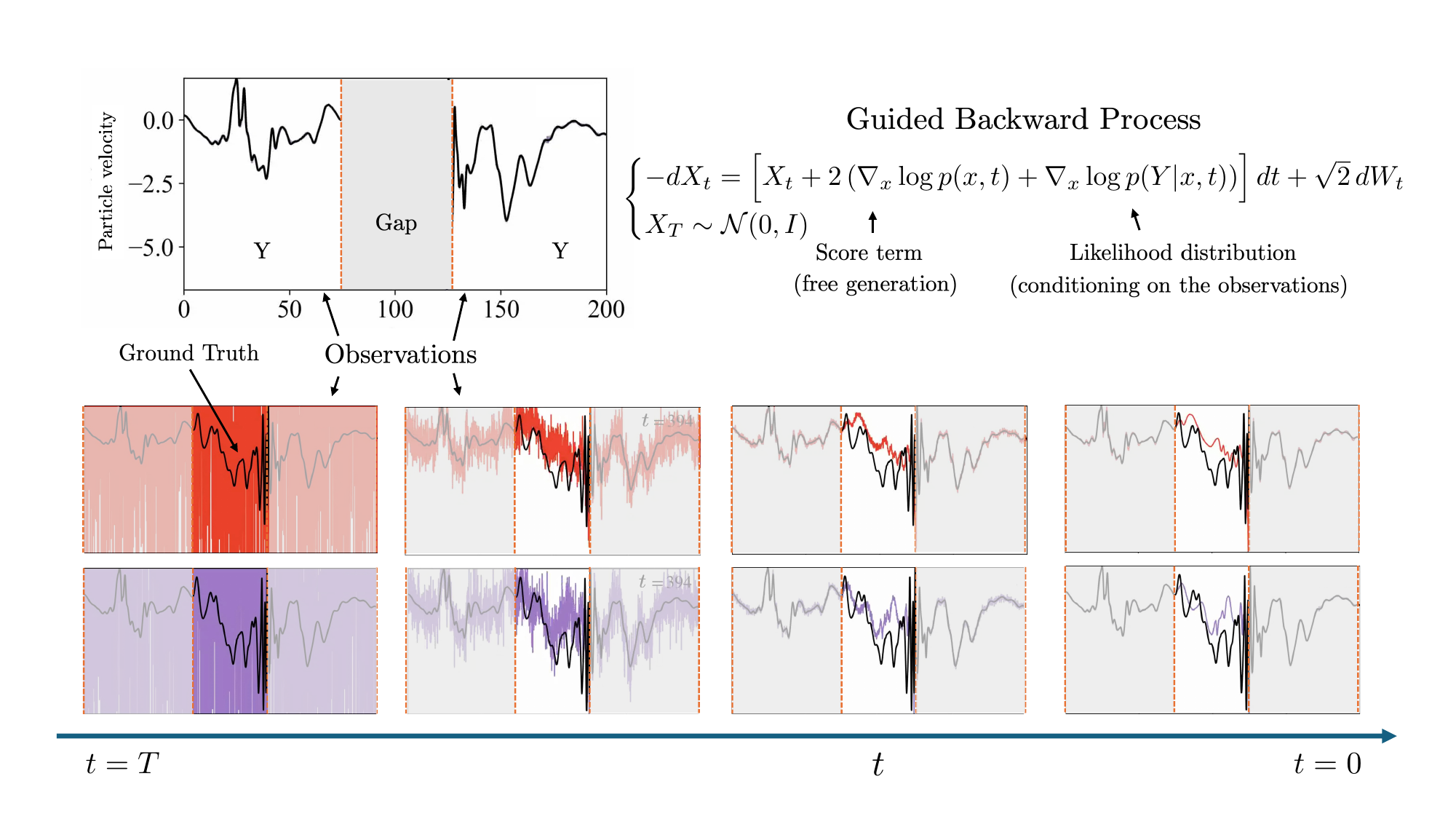}
      \caption{\textbf{Guided diffusion for reconstruction of missing turbulent data.} Top left: example of a Lagrangian velocity trajectory along tracers in which a central time interval is unobserved (gap), while the measurements $Y$ are available on the two sides of the gap. Top right: schematic of the guided backward dynamics, where the usual score term driving unconditional generation is supplemented by the gradient of the likelihood, $\nabla_x \log p(Y|x,t)$, which enforces consistency with the available observations \citep{li2025stochastic}. Bottom: evolution of two independent guided samples during the backward process, from two different initial Gaussian statse to the final reconstructed trajectories. {Here, the backward process is parametrized in terms of the forward diffusion time $t$, spanning in the reverse direction from $t=T$ to $t=0$ during generation.} The observed regions progressively constrain the generated signal, while the missing interval is filled with statistically plausible fluctuations compatible with both the learned prior and the measurements. The black curve denotes the reference ground truth, while the colored curves show the generated trajectories during the reconstruction process. }
      \label{fig:guided_diffusion}
  \end{figure*}

One possibility consists of training conditional generative models, where the conditioning variables are explicitly incorporated during the learning phase. Representative examples include image-to-image diffusion models, such as~\citep{saharia2022palette}, which learn the conditional distribution associated with a specific task. Although highly successful, these approaches generally require retraining whenever the observation operator or the conditioning strategy is modified.\\
A related strategy is \emph{classifier-free guidance}~\citep{ho2022classifier}. During training, the conditioning information is randomly removed for a fraction of the training samples, such that the same diffusion model learns both the conditional and unconditional scores. At inference time, the difference between these two score estimates isolates the contribution associated with the conditioning information. By amplifying this contribution during the reverse diffusion process, the generated samples can be driven toward the desired condition without requiring a separately trained classifier. Amplifying the conditioning contribution provides control over sample fidelity and diversity but generally does not preserve the target conditional distribution. Its use for uncertainty quantification therefore requires a separate assessment of ensemble variability and calibration.
A conceptually different approach consists of conditioning an already trained unconditional diffusion model during the reverse sampling process, without retraining the generative model for a specific observation operator. An early example is RePaint~\citep{lugmayr2022repaint}, which uses a pretrained unconditional diffusion model for image inpainting and repeatedly enforces consistency with the observed regions during the reverse diffusion process while generating the missing content. 
Similar ideas have been extended to physical systems. 

For instance, \citet{gao2025generative} combine gradient guidance with virtual observations to condition the generation of solutions to parametric PDEs, demonstrating the approach for both cylinder flow on an unstructured mesh and turbulent channel flow on a structured mesh, without requiring retraining for different parameter values.
More generally, several guidance strategies have since been proposed for conditioning pretrained diffusion models on observations, as systematically discussed and compared within the framework of \emph{training-free guidance} by \citet{ye2024tfg}. Among these approaches, Diffusion Posterior Sampling (DPS)~\citep{chung2022diffusion} provides a particularly elegant and general framework, as it can accommodate a broad class of observations $Y$ and observation operators without requiring additional training of the diffusion model (see Fig.~\ref{fig:guided_diffusion}).

\begin{figure*}
      \centering      \includegraphics[width=1.0\linewidth]{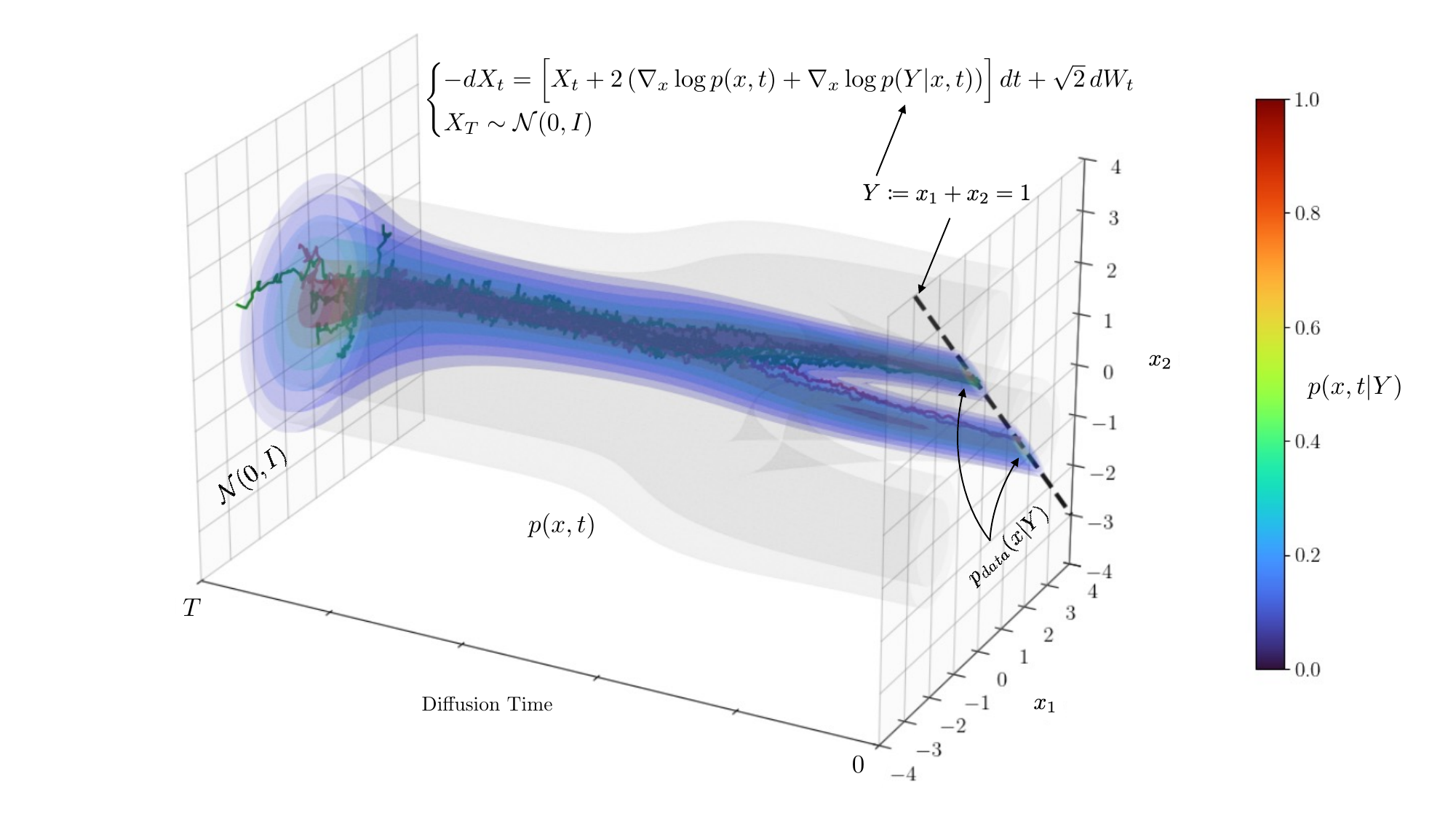}
      \caption{Schematic representation of conditional reverse diffusion in probability space. {Following the notation adopted in this section, the reverse dynamics is parametrized using the forward diffusion time $t$, which evolves from $T$ to $0$ during generation, rather than introducing the backward time variable $\tau=T-t$.} The gray surfaces represent the unconditional marginal probability densities $p(x,t)$ along the diffusion path (as shown in Fig.~\ref{fig:DM_trajectories}), while the colored surfaces show the corresponding conditional densities $p(x,t|Y)$ obtained after imposing the linear constraint $Y:=x_1+x_2=1$. Starting from realizations sampled from the Gaussian prior $\mathcal{N}(0,I)$, the additional likelihood score $\nabla_x\log p(Y|x,t)$ modifies the reverse-time drift and guides the stochastic trajectories toward regions of probability space compatible with the prescribed constraint, see \citep{li2026physics} for an application to 3D rotating turbulence, imposing incompressibility.  As the reverse dynamics proceeds, the conditional probability progressively concentrates around the  manifold $x_1+x_2=1$, indicated by the dashed line, ultimately sampling the conditional data distribution $p_{\mathrm{data}}(x|Y)$. The comparison between the gray and colored probability paths illustrates how likelihood guidance restricts the unconditional generative process to the subset of the learned data distribution consistent with the imposed constraint.}
      \label{fig:DM_trajectories_cond}
  \end{figure*}

{Throughout this section, for notational convenience, we label all intermediate states using the forward diffusion time $t$, rather than introducing the backward time variable $\tau=T-t$, even when discussing the reverse generative process. In this convention, $t=0$ always corresponds to the clean data distribution, so that $X_0\sim p_{\mathrm{data}}$, while increasing $t$ denotes progressively noisier states along the forward diffusion process.}
For the sake of clarity, let $Y$ denote the available observations which are assumed to be related to the unknown data sample $X_0$ through the forward measurement model
\begin{equation}
Y=\mathcal{A}(X_0)+\eta,
\end{equation}
where $\mathcal{A}$ denotes the measurement operator and $\eta$ the errors on the measurements. In general, $\mathcal{A}$ may be nonlinear and $\eta$ may follow an arbitrary probability distribution, leading to a broad class of Bayesian inverse problems~\citep{burns2026and}. For simplicity, and to illustrate the underlying principles of conditional diffusion models, we restrict our discussion to the common inpainting setting, where $\mathcal{A}$ is a linear projection (mask) operator and we assume the observational errors are zero mean Gaussian variables:
\begin{equation}
Y=MX_0 + \eta, \quad\eta\sim\mathcal{N}(0,\sigma^2_\eta)
\label{eq:observation}
\end{equation}
where $M$ is a binary masking matrix selecting the observed degrees of freedom.
The key observation is that the  sampling process should be driven towards the measured $Y$ when available and freely provide new data when not restricted. This is achieved using the score of the posterior distribution, $p(x,t|Y)$,  at difference from the score of the prior unconditioned distribution alone, $p(x,t)$, as done when the data are not available and generation is free. The two are related, for any noisy  realization, via Bayes' theorem using the likelihood of making an observation $Y$ conditioned on the entire signal realization at that given time during the generation:
\begin{equation}
\underbrace{p(x,t\,|\,Y)}_{\text{posterior}}
=
\frac{
\overbrace{p(Y\,|\,x,t)}^{\text{likelihood}}
\;
\overbrace{p(x,t)}^{\text{prior}}
}{
\underbrace{p(Y)}_{\text{evidence}}
},
\label{eq:bayes_guidance}
\end{equation}
 yielding for the conditional score:
\begin{equation}
\nabla_{x}\log p(x,t|Y)
=
\nabla_{x}\log p(x,t)
+
\nabla_{x}\log p(Y|x,t).
\label{eq:dps_score}
\end{equation}
The first contribution to the RHS is precisely the score learned during unconditional training, while the second represents a guidance term that drives, if known,  the reverse process toward samples compatible with available observations as shown in Fig. \ref{fig:guided_diffusion}.  The balance between these two contributions determines the posterior samples. 
Consequently, the same pretrained diffusion model can be, in principle,  employed for a broad range of inverse problems simply by modifying the likelihood function associated with the observations, without altering the learned prior. More precisely, the reverse diffusion process conditioned on the observations $Y$ is governed by
\begin{equation}
\begin{cases}
-dX_t
=
\Bigl[
X_t
+
g^2(t) \left(\nabla_x\log p(x,t) + \nabla_x\log p(Y|x,t) \right)
\Bigr]\,dt
+
g(t)\,dW_t \\[0.3em]
X_T\sim\mathcal{N}(0,I)
\end{cases}
\label{eq:reverse_conditioned}
\end{equation}
{Consistently with the notation adopted throughout this section, we retain the forward diffusion time $t$ also when describing the backward process. The reverse direction of the dynamics is therefore indicated explicitly through $-dX_t$, rather than by introducing the backward time variable $\tau=T-t$ used in Sec.~\ref{sec:backward_proc}. In this notation, the generative process evolves from $t=T$ toward $t=0$, with $X_T$ sampled from the Gaussian prior and $X_0$ corresponding to the reconstructed clean data.} Here, as in Eq.~\eqref{eq:reverse_sde}, the dependence of the score fields on the current process realization is left implicit in both score terms. Sample generation is therefore guided toward trajectories that are simultaneously likely under the learned data distribution and consistent with the available measurements. Fig.~\ref{fig:DM_trajectories_cond} presents how the evolution of the two-dimensional trajectories shown in Fig.~\ref{fig:DM_trajectories} for the unconditional setting is modified by the presence of conditional guidance.
The shaded gray regions represent the evolving unconditional prior distribution, while the colored regions show the corresponding time-dependent posterior probability density constrained by the observations $Y$, according to Eq.~\eqref{eq:bayes_guidance}.\\
Nevertheless, even if there is not need to retrain the score-based model for different guidance (different sets of observations) the sampling is still not doable. The main difficulty lies in evaluating the likelihood of the observation, given the noisy realization of the entire signal at intermediate diffusion step  $p(Y|x,t)$, i.e. we need to have a good guess of the probability to reach the observation by further denoising.
Indeed, the law of total probability gives: 
\begin{equation}
p(Y|x,t)
=
\int
p(Y\,|\,x_0,0)\,
p(x_0,0\,|\,x,t)
\,dx_0,
\label{eq:likelihood_marginal}
\end{equation}
where $p(x_0,0\,|\,x,t)$ denotes the probability to have a final generation $x_0$ given the noisy intermediate sample $x_t$, the so-called   posterior probability density, which is unknown, unlike the forward transition density 
\eqref{eq:ou_transition}.  Consequently, the likelihood in Eq.~\eqref{eq:likelihood_marginal} cannot be evaluated analytically. Moreover, evaluating the multidimensional integral in Eq.~\eqref{eq:likelihood_marginal} at every reverse diffusion step would be computationally prohibitive, making its direct numerical computation impractical for high-dimensional problems~\citep{burns2026and}.
Diffusion Posterior Sampling circumvents these difficulties by approximating the denoising distribution with a Dirac delta concentrated at its conditional mean,
\begin{equation}
p(x_0,0\,|\,x,t)
\approx
\delta\!\left(
x_0-
\mathbb E[X_0\,|\,X_t=x]
\right),
\label{eq:dps_delta}
\end{equation}
thereby replacing the full posterior distribution of the clean sample by its conditional mean. This avoids an intractable marginalization but neglects the remaining conditional uncertainty, so its accuracy and uncertainty calibration must be assessed case-by-case \citep{chung2022diffusion}. 
As discussed in Sec.~\ref{sec:diffusion_models}, the learned score can be use to estimate  this conditional mean through Tweedie's formula,
\begin{equation}
\mathbb E
\left[
X_0
\,\middle|\,
X_t=x
\right]
\simeq
\hat{x}_{0,\theta}(x,t)
=
\frac{1}{\sqrt{\bar\alpha_t}}x
+
\frac{\sigma_t^2}{\sqrt{\bar\alpha_t}}
\,s_\theta(x,t),
\label{eq:tweedie_dps}
\end{equation}
which is evaluated at every reverse diffusion step. Equation~\eqref{eq:tweedie_dps} is simply Eq.~\eqref{eq:tweedie} rewritten using the notation of discrete diffusion models, where the variance of the forward process satisfies
$\sigma_t^2=1-\bar{\alpha}_t$.
The pretrained diffusion model therefore provides, at each reverse step, an estimate of the clean sample that is most consistent with the current noisy realization.
Substituting this estimate into Eq.~\eqref{eq:likelihood_marginal} yields the DPS approximation of the intermediate-time likelihood,
\begin{equation}
p(Y|x,t)
\approx
p\!\left(
Y\,\middle|\,
\hat{x}_{0,\theta}(x,t),0
\right),
\end{equation}
which replaces the intractable marginalization over all possible clean samples by the likelihood evaluated at the estimated posterior mean.\\
\noindent
In practise, to evaluate the latter, one assumes that the observation model in Eq.~\eqref{eq:observation}, admits an energy-based representation
\begin{equation}
p(Y|\hat{x}_{0,\theta},0)
\propto
\exp\!\left[-E(\hat{x}_{0,\theta},Y)\right],
\end{equation}
where the energy function $E$  quantifies the mismatch between the measured data, $Y$ and the best estimation of the reconstructed signal at the step $t$ during the sampling.  Under the assumption \eqref{eq:observation}  we have:
\begin{equation}
E(\hat{x}_{0,\theta},Y)
=
\frac{1}{2\sigma_\eta^2}
\left\|
M\hat{x}_{0,\theta}-Y
\right\|^2.
\end{equation}
 More general observation operators or non-Gaussian noise models can be incorporated simply by replacing the quadratic energy with the corresponding log-likelihood.\\
\noindent
The likelihood score entering the reverse diffusion dynamics is then obtained by applying the chain rule,
\begin{equation}
\nabla_x\log p(Y|\hat{x}_{0,\theta},0)
=
-
\left(
\frac{\partial\hat{x}_{0,\theta}}{\partial x}
\right)^T
\nabla_{\hat{x}_{0,\theta}}
E(\hat{x}_{0,\theta},Y),
\label{eq:energy_guidance}
\end{equation}
where the Jacobian $\partial\hat{x}_{0,\theta}/\partial x$
is efficiently evaluated by automatic differentiation during the reverse sampling process.

\noindent
The reverse dynamics is therefore modified by adding the likelihood guidance to the unconditional score, leading to
\begin{equation}
\begin{cases}
- dX_t
=
\left[
X_t
+g^2(t) s_\theta(X_t,t)
+
\zeta(t)
\nabla_x \log p(Y|\hat{x}_{0,\theta},0)
\right] dt
+g(t)dW_t,\\
X_T \sim \mathcal{N}(0,I)
\end{cases}
\label{eq:dps_backward}
\end{equation}
where $\zeta(t)$ is a guidance coefficient introduced to empirically control the relative strength of the observational constraint during the reverse diffusion process. So $\zeta(t)$ is an important tunable parameter to balance consistency with observations and sample realism.

\noindent
Importantly, all the information associated with the conditioning enters exclusively through the likelihood guidance, while the pretrained score network remains completely unchanged. The diffusion model therefore continues to represent the unconditional data prior, whereas the observations influence the generated trajectory only through the additional guidance term. This separation between prior modelling and observational constraints is the key feature that makes Diffusion Posterior Sampling a training-free conditional generation strategy.
At the same time, the quality of the generated samples ultimately depends on the approximation
$$
p(x_0,0\,|\,x,t)
\approx
\delta\!\left(x_0-\hat{x}_{0,\theta}(x,t)\right),
$$
which replaces the full posterior distribution of the clean sample by its conditional mean. As recently discussed in \citep{burns2026and}, this approximation may become inaccurate when the posterior distribution is broad, multimodal, or strongly non-Gaussian, potentially leading to incorrect uncertainty quantification and sampling artifacts. Understanding how to improve this approximation while preserving the flexibility and efficiency of training-free guidance remains an active area of research.

\noindent
The training-free nature of posterior guidance makes this framework particularly attractive for scientific applications. Unlike conditional diffusion models, a single pretrained generative model can be reused for arbitrary observation operators differentiable on the unknown fields, as, e.g., binary masks or the total flow energy, with the conditioning entirely encoded in the likelihood term. From the perspective of turbulence, this flexibility opens the possibility of reconstructing partially observed flow fields, assimilating sparse experimental measurements, performing super-resolution, or generating physically consistent flow realizations constrained by heterogeneous sources of information. More generally, posterior guidance provides a principled Bayesian framework for combining prior knowledge learned from large simulation datasets with incomplete observations, without requiring the costly retraining of task-specific generative models.

\subsection{Embedding physical constraints}
\label{sec:pcdm}
The posterior guidance strategy described above incorporates observational information during the sampling procedure by modifying the reverse dynamics. An alternative philosophy consists of embedding the desired physical constraints directly into the generative model, so that every generated realization automatically satisfies the prescribed properties. This idea naturally leads to \emph{Physics-Constrained Diffusion Models} (PCDM), where the constraints are enforced during training.
Physics-informed diffusion models have also been developed for flow-field reconstruction. For instance, \citet{shu2023physics} incorporated information from the governing equations as additional conditioning for reconstructing turbulent flows from sparse observations, while \citet{amoros2026guiding} combined physics-informed training with a masked diffusion strategy that directly enforces consistency with sparse measurements during sampling. A more direct integration of physical constraints into the training objective was subsequently proposed by \citet{bastek2025physics}, where the standard diffusion training objective is augmented with a first-principles loss penalizing violations of the governing equations. In this way, the model is explicitly encouraged during training to generate samples satisfying the prescribed physical constraints, with the formulation accommodating equality and inequality constraints as well as auxiliary optimization objectives. 
More generally, physical constraints can be incorporated at different stages of the diffusion framework, either during training or during sampling, and through different enforcement strategies. These range from \emph{soft} constraints, introduced as penalization terms in the training objective, to \emph{hard} constraints, embedded directly into the model architecture or enforced through suitable projections. In the context of incompressible flows, \citet{genuist2026divergence} systematically investigate and compare several score-based diffusion formulations based on the Leray projection, illustrating different strategies for enforcing the divergence-free constraint during turbulent-flow generation.
A recent example of hard constraint enforcement in fully three-dimensional turbulence is provided by \citet{li2026physics}, who incorporate incompressibility and zero-mean momentum directly into the diffusion dynamics, ensuring that these constraints are satisfied by construction throughout the generative process. \\
The construction follows a remarkably similar argument to the one employed in Diffusion Posterior Sampling.
As discussed in Eq.~\eqref{eq:tweedie_dps}, Tweedie's formula provides an estimate $\hat{x}_{0,\theta}$ of the conditional mean of the clean sample given its noisy realization. Suppose now that the desired physical property can be expressed through a {\it linear} projector $P$,  as, e.g.,  the divergence-free projection for incompressible turbulence, the imposition of Dirichlet boundary conditions, or the projection onto zero-mean fields. Because the projection is linear, it commutes with the conditional expectation. Consequently, instead of  applying the physical constraint to every possible realization of the unknown clean sample, it is sufficient to apply it directly to the posterior mean predicted by the diffusion model,
\begin{equation}
P\!\left(
\mathbb E[X_0\,|\,X_t=x]
\right)
=
\mathbb E
\left[
P(X_0)
\,\middle|\,
X_t=x
\right].
\label{eq:projection_expectation}
\end{equation}
This identity shows that the conditional expectation of the physically projected field is exactly the projection of the conditional expectation. Therefore, the estimate of the clean sample satisfying the prescribed physical constraint is obtained simply by projecting the network prediction,
\begin{equation}
P(\hat{x}_{0,\theta}).
\end{equation}
Substituting this projected estimate back into Tweedie's formula yields the corresponding physically constrained score,
\begin{equation}
s_\theta^*(x,t)
=
\frac{1}{\sigma_t^2}
\left(
\sqrt{\bar\alpha_t} \,
P(\hat{x}_{0,\theta})
-
x
\right),
\label{eq:projected_score}
\end{equation}
which replaces the unconstrained score throughout both the training and sampling procedures. Remarkably, the empirical score-matching objective remains completely unchanged. In contrast to conventional physics-informed neural networks, where the governing equations typically appear as additional penalty terms in the loss function, the physical constraints are here enforced in a strong sense, i.e.  ``exactly'', through the projection operator while leaving the optimization objective unchanged.
The only modification consists of replacing the denoised estimate $\hat{x}_{0,\theta}$ by its projected counterpart before evaluating the loss, thereby embedding the desired physical constraint directly into the denoised prediction produced by the network avoiding to  introduce it as an additional penalty term~\citep{li2026physics}. Since the projection is performed at every optimization step, the neural network progressively learns a score whose denoised prediction already belongs to the prescribed physical manifold. As a consequence, the generated samples automatically satisfy the imposed constraints without requiring any correction during the reverse diffusion process.
From the perspective of turbulence, this strategy is particularly attractive because many fundamental physical properties naturally admit linear projection operators. 
Beyond its practical implementation, this formulation reveals a deeper conceptual connection with posterior guidance. In both approaches, the denoised estimate obtained through Tweedie's formula plays the central role. Posterior guidance enriches this estimate with observational information during sampling, whereas physics-constrained diffusion models modify it through suitable projection operators during learning. From this viewpoint, Tweedie's formula becomes the natural interface between data-driven generative modelling and external sources of information, whether arising from observations, physical constraints or prior knowledge. This perspective suggests a unified framework in which increasingly sophisticated physical principles may be incorporated without fundamentally altering the underlying generative architecture.

%% file: chapters/Applications.tex
\section{Applications to Turbulent Flows}
\label{sec:applications}
\begin{figure*}
      \centering
\includegraphics[width=1.\linewidth]{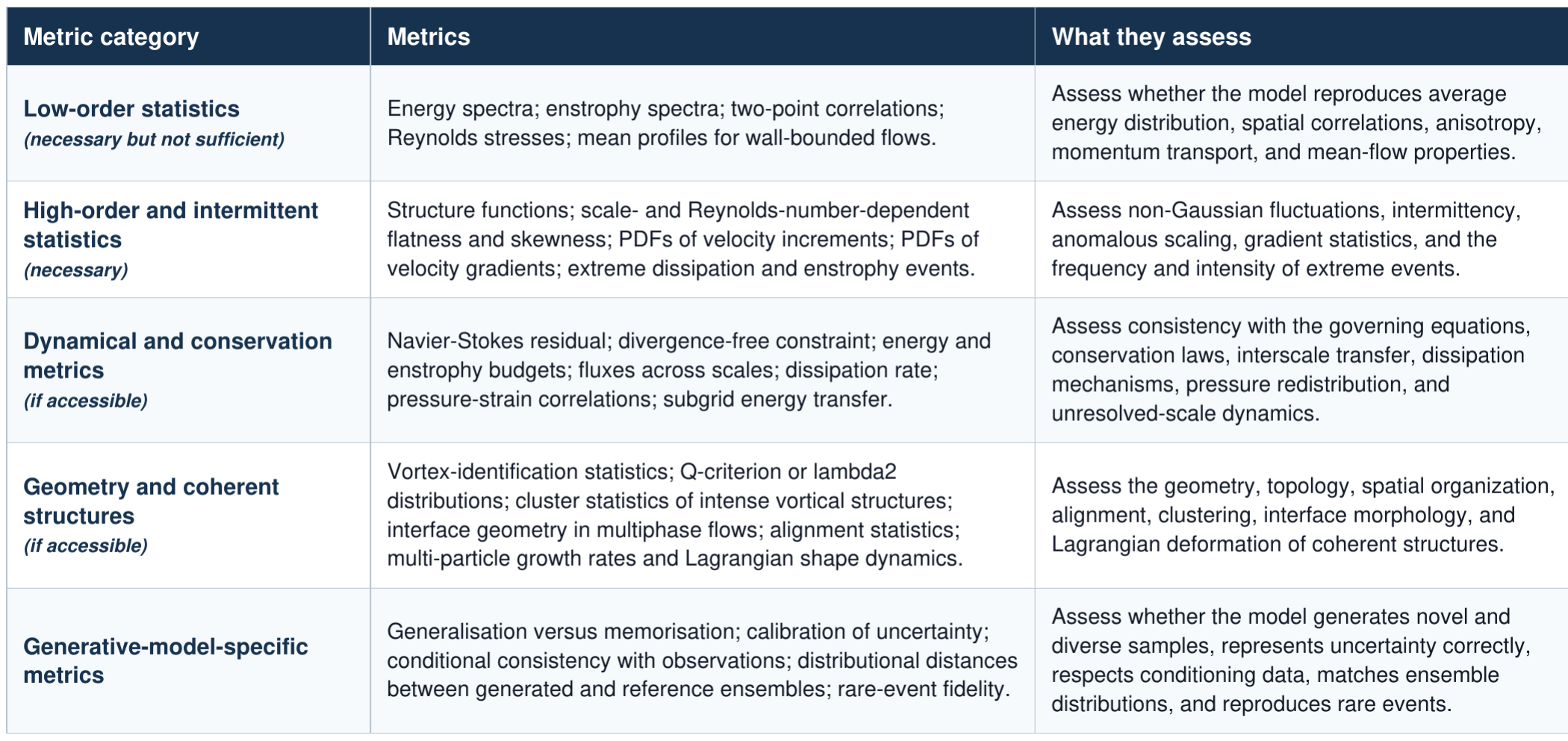}
      \caption{Physics-aware metrics: beyond visual plausibility.
      }
      \label{fig:metrics}
  \end{figure*}

The rapid development of diffusion models and, more broadly, generative artificial intelligence has recently stimulated an equally rapid expansion of applications to turbulence. Within only a few years, generative models have been proposed for problems ranging from synthetic turbulence generation and super-resolution to subgrid-scale modelling, data assimilation, uncertainty quantification and reduced-order modelling. The resulting literature is already vast and continues to grow at an exceptional pace (Sec. \ref{sec:biblio}). The objective of the present section is therefore not to provide a comprehensive review of the existing literature.  We aim to present a few cherry-picked examples to highlight the main classes of applications, the scientific questions they address, and the challenges that remain open. \\
It is important to stress again the main message here, it is not  the novelty of the algorithm to implement the data-driven task of interest here, which will become obsolete soon anyhow, but the set of specific physics targets and metrics that must be used to assess its performance for applications to turbulence, that will probably remain the same being deeply rooted to the  physical backbones (Fig. \ref{fig:metrics} and Sec. \ref{sec:bestpractise}).
\subsection{Eulerian, Lagrangian  and Mesh-free turbulence}
\label{sec:applications_eul_lag_mesh}
From the viewpoint of turbulence, the first natural distinction concerns the type of physical data that the generative model is designed to represent and study. Broadly speaking, current applications may be grouped into three categories. The first concerns 
\emph{Eulerian} fields, instantaneous snapshots defined on fixed spatial grids, and eventually their multi-time evolution. Early developments have primarily considered two-dimensional turbulent flows in simplified set-ups, not necessarily representing real benchmarks or open questions in turbulence. More  recent advances have begun extending these methodologies to fully three-dimensional turbulence -at small/moderate resolutions-, where, nevertheless,  both the computational cost and the complexity of the underlying statistics increase dramatically. A second class of turbulent data is  given by  the \emph{Lagrangian} description, where flow properties are studied  along trajectories of particles/objects moving in the underlying flow, and  where the typical target  is to generate, control and  predict  the temporal fluctuations of single or multi-particles configurations \citep{toschi2009lagrangian}, (Fig. \ref{fig:bigfour4}, left panel). Finally, a third class concerns \emph{mesh-free} representations of turbulence, where the data consist of instantaneous field distributions on an irregular point cloud, as, e.g.,  particle ensembles (Fig. \ref{fig:bigfour4}, right panel) or adaptive discretizations that are not associated with a structured computational mesh. Compared with grid-based representations, these approaches remain considerably less explored owing to the additional challenges posed by their irregular geometry and the presence of a huge scale separation as, e.g, the one that characterizes the preferential concentration of inertial particles in the  sub-Kolmogorov and inertial ranges~\citep{bec2007heavy, toschi2009lagrangian}.

Although these three classes often rely on similar generative methodologies, they represent progressively more challenging learning problems. Lagrangian trajectories are inherently complex because of their strongly intermittent temporal dynamics and long-range correlations, yet each realization remains relatively low-dimensional and possesses a simple sequential structure. Eulerian fields, by contrast, involve orders of magnitude more degrees of freedom. 3D  turbulence combines extreme dimensionality with the full complexity of multiscale interactions (Figs.~\ref{fig:bigfour1}-\ref{fig:bigfour4}), making it one of the most demanding benchmarks for modern generative models. Mesh-free representations add another layer of complexity. The lack of spatial structure prevents the direct use of many of the convolutional and spectral architectures that have driven recent advances in generative modelling, requiring new learning strategies capable of handling both geometric irregularity and turbulent multiscale statistics. These increasing levels of complexity have naturally led to the development of distinct modelling strategies, which we briefly  discuss separately in the following.

\subsubsection{Eulerian Turbulence}
Eulerian fields constitute the description employed in most computational and experimental studies of turbulence and therefore lie at the heart of numerous scientific and engineering applications. It is thus not surprising that most of the works focus on Eulerian fields generation.

An interesting step towards the application of diffusion models to the generation of spatio-temporal Eulerian turbulence is based on Conditional Neural Field Latent Diffusion (CoNFiLD)  \citep{du2024conditional}. CoNFiLD should not be interpreted as a LES formulation or as a solver that explicitly advances the primitive Navier-Stokes equations. It is a data-driven generative surrogate trained on flow sequences obtained from high-fidelity numerical simulations, predominantly DNS, and learns the probability distribution of their spatio-temporal evolution. The training and validation cases include medium intensity turbulent channel flow, flow over periodic hills, wall-bounded turbulence over rough elements, and an irregular pipe flow subject to stochastic forcing.
CoNFiLD does not operate directly on the full flow field, it first compresses each instantaneous turbulent state into a compact latent representation using a conditional neural field based on the SIREN architecture~\citep{sitzmann2020implicit}. The conditional field acts jointly as an encoder and decoder: during encoding, a latent vector associated with each instantaneous field is inferred through an auto-decoding procedure, while the SIREN represents the corresponding continuous spatial field. A latent diffusion model is then trained on sequences of these latent vectors and generates new spatio-temporal latent trajectories by stochastic denoising. The generated sequence is finally passed through the trained conditional neural field to reconstruct the corresponding velocity and pressure fields in physical space. 
From the perspective of turbulence modelling, this work addresses one of the principal obstacles to applying generative models to realistic computational fluid-dynamics problems: the huge dimensionality of three-dimensional turbulent flow fields (Fig. \ref{fig:bigfour1}), which can rapidly render diffusion directly in physical space computationally prohibitive\footnote{State-of-the-art DNS of fully resolved Navier-Stokes turbulence can involve grids containing tens of thousands of points in each spatial direction.}. By transferring the generative process to a learned low-dimensional manifold (latent space), the computational and memory requirements of both training and sampling can be substantially reduced, making four-dimensional (3D$+$time) generative modelling  more tractable.
  \begin{figure*}
  \centering
  \includegraphics[width=1.\linewidth]{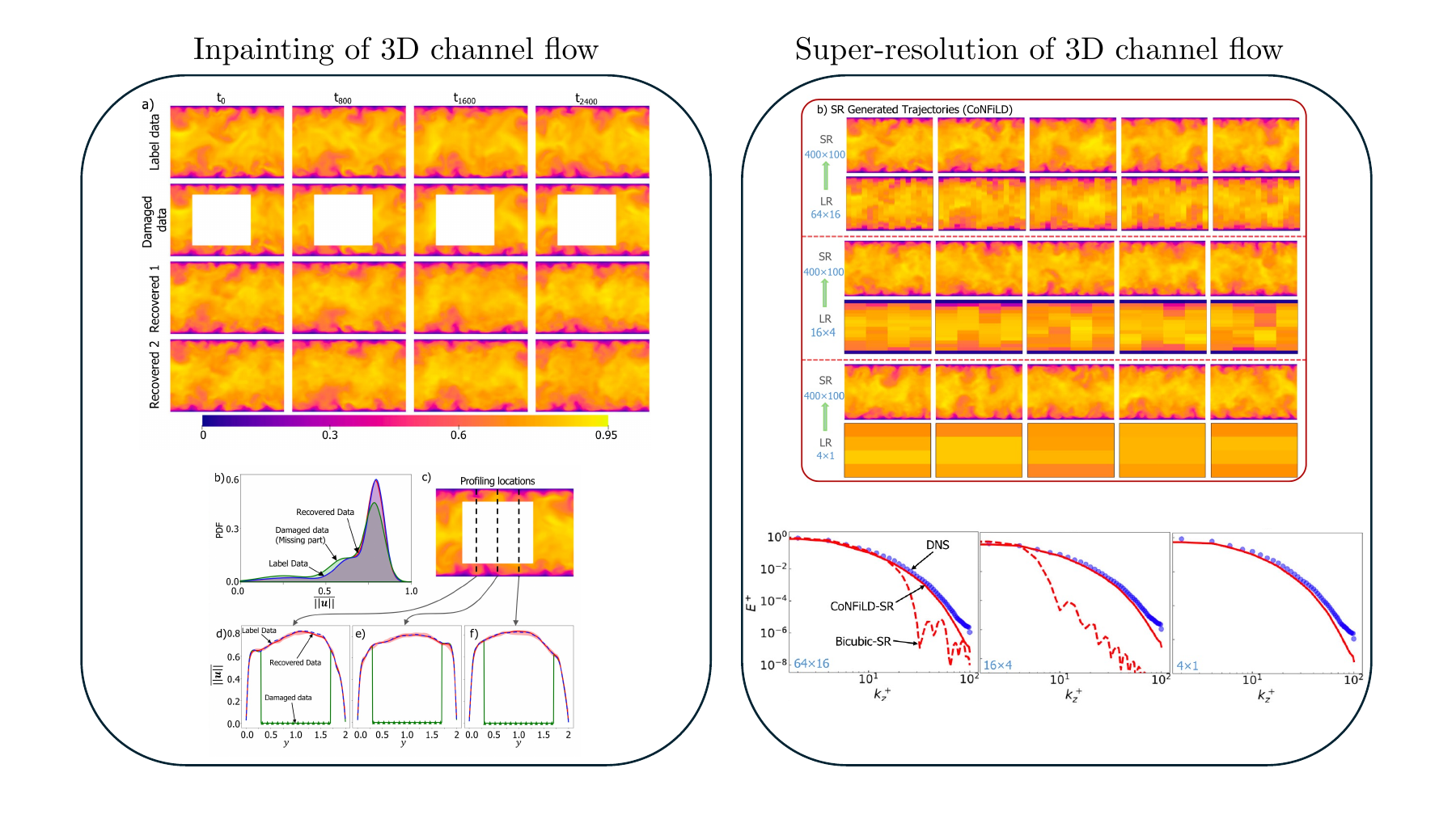}
      \caption{ Conditional generation of three-dimensional turbulent channel flow with CoNFiLD.
      {Left:} Inpainting of missing flow information. Instantaneous velocity-magnitude fields are shown for the reference DNS (label data), artificially damaged inputs containing a masked region, and two independent CoNFiLD reconstructions at several time instants. The lower panels compare the probability density function of the velocity magnitude and representative wall-normal profiles, showing that the recovered fields reproduce the instantaneous structures and the single-point statistics of the reference flow despite the missing data.
      {Right:} Super-resolution of turbulent channel flow from progressively coarsened inputs. CoNFiLD reconstructs high-resolution spatio-temporal fields from low-resolution observations going from $64\times16$ down to $4\times1$. The corresponding energy spectra show that the generated fields recover a substantial fraction of the unresolved small-scale content and remain closer to the DNS spectra than standard bi-cubic interpolation. The figure illustrates the application of conditional latent diffusion to different inverse problems, including data restoration and super-resolution, within the same generative framework. In perspective, future benchmarking against more specific turbulent metrics (see Fig. \ref{fig:metrics}) will unlock the tool’s potential across a broader range of applications, if successful. {Reproduced from Fig. 7  and Fig. 8 (panels b and c) of 
Du, Pan; Parikh, Meet Hemant; Fan, Xiantao; Liu, Xin-Yang;  Wang, Jian-Xun. 2024 {\it CoNFiLD: Conditional neural field latent diffusion model generating spatiotemporal turbulence.}  arXiv:2403.05940v2
 under CC BY license (http://creativecommons.org/licenses/by/4.0/), see \citep{du2024confildarx}.
    }
    }
      \label{fig:CoNFiLD}
  \end{figure*}
 The primary contribution of the work is methodological, and its numerical examples are mainly application-oriented demonstrations, without  attempting to resolve open questions in turbulence physics. Validation is based primarily on comparison with the DNS datasets from which the model learns and includes instantaneous flow structures and some conventional statistical benchmarks. For the turbulent-channel configuration, for example,  the comparison to  mean velocity profiles, root-mean-square (RMS) velocity fluctuations, two-point velocity correlations, and energy spectra is shown; for the more  complex configurations, the analysis relies to first- and second-order statistics and  instantaneous structures, see e.g. Fig.~\ref{fig:CoNFiLD}, where CoNFiLD is applied to two conditional-generation tasks for three-dimensional turbulent channel flow: inpainting of missing flow regions from incomplete observations and super-resolution from spatially coarse inputs. 
These results provide a first evidence that the generated fields reproduce some statistical and spectral properties of the reference turbulence.
The big question that remains open is whether compression is loss-less, and, if not, what is missed in the reconstruction, in particular concerning extreme events, small-scales and high-frequencies turbulent fluctuations and all those properties that make 3D turbulence a genuine anomalous out-of-equilibrium statistical system.
The introduction of latent-space diffusion, namely diffusion performed on a low-dimensional compressed manifold learned by a neural network, represents an important conceptual advance, suggesting a promising route towards scalable generative modeling of 3D turbulent flows, for which operating directly in the original physical state space rapidly becomes computationally prohibitive.

A complementary use of latent-space generative modelling is based on the \emph{Generative Learning of Effective Dynamics} (G-LED) framework for forecasting high-dimensional nonlinear dynamical systems \citep{gao2024gled}. As in CoNFiLD, the central idea is to exploit a compact representation of the system state in order to reduce the computational burden associated with the evolution of high-dimensional fields. The role played by the diffusion model is, however, different. Whereas in CoNFiLD the diffusion process generates the temporal evolution of the latent representation, G-LED evolves the effective dynamics directly in the low-dimensional space through an autoregressive attention-based model~\citep{vaswani2017attention}. A diffusion model is then employed as a probabilistic decoder, mapping the predicted latent dynamics back to the corresponding high-dimensional physical states. In this sense, the diffusion model plays a role analogous to that of the conditional neural field used for reconstruction in CoNFiLD, but here provides a probabilistic mapping from the latent representation to the full flow field.
This separation between latent evolution and probabilistic reconstruction is particularly interesting from the perspective of turbulence modelling.
Learning the temporal dynamics on a compact manifold can substantially reduce the dimensionality of the forecasting problem, while a generative decoder provides a mechanism for representing the distribution of fine-scale fluctuations associated with a given coarse latent state. This is potentially advantageous over deterministic decoders trained with pointwise reconstruction losses, such as the mean-squared error risking to smooth fluctuations that are not uniquely determined by the latent representation. A diffusion-based decoder can instead represent a conditional distribution over possible high-dimensional states and therefore retain stochastic variability that would otherwise be lost by a deterministic reconstruction.

G-LED was demonstrated on systems of increasing complexity, from the Kuramoto-Sivashinsky equation, up to three-dimensional turbulent channel flow. 
Although this idea still requires extensive validation in canonical turbulent flows, it illustrates how generative models may become one component of hybrid reduced-order frameworks, where deterministic latent dynamics are combined with probabilistic reconstruction of the unresolved scales, offering a promising route towards computationally efficient evolution  of high-dimensional turbulent flows. 
At the same time, computational efficiency cannot be considered independently of physical and statistical fidelity. As discussed further in Sec.~\ref{sec:open}, the usefulness of such latent representations ultimately depends on whether they preserve the dynamically relevant information contained in turbulent flows. In particular concerning the well known difficulty to {\it compress} turbulence information in a lossless way and/or without missing important-but-rare  events (Fig.s~\ref{fig:bigfour1}-\ref{fig:bigfour4}).

A different strategy is to attack the problem by generating directly the flow evolution (see, e.g. the Auto-regressive Conditional Diffusion Models (ACDM) proposed by  \citet{kohl2026benchmarking}). In contrast to the latent-space approaches discussed above, the diffusion model operates directly on the physical flow fields and predicts their temporal evolution auto-regressively, without introducing an intermediate reduced representation. At each forecasting step, the model samples the next flow state conditioned on the previously generated one, thereby using diffusion itself as the time-evolution mechanism. A closely related use of conditional diffusion for flow-field evolution has also been explored by \citet{li2024learning}, which will be discussed later in the context of conditional generation and flow-field reconstruction.
The ACDM approach was systematically benchmarked on a few reduced  two-dimensional flow configurations, including incompressible wake flow and transonic flow behind a cylinder as well as 2d planes extracted from 3d isotropic turbulence. 
\begin{figure*}
\centering
\includegraphics[width=1.\linewidth]{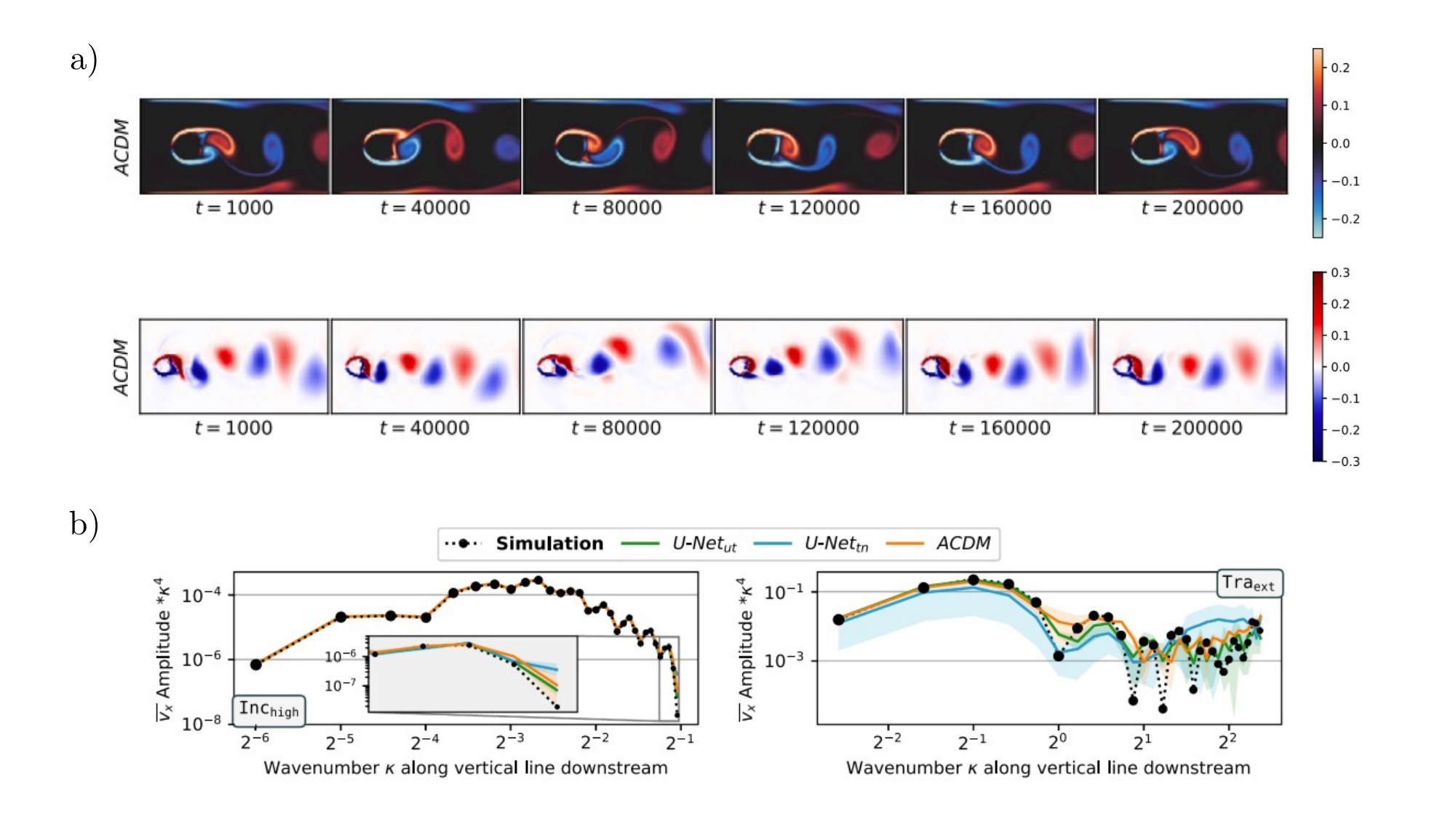}
\caption{ {(Panel a)} Long-horizon vorticity predictions obtained with the autoregressive DM. The top row correspond to a sequence from 2D incompressible wake flow at $Re=1000$, while the bottom  row correspond to a sequence from the transonic flow at $Re=10k$ and Mach number $Ma=0.5$. {(Panel b)} Spatial-frequency spectra evaluated along a vertical line located one cylinder diameter downstream, for (left) incompressible wake flow and for the transonic flow  (right). The predicted spectra are computed and compared against baselines from the mean flow obtained by averaging over a long rollout horizon. View in perspective, successful benchmarking against more specific turbulent metrics (see Fig. \ref{fig:metrics}) will potentially unlock the tool for a broader set of empirical applications. { Reproduced from Fig. F1 (data for ACDM only) and Fig. F2 of Kohl, Georg; Chen, Li-Wei;  Thuerey, Nils. 2026 {\it Benchmarking autoregressive conditional diffusion models for turbulent flow simulation}. Neural Networks p. 108641  under CC BY license (http://creativecommons.org/licenses/by/4.0/), see \citep{kohl2026benchmarking}. }}
\label{fig:Auto_Reg_CDM}
\end{figure*}
The analysis presented in~\citet{kohl2026benchmarking} is primarily focused on benchmarking the predictive accuracy and, in particular, the long-term stability of different autoregressive forecasting strategies, without any exhaustive assessment of high-order turbulence statistics. Within this setting, diffusion-based models exhibit competitive accuracy and notably stable long-time rollouts compared with several established neural forecasting architectures (see, e.g.,  Fig.~\ref{fig:Auto_Reg_CDM}, which compares long-horizon vorticity predictions and the corresponding spatial-frequency spectra for the two-dimensional wake-flow and transonic-flow applications). Their probabilistic character also enables multiple future realizations to be sampled from the same initial condition, providing a natural representation of the forecast uncertainty.

Taken together, CoNFiLD, G-LED, and autoregressive conditional diffusion illustrate three complementary ways of incorporating generative models into the evolution of turbulent flows. Diffusion can be used to generate the dynamics in a compressed latent space before reconstruction in physical space; the latent dynamics can instead be evolved by an autoregressive model, with diffusion acting as a probabilistic decoder; or diffusion can operate directly in physical space and serve as the autoregressive evolution model itself.
Which of these paradigms, or which combination of them, will ultimately prove most effective for fully three-dimensional turbulence remains an open question and will likely depend on the required balance between computational efficiency, temporal stability, and statistical fidelity, as well as on the specific application and the physical quantities of interest. 

A step towards instantaneous 3D generation was recently taken by \citet{li2026physics}, who investigated the synthesis of three-dimensional rotating turbulence. Rotation provides a particularly demanding benchmark because it introduces strong anisotropy over a broad range of scales while retaining the multiscale and intermittent character of fully developed turbulence. The study shows that reproducing a broad hierarchy of turbulent statistics can benefit substantially from the explicit incorporation of physically motivated constraints into the generative process. Building upon the Physics-Constrained Diffusion Model framework discussed in Sec.~\ref{sec:pcdm}, the proposed approach imposes {\it hard} physical constraints through projection operators acting on the denoised estimate, following a philosophy closely related to the projection-based conditioning strategies introduced in Sec.~\ref{sec:cond_gen}. 
These recent developments point toward a direction that is likely to become increasingly important as generative models are applied to more realistic turbulent flows. As the complexity of the underlying physics increases, purely data-driven optimization may no longer be sufficient to recover the correct statistical behavior across all dynamically relevant scales.  These ideas are not specific to diffusion models, but extend naturally to the broader class of generative approaches based on continuous transport and stochastic interpolants, discussed in Sec.~\ref{sec:stoch_interp}. Applications of flow matching to turbulence are beginning to emerge, see, e.g.  \citep{parikh2026conditional} for a conditional flow-matching framework for reconstructing instantaneous near-wall velocity-fluctuation fields from wall-based measurements while explicitly quantifying predictive uncertainty. The method combines continuous-time flow matching with a second learned probabilistic observation operator that maps candidate velocity fields to the corresponding wall measurements while quantifying the uncertainty of this mapping, enabling zero-shot conditional generation without retraining the generative model.
The approach is tested under progressively degraded observability, including sparse, incomplete, and low-resolution wall measurements. 
Applications of flow matching and other stochastic-interpolant methods to turbulence remain considerably less explored than diffusion models. The relative advantages in terms of computational efficiency, conditioning flexibility, and, most importantly, faithful reproduction turbulence metrics still require a systematic assessment of common benchmark problems.
A  common limitation of many recent studies is the difficulty of comparing different approaches to canonical turbulence benchmarks using a common set of statistical diagnostics (see also Sect.~\ref{sec:open}). As a consequence, several important questions remain largely open, including the robustness of these models across different turbulent regimes and their ability to generalize, even after retraining, to flow configurations beyond those for which they were originally developed.\\

\subsubsection{Lagrangian Turbulence} 
\begin{figure*}
\centering
\includegraphics[width=1.\linewidth]{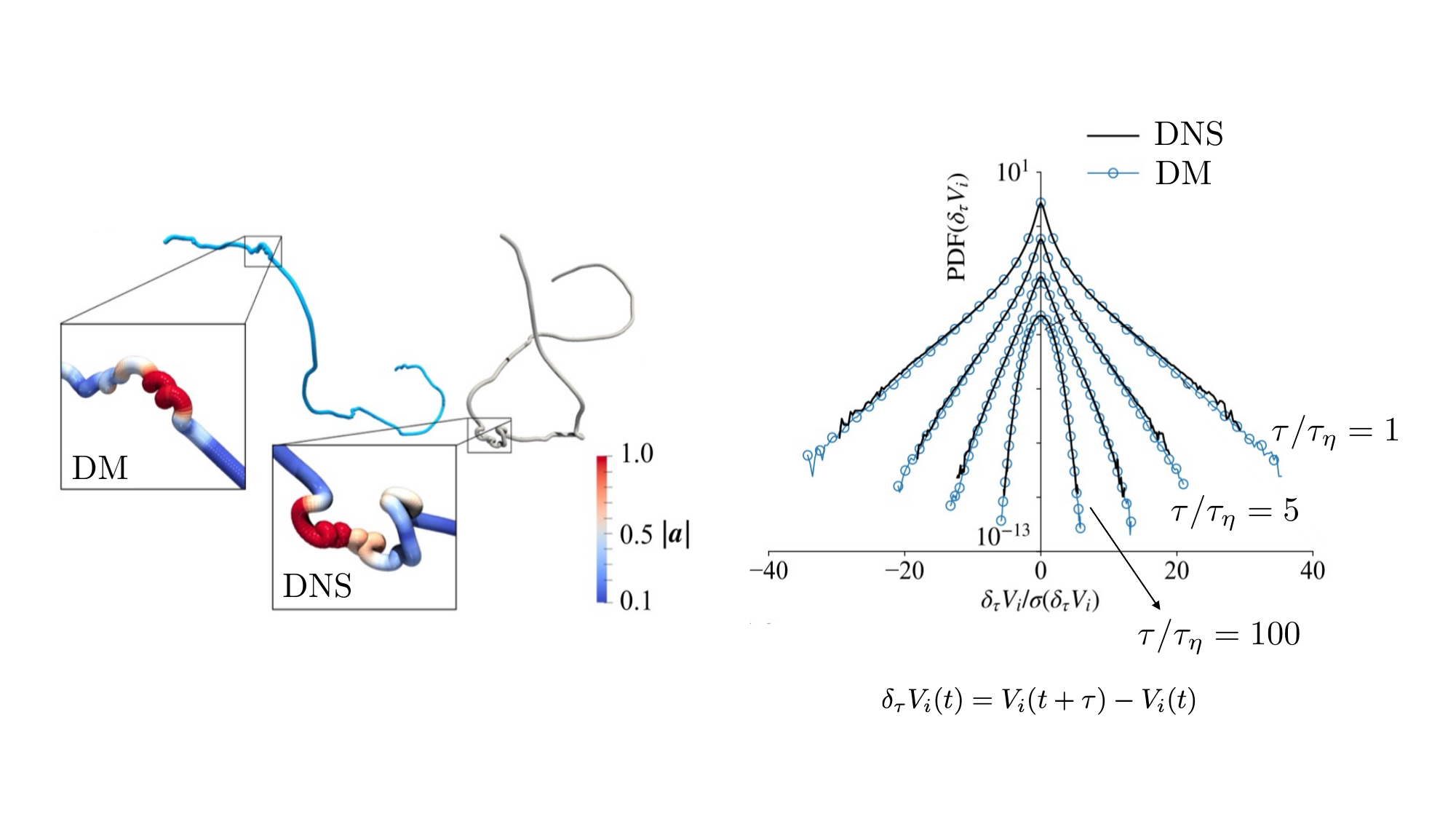}
\caption{ {Generative modelling of Lagrangian turbulence}.
Comparison between DNS and diffusion-model (DM) results for single-particle Lagrangian statistics and trajectories. (Left) Representative three-dimensional particle trajectories, with the local acceleration magnitude highlighted along selected portions of the paths. (Right) Probability density functions of normalized Lagrangian velocity
increments, $\delta_\tau V_i(t)=V_i(t+\tau)-V_i(t)$, for different time lags from $\tau/\tau_\eta=100$ down to $1$. {Reproduced from Fig 1 (panels a and f) of Li, Tianyi; Biferale, Luca; Bonaccorso, Fabio; Scarpolini, Martino Andrea;  Buzzicotti, Michele. 2024  {\it Synthetic lagrangian turbulence by generative diffusion models}. Nature Machine Intelligence 6 (4), 393–403 under CC BY license
(http://creativecommons.org/licenses/by/4.0/), see \citep{li2024synthetic}.} }
\label{fig:DM_Lagr}
\end{figure*}

Lagrangian turbulence constitutes one of the earliest testing grounds for diffusion-based generative models. They provide an ideal setting in which to investigate whether modern generative models can learn the complex statistics of turbulent transport before addressing the substantially larger datasets associated with Eulerian descriptions.
Earlier physics-informed approaches already demonstrated the potential of learning turbulence directly in a mesh-free Lagrangian representation \citep{Tian2023}, proposing a Lagrangian LES in which interacting-particle dynamics are learned from DNS while incorporating physical constraints. Their validation emphasizes statistical fidelity  (not trajectory-wise), including Lagrangian velocity and acceleration distributions, velocity-gradient statistics and energy spectra.  On the other hand,  DM have been able to pass all tests (Fig. \ref{fig:metrics}) for the generation of synthetic single-particle Lagrangian trajectories,  with training from direct numerical simulations of 3D HIT \citep{li2024synthetic}. The model reproduced a broad hierarchy of multiscale observables, including the non-Gaussian probability distributions of velocity increments, anomalous scaling exponents, and the enhanced intermittency observed around the dissipative time scale. Representative generated trajectories and the corresponding scale-dependent velocity-increment statistics are shown in Fig.~\ref{fig:DM_Lagr}, illustrating the close agreement with the reference DNS across widely separated temporal scales. Remarkably, the generated trajectories also contained extreme fluctuations rarer and more intense than those present in the training set, while retaining statistically consistent probability tails. This study therefore provided an early demonstration that diffusion models can generate statistically faithful synthetic Lagrangian datasets without explicitly simulating the underlying Eulerian flow. Importantly, it also remains one of the relatively few studies in which the quality of the generated turbulence was assessed against a broad set of higher-order statistical diagnostics, including probability density functions of velocity increments over multiple time lags and the corresponding higher-order moments characterizing multifractal intermittency (Fig.~\ref{fig:bigfour2}), see also generalization to study charged particles \citep{martin2025generation}.\\
In the high-Reynolds-number limit, turbulent relative dispersion is intimately connected to spontaneous stochasticity, whereby arbitrarily small perturbations can be amplified to macroscopic separations within a few eddy-turnover times \citep{thalabard2020butterfly,bandak2024spontaneous}. Generative models should therefore be tested on their ability to reproduce not only single-particle statistics but also the intrinsically stochastic nature of turbulent pair and multiparticle dispersion \citep{pantea2026turbulent}.

The extent to which such models genuinely generalize beyond the training distribution remains unknown. Likewise, the physical mechanisms implicitly encoded by the denoising process remain difficult to interpret. These limitations connect directly to the broader questions of data efficiency, out-of-distribution generalization, and interpretability, discussed in Sec.~\ref{sec:mem_gen} and Sec.~\ref{sec:mgd}, respectively.

While several subsequent studies have extended diffusion models to more complex Lagrangian settings, including inertial particles and particle-pair dynamics~\citep{li2024generative,pantea2026turbulent}, the generative modelling of Lagrangian turbulence remains comparatively less explored than its Eulerian counterpart. At the same time, related developments have emerged outside the framework of generative artificial intelligence, where effective stochastic models are inferred directly from Lagrangian observations. Recent examples include data-driven Mori-Zwanzig approaches~\citep{de2026data, freitas2026learning}, which aim to learn reduced stochastic equations describing the influence of unresolved degrees of freedom on Lagrangian particle dynamics. Although conceptually distinct from generative diffusion models, these approaches share the broader objective of constructing data-driven probabilistic descriptions of turbulent transport. 

\subsubsection{Mesh-Free Turbulence} 
Among the three classes of turbulent data discussed above, mesh-free and unstructured representations remain by far the least explored from the perspective of generative modelling. Unlike structured Eulerian grids, where convolutional operators and spectral representations naturally exploit the underlying lattice structure, mesh-free data are represented by irregular sets of points or by computational meshes with non-uniform connectivity. Learning in such domains therefore requires architectures that are insensitive to the arbitrary ordering of the points while still exploiting the local geometric relations between neighboring degrees of freedom.

Much of the progress in this direction initially emerged outside the generative-AI community. Contributions are mainly based on Graph Network-based Simulators (GNS) to learn the temporal evolution of physical systems represented by particles \citep{sanchez2020learning}. In this framework, the particles are not generated by the neural network: they constitute the discrete representation of the physical state and are encoded as nodes of a graph, while edges connect neighboring particles and carry information about their relative geometry. The dynamics is then learned through repeated message-passing operations between connected nodes. The same framework was demonstrated across a broad range of systems involving fluids, rigid bodies, deformable materials, and their interactions, and was shown to generalize from one-step training to long autoregressive rollouts and to systems containing substantially more particles than those encountered during training~\citep{sanchez2020learning}. 
Only more recently generative approaches began to emerge in the context of mesh-free and unstructured representations. For example, \citet{gao2025generative} extended diffusion-based generative learning to PDEs discretized on unstructured meshes by combining graph neural-network encoders and decoders with latent-space diffusion and gradient-based guidance. The use of a graph-based latent representation partially alleviates the computational cost associated with applying diffusion directly to the full high-dimensional state. In the unstructured benchmark considered by \citet{gao2025generative}, a mesh containing approximately $1.2\times10^4$ cells is compressed to a latent graph with $1024$ nodes. In principle, this separation allows the physical mesh to be considerably larger than the representation on which the diffusion model operates. The scalability bottleneck is therefore shifted from the diffusion process itself to the graph encoder and decoder, which must still process the full mesh, and to the ability of the compressed representation to retain dynamically relevant small-scale information. The actual demonstration is, however, restricted to a two-dimensional, low-Reynolds-number cylinder flow. Whether such compression remains effective for fully developed three-dimensional turbulence, where the number of degrees of freedom can be orders of magnitude larger and statistically relevant information extends down to intermittent dissipative scales, remains largely unexplored.
\begin{figure*}
\centering
\includegraphics[width=1.\linewidth]{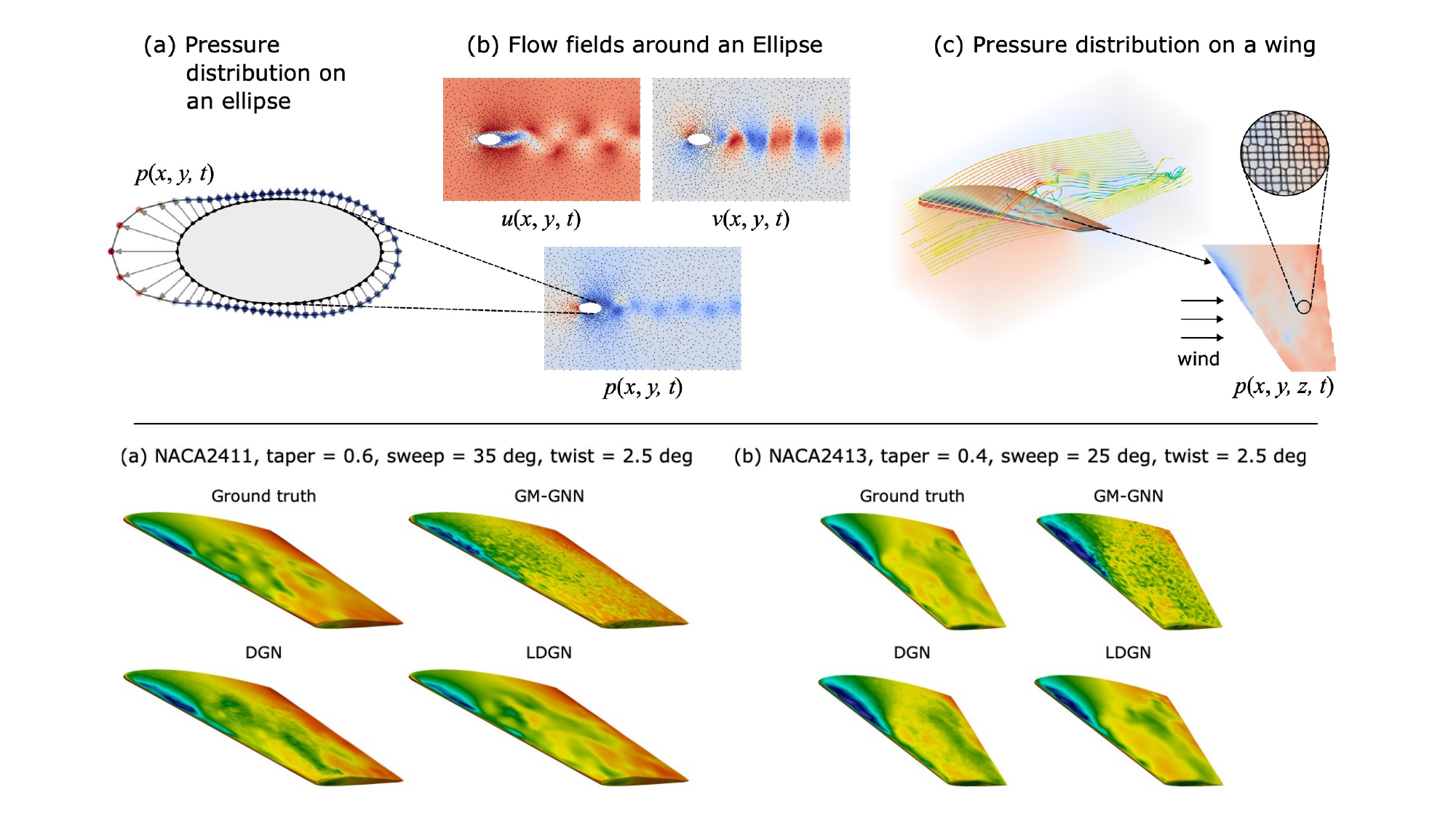}
\caption{ {Generative modelling of mesh-based simulations with diffusion graph networks.}. Top: examples illustrating how diffusion graph networks (DGN) and latent diffusion graph networks (LDGN) can generate distributions of states for mesh-based simulations conditioned on the mesh geometry and physical parameters. The benchmarks include (a) the pressure distribution on an ellipse in two-dimensional flow, (b) the corresponding velocity and pressure fields, and (c) the pressure distribution on a wing in three-dimensional turbulent flow.
Bottom: representative samples for two configurations from the
\textsc{Wing-Test} dataset, comparing the ground truth with predictions generated by the graph-mesh GNN (GM-GNN), DGN, and LDGN models. The comparison illustrates the ability of the generative graph-based approaches to reproduce spatially heterogeneous pressure distributions on complex three-dimensional geometries.  In perspective, more extended benchmarking against turbulent  specific metrics
(see Fig. \ref{fig:metrics}) will make these tools available for a broader range of applications, if successful. 
{Reproduced from Fig. 3 (top panels a,b,c) and Fig. 14 (bottom panels a,b) of 
Lino, Mario; Pfaff, Tobias;  Thuerey, Nils. 2025. {\it  Learning distributions of complex fluid simulations with diffusion graph networks} arXiv: 2504.02843v1, 
under CC-BY license
(http://creativecommons.org/licenses/by/4.0/), see \citep{linoarxive}.}}
\label{fig:Lino2025}
\end{figure*}
A central difficulty of conventional graph neural networks is that message passing is intrinsically local: information propagates only between neighboring nodes at each layer, so that communicating correlations across increasingly large domains requires many  operations and rapidly becomes computationally inefficient. To alleviate this limitation, one can introduce a multiscale graph architecture in which information can propagate through progressively coarser graph representations, thus increasing the effective receptive field while limiting the number of message-passing operations \citep{Lino2025}. A Latent Diffusion Graph Network (LDGN) can also be introduced, in which the full-resolution graph is first compressed and the diffusion process is performed on the resulting latent graph, not  directly on the original mesh. Hence,  the raw graph-scaling computational problem is mitigated through graph coarsening and latent-space generation.

Recent approaches show that for problems containing only approximately $O(100)$ nodes, a single-scale graph network remains feasible, whereas for larger configurations containing $O(10^3)$  the single-scale message passing becomes inadequate, and increasing the number of graph scales substantially improves performance \citep{Lino2025}. 

Representative examples are shown in Fig.~\ref{fig:Lino2025}, illustrating both the latent graph compression and diffusion-based generation introduced by \citet{Lino2025}. The \textsc{Wing} examples also illustrate the potential advantage of diffusion-based graph models over a standard graph-neural-network approach, with the generated pressure fields more closely reproducing the spatial structure of the corresponding ground truth.

Although these studies are not primarily aimed at canonical turbulence, they demonstrate that modern generative frameworks can be naturally combined with geometric deep learning, opening the possibility of applying generative AI directly to irregular and unstructured discretizations.

Despite these encouraging developments, mesh-free and unstructured generative modelling remains considerably less mature than its structured-grid counterpart. The difficulty is algorithmic and also computational. Existing point-cloud and graph architectures~\citep{qi2017pointnet, qi2017pointnet++} are commonly developed for systems containing between $\mathcal{O}(10^3)$ and $\mathcal{O}(10^4)$ points, whereas even a minimal three-dimensional particle distributions in turbulence may already require a much larger number of degrees of freedom to span the huge range of scales separations necessary to describe small-scale clustering, collision, and nucleation effects as well as turbo-phoresis, large scale particles distributions, break-ups etc.. 

Multiscale and latent graph representations can substantially alleviate the direct computational scaling with mesh resolution, but they do not eliminate the latter underlying challenge. The graph encoder and decoder, together with the construction of geometric neighborhoods, must still process the high-resolution representation, while increasingly aggressive compression raises a complementary question: how much physically relevant information can be discarded before the statistics of the generated field are altered? Current generative demonstrations remain at resolutions of order $10^4$ mesh elements or below, and scalability to the $10^5$-$10^6$ degrees of freedom, or beyond, relevant to fully three-dimensional turbulent flows remains largely unexplored.

This limitation is particularly relevant for applications in particle-laden turbulence, rain initiation, spray combustion, reactive suspensions, aerosol coagulation, plankton dynamics, and even planet formation in protoplanetary disks. In these systems, physically relevant observables depend on collective effects such as preferential concentration, caustic formation, and multiparticle distributions conditioned on turbulent fluctuations spanning widely separated spatial and temporal scales. Developing scalable generative architectures that combine geometric flexibility, computational efficiency, and faithful reproduction of multiscale multiphase statistics therefore remains an important open direction for future research.

\subsection{Generative Models for Data Assimilation and Reduced Order Modeling}
\label{sec:data_assimilation}
\begin{figure*}
      \centering
\includegraphics[width=1.\linewidth]{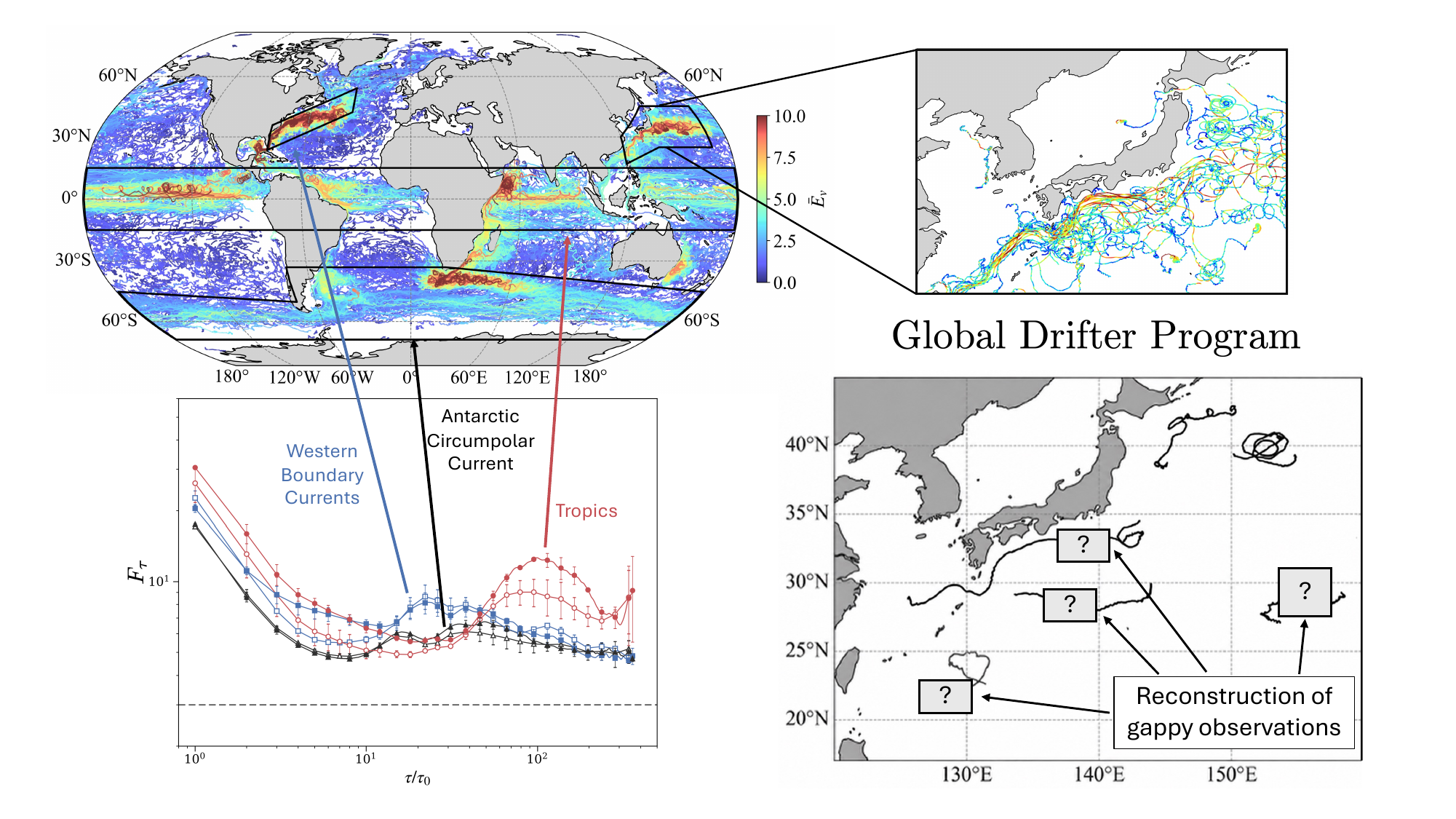}
      \caption{Reconstruction of ocean-drifter dynamics with a pretrained diffusion model. Application of the conditional diffusion model (C-DM) discussed in Sec.~\ref{sec:cond_gen} to velocity records from the NOAA Global Drifter Program~\citep{ldl_global_drifter_program}, following \citep{li2025stochastic}. The pretrained generative model is conditioned on the observed portions of gappy drifter trajectories to reconstruct the missing dynamics. Top: spatial distribution of the drifter dataset, with representative regions including the Tropics, Western Boundary Currents, and the Antarctic Circumpolar Current. Bottom left: scale-dependent Lagrangian flatness in the three oceanic regions, highlighting their distinct intermittent statistics. Solid symbols correspond to the ground-truth observations, while open symbols report the statistics of the reconstructed data in the corresponding geographic regions, identified by the same colors. The reconstructed statistics are computed using velocity increments for which at least one of the two points lies within the reconstructed gap. Bottom right: examples of trajectories containing unobserved temporal gaps.}
      \label{fig:drifters}
\end{figure*}

Beyond the unconditional generation of synthetic turbulent datasets, one of the most promising applications of modern generative models lies in the reconstruction of missing information from incomplete observations. Broadly speaking, these problems fall within the framework of \emph{data assimilation}~\citep{evensen2022data}, whose objective is to combine partial measurements, prior physical knowledge and statistical models in order to estimate the unknown state of a complex dynamical system. 
Several of the architectures discussed above, such as CoNFiLD, were developed  for the generation of synthetic turbulent flows and also for the reconstruction of incomplete observations, while latent-space evolution models naturally connect to reduced-order modelling and dynamical reconstruction. The boundaries between generation, forecasting and data assimilation are therefore often blurred, reflecting the fact that many modern generative frameworks can be adapted to multiple tasks with only minor modifications.
In the following, we shift the emphasis from the generation of physically plausible turbulent realizations to the reconstruction of missing information from partial observations. In this context, the probabilistic nature of modern generative models provides a fundamental advantage over traditional deterministic reconstruction methods. Since multiple physically plausible flow realizations may be compatible with the same set of observations, the objective is to characterize the posterior distribution of all admissible reconstructions (see Sec.~\ref{sec:cond_gen}). 
  \begin{figure*}
      \centering
\includegraphics[width=1.\linewidth]{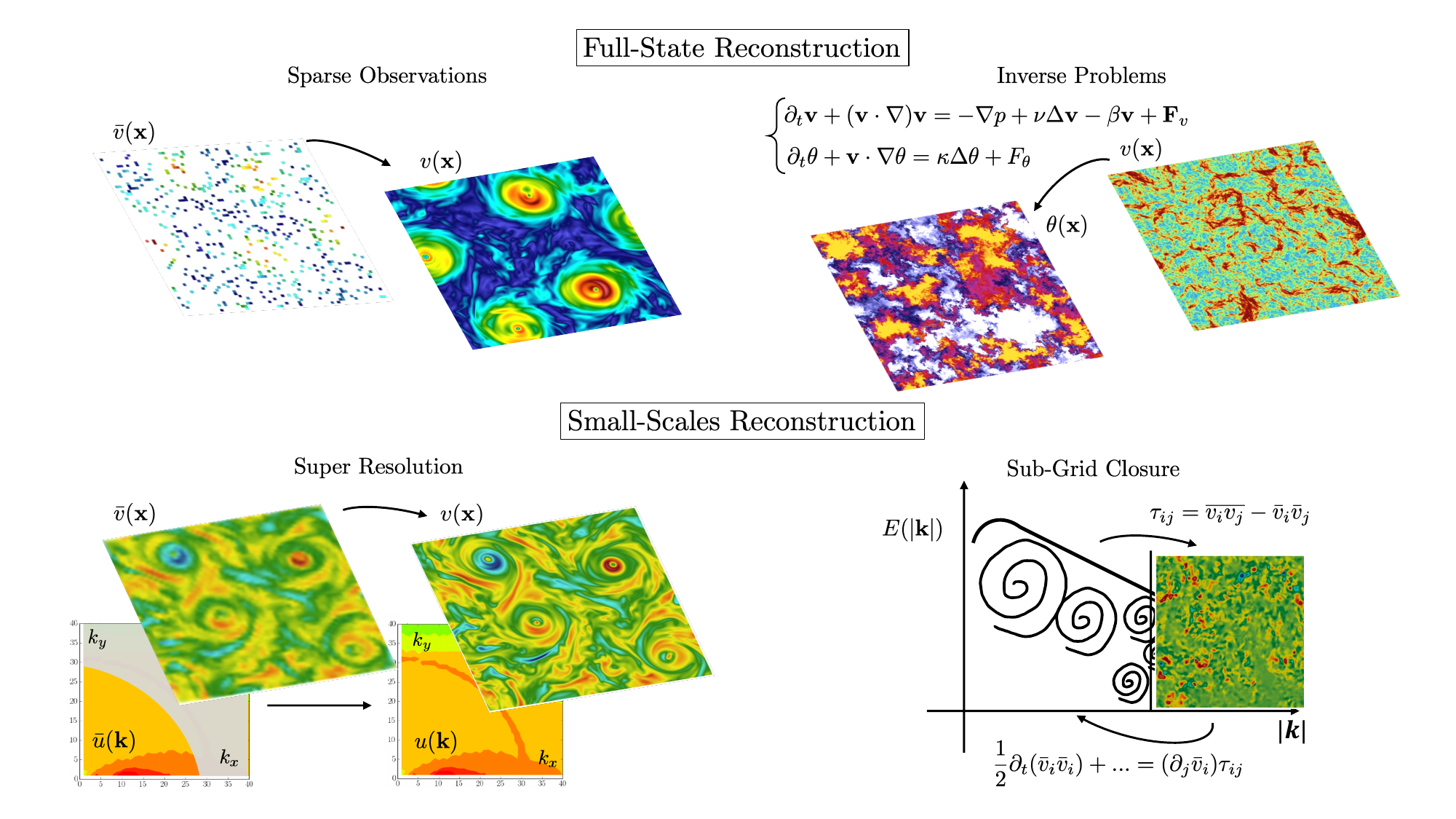}
      \caption{Representative conditional generation and data-assimilation problems in turbulent flows. The reconstruction tasks can be broadly separated into two classes.  \textit{Full-state reconstruction.} Sparse observations can be used to reconstruct the complete turbulent field, while more general inverse problems aim at inferring unobserved fields or variables from partial information on dynamically coupled quantities, as illustrated here for the reconstruction of the velocity and passive-scalar fields. \textit{Small-scale reconstruction.} Super-resolution aims at recovering the unresolved high-wavenumber content of a coarse-grained turbulent field, whereas sub-grid closure seeks to infer the effect of unresolved scales on the resolved dynamics through the corresponding sub-grid stresses. These examples illustrate applications where generative models can be conditioned on incomplete, coarse-grained, or indirect observations to reconstruct statistically and dynamically consistent turbulent fields.}
      \label{fig:dataassimilation}
  \end{figure*}
A useful classification of data-assimilation distinguishes four broad overlapping categories \citep{buzzicotti2023data} (see also Fig. \ref{fig:dataassimilation}): (i) \emph{full-state reconstruction}, where the objective is to recover the complete flow fields from sparse or incomplete observations; (ii) \emph{inverse problems}, where physical quantities that cannot be measured directly are inferred from other observed variables to which they are dynamically coupled; (iii) \emph{super-resolution}, which aims at reconstructing the unresolved small-scale content of a flow from coarse measurements; and (iv) \emph{dynamical reconstruction}, where the objective is to model the effect of unresolved scales on the evolution of the resolved ones, leading naturally to closure modelling and reduced-order simulations. 
Many of the methodologies introduced for unconditional generation can be transformed into powerful reconstruction algorithms by conditioning the generative process on partial observations, as discussed in Sec.~\ref{sec:cond_gen}. An illustrative example is provided by \citet{li2024learning}, who pretrain an unconditional score-based model of spatiotemporal dynamics and subsequently condition the sampling process on sparse measurements, enabling zero-shot reconstruction under different observation settings.
The approach is demonstrated across distinct applications, ranging from two-dimensional Kolmogorov flow to multivariate climate data and experimental PIV measurements. In the climate application, a model trained on historical data from 1979-2022 is used to reconstruct climate states in 2023 from incomplete and noisy observations. Only a sparse fraction of the two horizontal wind components, $u$ and $v$, is observed, while temperature $T$ and surface pressure $P$ are completely unobserved and are inferred by the model by exploiting the learned cross-variable and spatiotemporal correlations.
The methodology is also tested on experimental PIV data, further illustrating its applicability beyond purely numerical settings.
Taken together, these examples highlight the flexibility of conditional score-based models as reconstruction tools across numerical simulations, geophysical datasets, and experimental flow measurements, while retaining a single pretrained generative prior.
As discussed earlier in Sec. \ref{sec:applications_eul_lag_mesh}, Lagrangian turbulence offers a more stringent test bed, combining extreme turbulent properties on a partially reduced 1D sequential data structure. 
Conditional guidance of a pretrained diffusion model, following the methodology discussed in Sec.~\ref{sec:cond_gen}, has been successfully applied  to reconstruct gappy Lagrangian velocity signals \citep{li2025stochastic}, both for  trajectories in 3D homogeneous isotropic turbulence and for signals obtained from real 2D observational data of the Global Drifter Program on the ocean surface \citep{ldl_global_drifter_program}. The latter provides a particularly stringent test, both because the available training data are significantly more limited and because oceanic Lagrangian statistics vary strongly with geographical location.
Remarkably, a single unconditional generative model trained on the heterogeneous drifter dataset can, once conditioned on the available portion of a trajectory, recover the distinct statistics associated with the geographical region in which the drifter evolves. During reconstruction, the model must therefore infer the appropriate statistical prior solely from a single, partially observed Lagrangian trajectory, without access to the surrounding velocity field or to explicit information about geographical location or season (Fig.~\ref{fig:drifters}). 
Thus, conditioning does not merely interpolate the missing measurements, but effectively selects from the learned prior the dynamical statistics compatible with the observed trajectory and its local oceanic environment. A data-driven prior learned exclusively from observations, and therefore without recourse to the governing equations, avoids directly inheriting biases associated with a specific numerical model and provides an independent statistical description of the observed dynamics. Such an approach is particularly appealing for geophysical and environmental applications, where observations are sparse and irregular, and where reducing dependence on imperfect numerical models is a major objective of data assimilation and quality control of the in-situ observations.
These examples clearly illustrate the remarkable breadth of modern data assimilation problems where generation, interpolation, forecasting and inverse problems should not be viewed as isolated tasks: they are different manifestations of the same underlying probabilistic inference problem. 
 At the same time, these encouraging preliminary results also raise one of the fundamental questions discussed earlier in this Perspective: to what extent do conditional generative models genuinely generalize, without memorizing the training data? Approaches based on posterior guidance, such as Diffusion Posterior Sampling (Sect. \ref{sec:cond_gen}), partially alleviate this issue by separating the learning of unconditional prior from the assimilation of observations at inference time. However, whether this separation ultimately leads to a better generalization remains an important open question that deserves more systematic investigation.

Physical priors can also be directly incorporated into conditional reconstruction frameworks  by using, e.g.,  a diffusion-based framework to reconstruct three-dimensional flow fields from sparse measurements and augmenting the training objective with soft physical constraints \citep{amoros2026guiding}. 

The rapid progress of diffusion-based data assimilation should be viewed in the broader context of the evolution of data-driven flow reconstruction. Early studies established super-resolution and reconstruction from sparse or coarse observations as important machine-learning benchmarks for fluid dynamics. In particular, deterministic convolutional neural-network approaches have been probed to recover high-resolution flow fields from strongly under-resolved inputs~\citep{Fukami2019,Kim2021}. These methods predate the widespread use of diffusion models in fluid mechanics and should therefore be regarded as deterministic regression-based reconstruction techniques.
Such approaches are typically trained by minimizing pointwise reconstruction losses, often based on $L_1$ or $L_2$ norms. Although highly effective for recovering dominant flow structures and low-order statistics, a deterministic point estimate generally does not represent the full conditional distribution of unresolved turbulent fluctuations.
More recently \citep{Oommen2026} showed that this limitation extends to neural operators, whose conventional $L^2$-based training may lead to a pronounced spectral bias and an under-representation of the small-scale turbulent fluctuations. By combining operator learning with generative training, they demonstrated that such spectral deficiencies can be potentially mitigated.

In under-determined reconstruction problems, several fine-scale realizations may be compatible with the same coarse or sparse observations, whereas pointwise regression losses favour a single representative prediction and may therefore smooth fluctuations that are not uniquely determined by the input. This limitation provides an important motivation for stochastic generative approaches, which aim instead to sample from the distribution of physically plausible flow fields conditioned on the available observations.

Early generative studies have already emphasized that minimizing a pointwise reconstruction error is not sufficient to guaranty a faithful recovery of turbulent statistics \citep{buzzicotti2021reconstruction}. An example is provided by using deep generative adversarial networks for semantic inpainting of turbulent shapshots,  where strategies based on the pixel-wise $L_2$ error have much less expressivity then adversarially informed generative reconstruction, showing the importance of evaluating the recovered fields through multiscale statistical metrics. Similarly, other  applications based on generative adversarial networks provided evidence of the benefits of such probabilistic formulations for reconstructing intermittent and multiscale turbulent dynamics, including rotating turbulence~\citep{li2023multi} and three-dimensional wall-bounded flows inferred from sparse wall measurements~\citep{cuellar2024three}.

These studies highlighted an important conceptual shift: the quality of a reconstruction for turbulent data should be assessed with a portfolio of different metrics (see Sec. \ref{sec:open} and Fig. \ref{fig:metrics}).
Besides providing improved reconstruction fidelity, DM offer a principled Bayesian framework in which uncertainty quantification is an intrinsic part of the reconstruction process, without entering as an  additional post-processing step. Initial applications focused on super-resolution and sparse-data reconstruction~\citep{shu2023physics}, while subsequent developments introduced multiscale-specific architectures, such as spectrally decomposed diffusion models designed to recover unresolved high-wavenumber fluctuations~\citep{sardar2024spectrally}, and systematic comparisons against GANs, POD-based methods  across a range of turbulent reconstruction tasks~\citep{li2023multi}. 
Recently, \citet{jamaat2026high} proposed a residual-based diffusion model for super-resolution of two-dimensional snapshots extracted from three-dimensional turbulent flows, studying the  reconstructions properties up to large upsampling factors and compared with deterministic and standard diffusion baselines. While promising, further validation against more stringent turbulence diagnostics would be valuable to assess whether the reconstructed fields reproduce the full multiscale statistics of turbulence.
The rapid emergence also of alternative generative paradigms, including stochastic interpolants~\citep{schiodt2026generative}, suggests that the underlying probabilistic formulation may ultimately prove more important than any particular sampling algorithm. 

\subsection{Generative Models for Anomaly Detection}
\label{sec:anomaly}
Most applications discussed so far exploit generative models primarily as samplers. Anomaly detection requires a complementary capability where one must determine whether an available observation is statistically consistent with a given distribution. The central challenge therefore shifts from generating samples to assigning them a meaningful measure of probability, typicality, or compatibility with the learned dynamics.

Diffusion models offer several possible routes toward this objective. Existing approaches generally construct anomaly scores from reconstruction errors, denoising trajectories, estimated diffusion times, scores, or approximations of the likelihood of the data~\citep{liu2025survey}.  Another possibility is to quantify how much diffusion is required before an observation becomes compatible with the learned distribution, associating anomalous inputs with larger inferred diffusion times \citep{livernoche2024diffusion}. Such developments suggest that generative models may be used to synthesize plausible states and also to assess the statistical consistency of an observed one.

However, for turbulence, a direct identification of anomalies with low-probability events would be fundamentally inadequate. Turbulent flows are intrinsically {\it anomalous}: intense velocity gradients, strong dissipative events and extreme fluctuations may occur with very small probability while remaining genuine. Conversely, an observation may not appear extreme according to any individual amplitude or one-point statistic and yet be anomalous (wrong) because of its physically inconsistent spatial organization, temporal evolution, or cross-scale correlations.
The relevant distinction is thus between a {\it black swans}, \emph{rare but admissible}, events and \emph{statistically or dynamically inconsistent} ones. An anomaly score should not depend exclusively on the absolute value of the probability density. Instead, anomaly detection should assess whether the observation is compatible with the appropriate turbulent ensemble, possibly conditioned on its physical context. This context may include the Reynolds number, forcing, geometry, boundary conditions, observation time, resolved large-scales or dynamical history.
This distinction also implies that no universal probability threshold can separate {\it normal} from {\it wrong}  turbulent states. Any decision rule must account for the natural variability of the reference ensemble and the scientific meaning assigned to an anomaly. A sensor failure, a transition between dynamical regimes, an unexpected change in forcing, and a physically admissible extreme event represent fundamentally different departures from typical behavior. Accordingly, useful detectors will likely need to combine several metrics (see Sec. \ref{sec:bestpractise} for a list of realistic turbulent examples).

This question connects directly to the discussion in Sec.~\ref{sec:mem_gen} on why generative models do not simply memorize their training samples. A model that reproduces only the empirical distribution of a finite dataset would interpret as anomalous any previously unseen realization. Reliable anomaly detection instead requires the model to infer a smooth statistical law that generalizes/interpolates  the training set. The same tension discussed there between memorization and generalization therefore becomes operationally decisive in this setting.
The maximum-entropy formulation introduced in Sec.~\ref{sec:mem_gen} provides a particularly transparent perspective. If (a big if) the turbulent ensemble is represented by the maximum-entropy distribution compatible with a prescribed set of statistical constraints, then, once the corresponding normalization is known, the full probability density can in principle be evaluated for any available realization. This offers a route from interpretable generative modelling to anomaly detection: one may assess not only whether a sample has low probability, but also which of the imposed statistical constraints, moments or multiscale observables make it incompatible with the reference ensemble. 
At present, these ideas remain largely unexplored in turbulence.

\subsection{Reinforcement Learning \& Generative Models}
The applications discussed throughout this Perspective treat turbulence primarily as the object to be generated, reconstructed, or forecast. An even more ambitious direction is turbulent-flow control, where the objective is  to predict the evolution of the flow, and also to identify control actions that optimize a prescribed objective. In this setting, one is naturally led to model the probabilistic evolution of the complete controlled system, comprising both the turbulent environment and the actions applied to it. The generative model therefore no longer serves only as a surrogate of the flow, but may instead act as a probabilistic world model that can be queried directly for planning and control.
This perspective naturally connects modern generative modelling with reinforcement learning (RL). In the standard RL framework, an agent repeatedly interacts with an environment by selecting an action, observing the resulting state transition, and receiving a reward~\citep{sutton1998reinforcement}. In turbulence-control applications, the environment is typically governed by the Navier-Stokes equations, making each interaction computationally expensive because it requires the evolution of a high-dimensional nonlinear system. Consequently, many current applications of RL to fluid mechanics remain limited by the cost of generating the large number of trajectories required for policy optimization, motivating the use of reduced-order, surrogate, or offline implementations.

Generative models suggest a different strategy, without repeatedly querying the numerical solver, one may first learn a probabilistic world model for the controlled dynamics, $p(\mathbf{x}_{t+1}\,|\,\mathbf{x}_t,\mathbf{a}_t),$ where $\mathbf{x}_t$ denotes the state of the turbulent flow and $\mathbf{a}_t$ the control action. More generally, one may learn the joint distribution of complete state-action trajectories from a sufficiently diverse ensemble of controlled ground-truth. Once such a model is available, optimization of a particular control objective can be formulated as conditional generation: the sampling process is biased toward trajectories with a large accumulated case-specific reward, in close analogy with the posterior-guidance strategies discussed in Sec.~\ref{sec:cond_gen}. In this way, the expensive Navier-Stokes solver can be largely removed from the inner optimization loop, while the learned generative model provides a probabilistic surrogate for the coupled flow-control dynamics.

This philosophy has recently emerged in the broader machine-learning community, where sequential decision making has been reformulated as a conditional generative modelling problem~\citep{ajay2022conditional}. A representative example is the diffusion-planning framework of \citet{janner2022planning}, in which a diffusion model generates complete state-action trajectories and planning is performed by guiding or conditioning the denoising process toward high-reward or goal-consistent trajectories. The method was demonstrated on standard control benchmarks, including locomotion tasks, Maze2D goal-reaching problems, and robotic block-stacking, illustrating how sampling and planning can effectively be merged into the same generative procedure.
Such approaches blur the traditional distinction between prediction, planning, and control, because trajectory generation itself becomes part of the optimization algorithm.

For turbulent flows, the implications could be profound. Instead of optimizing control actions through repeated direct numerical simulations, one could first learn the joint probability distribution of the coupled flow and control dynamics from a database of controlled simulations or experiments. The optimization could then be performed predominantly within the learned generative model, with only the most promising candidate strategies validated using the full Navier-Stokes equations or the experimental set-up.

Beyond computational efficiency, generative world models may also enable qualitatively different control strategies. The seminal work of \citet{ha2018world} introduced a learned latent model of the environment in which an agent could be trained largely inside imagined rollouts generated by the model itself, and demonstrated the idea on visual-control tasks including car racing and a DOOM environment. In a turbulence setting, an analogous probabilistic world model could support uncertainty-aware planning, risk-sensitive optimization, and robust decision making under incomplete or noisy observations. Because the model represents a distribution over possible future trajectories control strategies could  be optimized for their expected performance and their sensitivity to rare or adverse flow realizations.
These ideas remain largely unexplored in turbulence, they represent a particularly promising direction for future research. A sufficiently accurate generative world model could, in principle, unify forecasting, data assimilation, uncertainty quantification, and control within a common probabilistic representation of the coupled agent-environment dynamics.

%% file: chapters/Bibliometric.tex
\section{Bibliometric footprint of AI in turbulence}
\label{sec:biblio}
\begin{figure*}
      \centering      \includegraphics[width=1.0\linewidth]{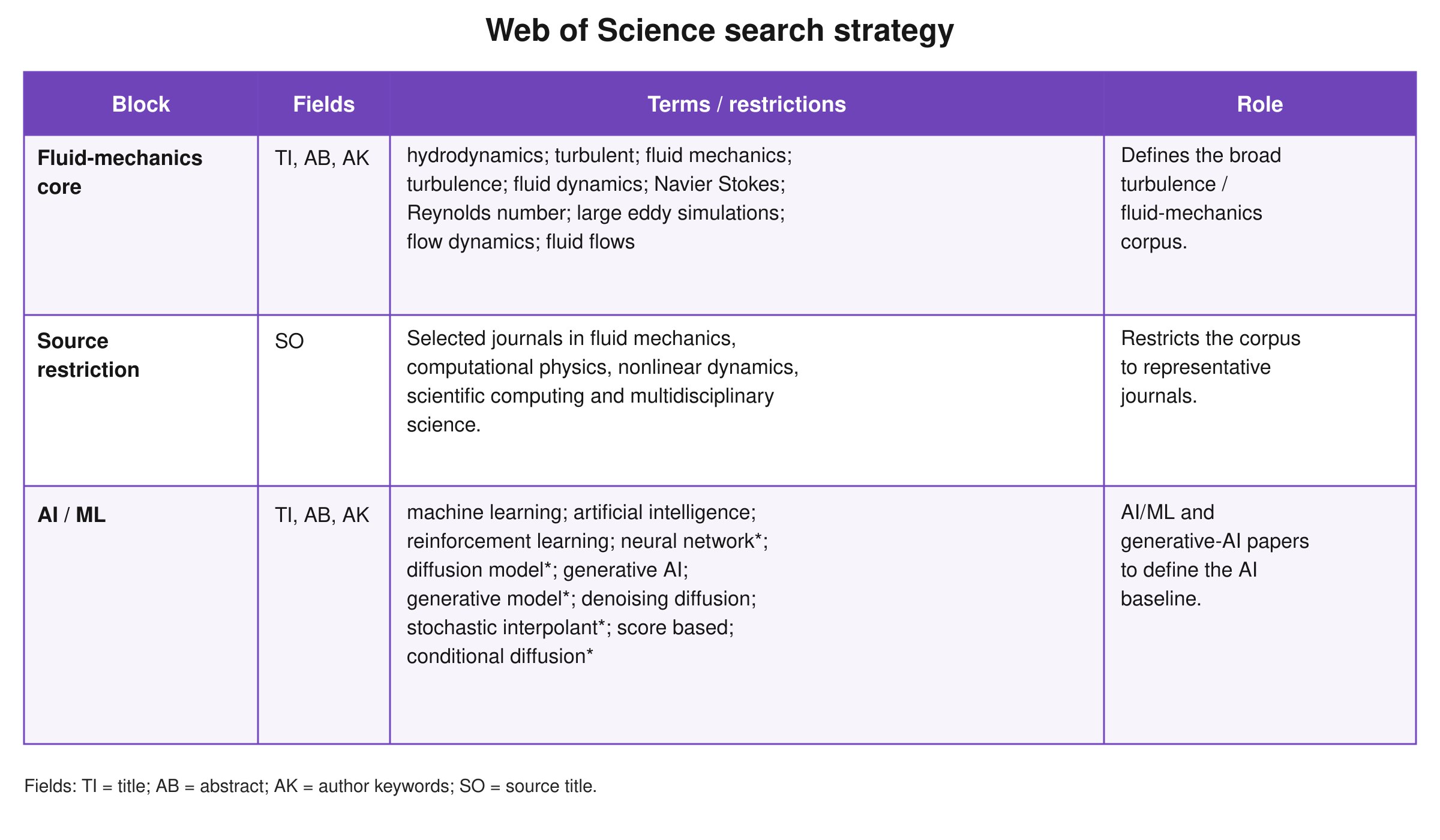}
      \caption{Summary of the Web of Science search strategy used to construct the non-AI turbulence baseline corpus. The query is restricted to the last 10 years 2016-2025 and  combines turbulence- and fluid-mechanics-related terms searched in titles, abstracts, and author keywords, restricts the results to a selected set of journals representative of fluid mechanics, computational physics and nonlinear dynamics, and excludes records containing machine-learning, artificial-intelligence or generative-AI terminology. This baseline is used for comparison with the AI-related turbulence set obtained by including at least one of the AI key-words listed in the third row.}
      \label{fig:WoS}
  \end{figure*}

To provide a first overview of the rapidly evolving field of AI applications to turbulence, and to characterize both the disciplinary background of the community and the venues in which this research is being published, we present a bibliometric analysis below. A qualitative assessment of the recent impact of artificial-intelligence methods on turbulence research can be obtained by comparing two bibliometric subsets extracted from Web of Science over the last decade: a broad turbulence corpus, constructed from generic turbulence-related queries without AI-specific terms, and a more restricted AI–turbulence corpus, obtained by complementing these queries with keywords associated with artificial intelligence, machine learning, and generative tools. The complete list of queries, together with the scientific topics and publication venues of interest for fluid mechanics considered, is reported in Fig. \ref{fig:WoS}.

\begin{figure*}
      \centering      \includegraphics[width=1.0\linewidth]{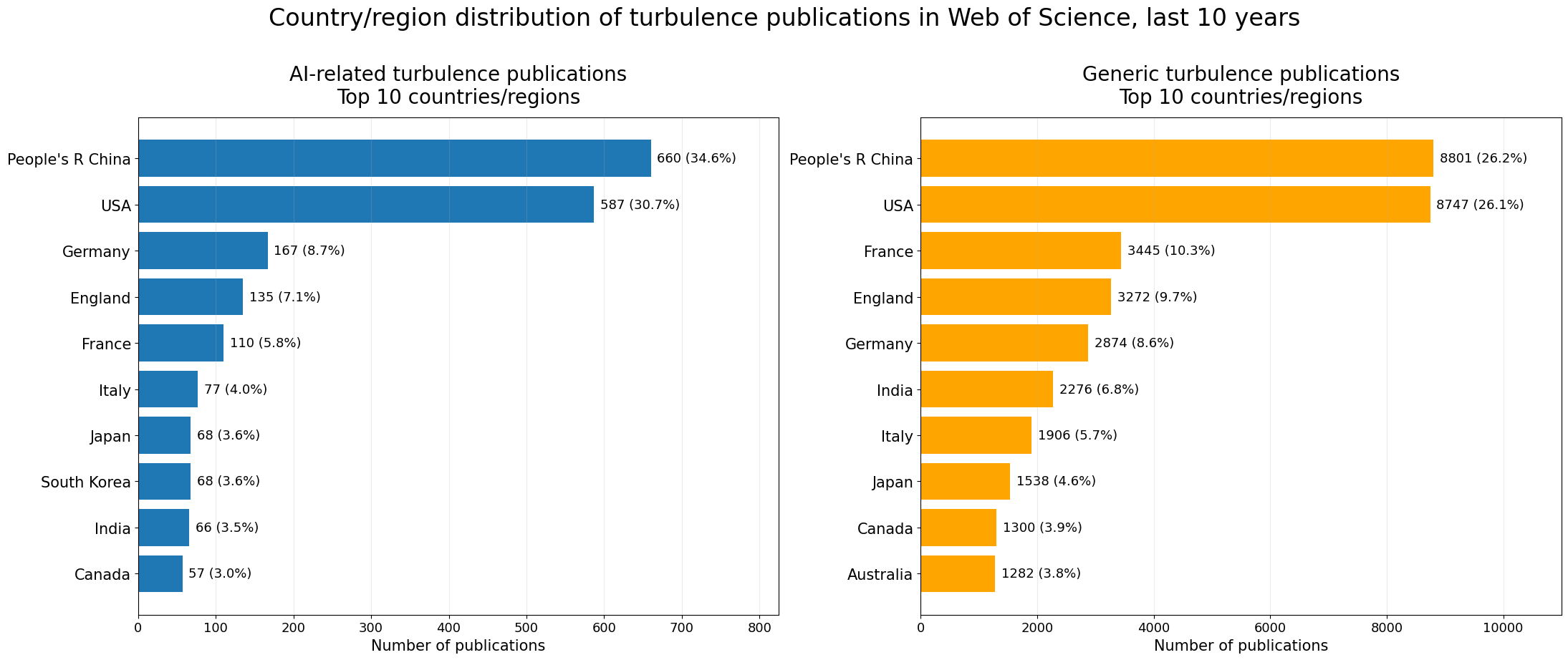}
       \caption{Country/region distribution of turbulence publications indexed in Web of Science over the last ten years. The left panel shows the AI-related turbulence subset, dominated by China and USA; the right panel shows the broader turbulence subset}
      \label{fig:chartcountry}
  \end{figure*}

\begin{figure*}
      \centering      \includegraphics[width=1.0\linewidth]{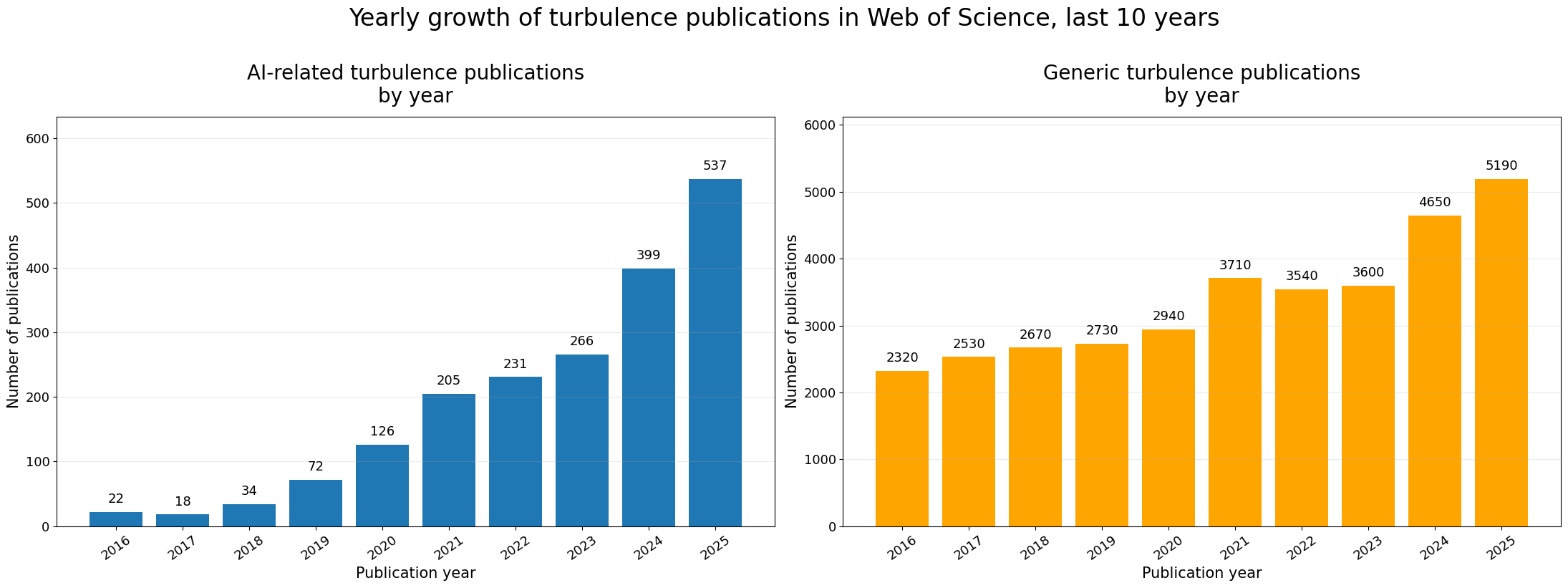}
      \caption{Comparison between the yearly number of publications retrieved using AI-related turbulence queries, left, and generic turbulence queries, right. The AI-related subset shows a strongly accelerating trend, increasing from only 22 records in 2016 to 537 records in 2025, with 2024–2025 alone accounting for almost half of the indexed records in this subset. The comparison highlights that AI-based turbulence research remains much smaller in absolute terms, but has expanded at a far larger relative rate.}
      \label{fig:chartyear}
  \end{figure*}

This comparison is necessarily imperfect. Web of Science records depend on indexing conventions, author affiliations, journal coverage, query design, and the ambiguity of some search key-words. The aim is to have a rough description of  where AI-based turbulence research is being produced, where it is being published, and how its disciplinary identity differs from that of the broader turbulence literature.
The first striking feature (Fig. \ref{fig:chartcountry}) is the much smaller size of the AI-turbulence subset. In the  distribution by the main leading countries, the generic turbulence corpus contains several thousands of records for the leading countries, whereas the AI one contains only hundreds. This confirms that, despite the rapid growth and visibility of AI tools, they still represent a relatively small fraction of the literature. The field is expanding, but it should not yet be confused with the mainstream of turbulence research. 
On the other hand, annual publication counts (Fig. \ref{fig:chartyear}) show that generative AI-related turbulence articles are growing much faster than the turbulence literature as a whole. Although the generic subset increases by only a factor of about two between 2016 and 2025, the AI-filtered subset grows from a few tens of papers per year to more than 500.  This rapid growth  strengthens the need for careful benchmarking, strong baselines, and turbulence-specific validation, since fast-growing methodological fields are especially vulnerable to over-claiming and positive-result bias discussed in Section \ref{sec:Reproducibility}.
  \begin{figure*}
      \centering      \includegraphics[width=1.0\linewidth]{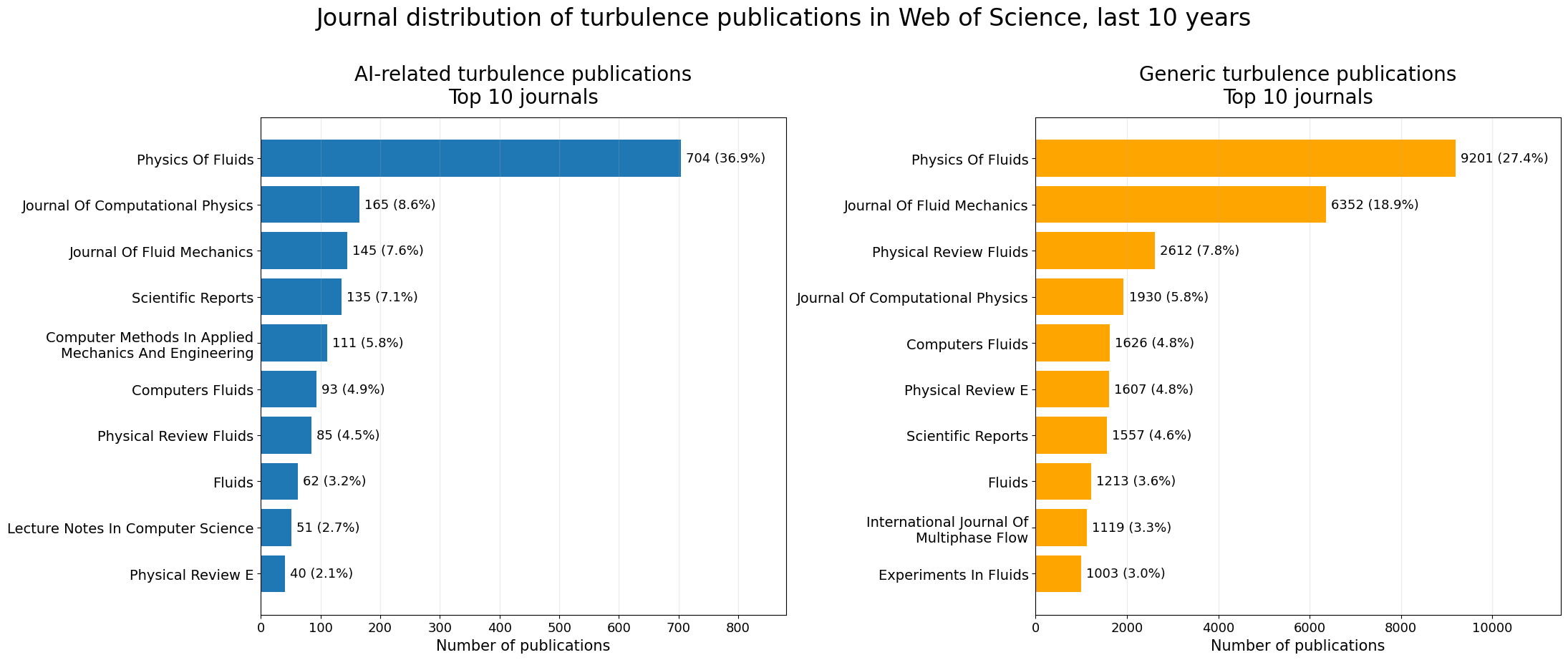}
      \caption{Distribution of the top 10 publication journals for turbulence-related papers indexed in Web of Science over the last decade, comparing the AI-related subset with the broader turbulence subset. The AI-related literature is more strongly concentrated in a few outlets, with Physics of Fluids alone accounting for nearly $37\%$ of the records.}
      \label{fig:chartjournals}
  \end{figure*}

The country-level comparison (Fig. \ref{fig:chartcountry}) shows a strong concentration of research on AI-turbulence in a few national systems. In both the generic and AI-filtered sub-sets, China and the USA dominate the output. Their relative weight appears even slightly more pronounced in the AI subset. 

The journal distribution (Fig. \ref{fig:chartjournals}) provides an equally interesting message. In generic turbulence papers, the leading journals are the traditional venues of fluid mechanics and applied physics.
By contrast, the AI-turbulence subset shows a different hierarchy. Physics of Fluids is  overwhelmingly dominant, followed by the Journal of Computational Physics, Journal of Fluid Mechanics, and others. The chart indicates that AI-turbulence papers are often framed  as contributions to numerical analysis, scientific machine learning, computational methods, and data-driven modeling more than to advance physical understanding and solving open physical problems. 
The bibliometrics comparison reinforces one of the main messages of this Perspective. Artificial intelligence for turbulence is not simply a new subfield that grows within the established turbulence community. It is a contact zone between fluid mechanics, scientific computing, data science, and machine learning. The dominance of a few countries should not only be interpreted as a measure of productivity; it also reflects where computational infrastructure, data availability, and algorithmic expertise are currently aligned.

A final observation concerns generative AI and diffusion models specifically. Their bibliographic footprint is still expected to be much smaller than that of machine learning for turbulence in general, mainly  because score-based and diffusion models entered the scientific-computing literature only recently (around 2020), and we probably still need to see their averaged asymptotic impact.

Overall, we can safely conclude that  AI-based turbulence research has become visible and internationally distributed, and it is published in both classical fluid-mechanics journals and (mainly) in computational-methods venues, confirming its interdisciplinary nature. This growth is promising, but it also strengthens the need for community standards: bibliometric growth alone will not guaranty scientific progress. 

%% file: chapters/Opendata.tex
\section{Toward an ImageNet for Turbulence}
\label{sec:open}
\noindent {\it Generative AI for turbulence will not progress through models alone; it needs shared data, shared tasks, shared metrics, and shared failure cases.}\\
{\sc Why turbulence needs an “ImageNet moment”.}
ImageNet did not transform computer vision simply because it was a large dataset; it provided a shared task, standardized labels, leader-boards, and a common language for comparing algorithms at scale. The ImageNet Large Scale Visual Recognition Challenge evaluated methods for classification and detection on large-scale visual data, enabling systematic comparison across many institutions and algorithms \citep{deng2009imagenet}.
For turbulence, the equivalent need is “more DNS and experimental data”, and a curated ecosystem where generative models can be tested on physically meaningful tasks: reconstruction, super-resolution, forecasting, closure modeling, data assimilation, uncertainty quantification, and out-of-distribution generalization.
Indeed, the next bottleneck for generative AI in turbulence may not be architectural, but communal. As in computer vision before the ImageNet era, progress is currently fragmented across datasets, flow configurations, numerical solvers, resolutions, Reynolds numbers, observables, and validation metrics. A turbulence analog of ImageNet should therefore not be understood as a single monolithic database but as a federated benchmark ecosystem: a set of curated open datasets, standardized tasks, physics-aware metrics, reproducible baselines, and community challenges designed to measure  visual or  low-order statistical plausibility, and -more importantly- physical fidelity.\\
Some resources  already exist  (Sect. \ref{sec:existing}). 
These initiatives are essential, but have not always been constructed or planned to be useful for ML applications. They do not constitute a complete benchmarking culture for generative AI. Existing repositories often focus on (few and big) data access, whereas generative modeling requires carefully designed train/test splits, hidden evaluation sets, agreed physical metrics, baselines, uncertainty protocols, and tests of extrapolation across Reynolds number, geometry, forcing, boundary conditions, and measurement sparsity. As a result, our community is mainly facing a series of bottlenecks preventing real breakthrough applications of data-driven tools for scientific discovery (Compare with the Nobel-prize winning Alpha-fold project in Chemistry): (i)  Existing datasets are often large but not benchmark-ready; (ii)  Benchmarks are often task-specific but not generative-model-specific; (iii)  There is no widely accepted leader-board for turbulence generation, reconstruction, or conditional sampling; (iv) Many papers use private DNS/LES data, making comparison difficult. \\
A very recent and conceptually valuable step in this direction is StocBench \citep{pfister2026stocbench}, which provides a controlled and reproducible comparison of different generative approaches, including diffusion samplers, flow matching, stochastic interpolants and few-step distilled models (Sec. \ref{sec:generative}), under common datasets, inference budgets, and statistically converged reference ensembles. This type of model-to-model benchmarking is precisely what the fluid-mechanics community needs. However, from a turbulence perspective, the physical benchmark remains relatively limited. All methods are tested on a low-resolution, stochastically forced two-dimensional Kolmogorov flow, with only a limited scale separation between forcing and dissipation and with comparatively smooth small-scale dynamics. The main diagnostics, conditional mean and variance errors, energy distance, and the enstrophy spectrum during autoregressive rollouts, primarily probe low-order distributional and spectral properties.  StocBench should therefore be regarded as an excellent benchmark for comparing generative algorithms on stochastic fluid dynamics, but not yet as a benchmark of fully developed turbulence in the stronger physical sense advocated here.

\subsection{Best practice for  turbulence benchmark ecosystems}
\label{sec:bestpractise}
A common best practice to develop benchmarks for turbulence should be specific with regard to four different aspects:  data, tasks, metrics, and protocols.\\
{\sc 1. Data layers}.\\
The repository should contain multiple levels of complexity, as, e.g., representative subsets of the following flow configurations. Ideal  turbulence systems: homogeneous and Isotropic Turbulence, both 3D direct cascade  and 2D inverse cascade turbulent regimes; Passive scalar turbulence; MHD turbulence;  Rotating/stratified turbulence;  Compressible turbulence.
 Wall-bounded turbulence:  Channel flow;   Boundary layers;  Pipe flow; turbulence in the Rough-wall;  Separated flows.
 Multiphase and reacting flows (both using Eulerian-Eulerian and Eulerian-Lagrangian descriptions): bubbly turbulence;  Particle-laden flows;  Droplet-laden turbulence; Turbulent fields relevant to combustion.  Geophysical and environmental flows:  Rotating stratified turbulence;  Atmospheric boundary layers;  Oceanic turbulence;  Convection.
 Experimental datasets: PIV/PTV data; Sparse sensor data;  Noisy partial observations.\\
It is very important to stress the synergetic nature of experimental (or observational) data and numerical data. The former are typically unavoidably messy and noisy, often the only way to assess extreme situations in the dimensionless control-parameters phase-space (high Reynolds, Mach, Rayleigh numbers etc...); similarly, in many cases, experiments are the only reliable means of establishing ground truth because the equations of motion governing certain systems are unknown or too complex to solve accurately, for example, in many multiphase or multi-component flows with high dispersed-phase volume fractions. On the other hand, computational tools based on primitive equations, when available, represent the only way to asses a {\it clean} ground-truth and therefore  a key factor in the data-driven training process. In our opinion, the ideal situation is when, for some window of the physical control-parameters,  both DNS and Experiments overlap as, e.g. in Channel Flows an Rayleigh-Benard convection,  in order to be able to mimic the aleatoric errors affecting the experimental data by deteriorating the DNS ground-truth and vice-versa, to assess the epistemic errors in the DNS modeling for those cases where the model equations are not known. Once the two tools are synchronized and bench-marked on common applications, they can be extended to the phase-space regions where only one of the two is exploitable. \\
{\sc 2. Benchmark tasks for generative AI in turbulence}.\\
Among the many important topics the community should focus on is the following   set of “challenge tracks”: \\
{\bf Unconditional generation}. Goal: generate physically realistic turbulent fields without conditioning.
Example tasks: Generate 1D string of data (e.g. Lagrangian trajectories); 2D (inverse energy cascade regime) or 3D velocity fields at changing the  Reynolds number and the spatial resolutions\footnote{As of now, data-driven tools are very far from being able to generate 3D configurations at resolutions needed for scientific advancements, and this is due to both technical limitations in the ML architectures and lack of training data for high resolution turbulent flows. To the best of our knowledge, state-of-the-art generative tools reach a maximum of $O(100^3)$ grid points, while moderate Reynolds number flows would require at least a factor $10^3$ more or larger, see also the sketched phase-diagram in Fig. \ref{fig:phasespace}}.  Generate scalar fields with prescribed spectra and anomalous scaling.  Generate multiphase configurations with realistic phase-conditioned statistics. {\bf Validation} should include spectra, structure functions, PDFs of velocity gradients, enstrophy, dissipation, and coherent-structure statistics. \\
{\bf Conditional generation and reconstruction}.
Goal: reconstruct missing or unobserved fields from partial information.
Examples:  Reconstruct 3D velocity from 2D slices;  gappy-Lagrangian Trajectories;  Reconstruct full velocity from sparse sensors; infer pressure from velocity;  reconstruct small scales from filtered LES-like fields;  infer bubble-phase or particle-phase velocity from liquid-phase measurements or vice-versa.This is especially relevant for experimental turbulence, where full 3D, time-resolved measurements are rarely available. \\
{\bf Super-resolution and subgrid generation}.
Goal: generate unresolved scales conditioned on coarse fields.
Examples: DNS-like fields from LES-resolution inputs; stochastic subgrid reconstruction; back scatter-consistent small-scale generation; scale-aware super-resolution across Reynolds numbers; multiple Lagrangian trajectories from  partial subset in the bunch. \\
  {\bf Temporal forecasting and probabilistic roll-out}.
Goal: test whether generative models can produce physically consistent future trajectories.
Examples:  Short-time probabilistic forecasting; ensemble generation; extreme-event prediction; long-time statistical stationarity; conditional roll-outs constrained by Navier–Stokes residuals.
The key here is to avoid evaluating only one-step prediction. The benchmark should test whether models preserve invariant measures, spectra, fluxes, and intermittency over time \citep{l1997temporal} and comparing with other tools that embeds the {\it solver-in-the-loop} exploiting a differentiable physics AI environment \citep{um2020solver}.  \\
{\bf Data assimilation and inverse problems}.
Goal: combine generative priors with sparse/noisy observations.
Examples: Recover full flow fields from sparse probes; assimilate PIV-like partial data; infer forcing or boundary conditions; reconstruct hidden variables in multiphase flows; estimate uncertainty in unobserved regions.
This is one of the strongest arguments for generative AI in turbulence: not replacing solvers but providing physically informed priors for ill-posed inverse problems. \\
 {\bf Out-of-distribution generalization}.
Goal: test whether models interpolate or genuinely learn transferable structure.
Examples: Train at lower Reynolds number, test at higher Reynolds number; train on one geometry, test on another; train on one forcing, test on another; train on numerical data, test on experimental data; train on homogeneous turbulence, test on weakly inhomogeneous turbulence.
This is essential because the ImageNet analogy also contains a warning: benchmarks can drive progress, but they can also encourage overfitting to the benchmark itself. Later, ImageNet reassessments showed that benchmark gains do not always translate directly into robust generalization \citep{beyer2020we}.  \\
{\sc 3. Physics-aware metrics: beyond visual plausibility}\\
  Fig. (\ref{fig:metrics}) summarizes a comprehensive evaluation framework of the turbulence-model that combines low- and high-order statistics, physical and conservation checks, coherent-structure diagnostics and generative-model metrics. Together, these criteria test whether the generated fields are  visually realistic, and also dynamically accurate, physically consistent, and faithful to  rare and extreme events. In general, the metric hierarchy is designed to prevent the models from succeeding through superficial similarity.  The strongest benchmark would report performance across all accessible categories and explicitly identify which physical properties cannot be evaluated because of limitations in resolution or access to the governing variables.\\
{\sc 4. Data standards and metadata.} \\
To  make  the data-set useful and satisfying the FAIR principle, it must also be curated, described and reusable. Best-practices concerning technical aspects require:
Using standardized file formats: HDF5, Zarr, NetCDF; supplying metadata as, e.g.,  Reynolds number, grid, timestep, forcing, boundary conditions, numerical scheme, filtering, non dimensionalization; identifying clear train/validation/test splits; providing hidden test sets for challenges; identifying standard pre-processing pipelines; providing  licensing and citation rules, versioning, DOIs and dataset cards; citing solver and simulation provenance. 
Each data set should be accompanied by a “flow card" that describes the governing equations, non-dimensional parameters, numerical or experimental protocol, resolution criteria, known limitations, recommended tasks and physically relevant validation quantities.

\subsection{Existing repositories and 
benchmarks: a critical perspective. }
\label{sec:existing}
Several recent initiatives already point toward the type of open-data and benchmark ecosystem needed for generative AI in turbulence. However, they also reveal that the turbulence community does not yet possess a shared benchmark infrastructure. Existing resources can be seen as important partial prototypes, each addressing a different part of the problem.
An example is the UK Turbulence Consortium database, which collects datasets produced in the context of high-performance turbulence simulations, including canonical and applied configurations. The UKTC database  \citep{UKTC_database} is  embedded in a national HPC ecosystem and reflects the fact that high-quality turbulence data are often the result of large-scale coordinated computational campaigns, not of isolated calculations.  However, like many HPC-derived repositories, it is primarily a data catalog, and not  a full machine-learning benchmark.
Another example is the curated dataset for data-driven turbulence modeling \citep{mcconkey2021curated}. This dataset was explicitly designed for machine-learning-augmented turbulence closure modeling and contains RANS simulations with corresponding DNS/LES labels across several  flows, including periodic hills, square duct, parametric bumps, converging-diverging channels and others. Its main strength is to translate turbulence-modeling data into a form immediately usable for supervised learning. This is extremely valuable for RANS correction and closure modeling, but its direct applicability to generative AI for fully developed turbulence is more limited: the dataset is closure-oriented, whereas generative models require access to high-dimensional instantaneous fields, temporal sequences, multiscale statistics and rare-event details. 
Another  useful example is the database for reduced-complexity modeling developed by the AIAA  and hosted through the University of Michigan. This collection includes time-resolved data for a few  datasets, spanning LES of a turbulent jet, DNS and PIV data for turbulent boundary layers, laminar and pitching airfoils, gust–airfoil interaction, and separated turbulent flow over an airfoil. It was designed to support the conception, training, comparison and evaluation of reduced-order-models. For fully developed turbulence, its relevance is intermediate: some datasets are turbulent and experimentally grounded. The main emphasis is on coarse grained ROM, not on universal high-Reynolds-number turbulence statistics. Still, it offers an important lesson: a useful repository should not simply store data, but should document flow setup, available variables, use cases and examples of model evaluation \citep{Towne2022ReducedComplexityDatabase}. 
Another recent benchmarks come from the broader machine-learning and robotics communities. FluidLab \citep{xian2023fluidlab}, for example, provides a differentiable environment for complex fluid manipulation tasks, with a GPU-accelerated differentiable simulator and benchmark tasks involving solid–fluid and multi-fluid interactions. Its relevance for turbulence is mostly methodological. It demonstrates how a benchmark can combine simulation, tasks, optimization loops and learning algorithms in a reproducible environment. However, it is not designed for fully developed turbulence: its target problems are manipulation, control and differentiable simulation of low-dimensional complex systems.
Similarly, MegaFlow2D \citep{Xu2023MegaFlow2D} provides a large parametric dataset for machine-learning super-resolution in CFD, with more than two million snapshots from 3000 two-dimensional internal and external flow configurations, including paired low- and high-resolution simulations.   Its applicability to fully developed turbulence is limited by the two-dimensional nature of most of the data,  missing  many essential phenomena of the Big-4 challenges   (Figs. \ref{fig:bigfour1}-\ref{fig:bigfour4}). 
A more recent initiative, FD-Bench \citep{wang2025fd}, addresses the lack of fair and modular evaluation protocols for data-driven fluid simulation. It evaluates many baseline models across multiple scenarios under a unified setup and explicitly separates spatial, temporal and loss-function components. This is relevant for the future design of turbulence benchmarks because it recognizes that progress cannot be assessed fairly if architectures, losses, datasets and roll-out protocols are entangled. FD-Bench  provides a valuable benchmarking philosophy, but the turbulence community must enrich such frameworks with physics-aware metrics.
Another relevant direction is represented by HydroGym \citep{lagemann2025hydrogym}, a reinforcement-learning platform for fluid-dynamics flow control. HydroGym provides a standardized  interface, scalable runtime infrastructure and validated flow-control environments, ranging from canonical benchmarks to turbulent-like scenarios.
Finally, application-oriented datasets such as DrivAerML \citep{ashton2024drivaerml}show how high-fidelity CFD data can be organized around realistic industrial challenges. DrivAerML provides an open dataset for road-car external aerodynamics based on 500 parametrically morphed variants of the DrivAer notchback geometry, using consistent scale-resolving CFD workflows and open-source formats. Its value is high for surrogate modeling, design optimization and geometry-conditioned engineering aerodynamics, not for knowledge-based applications 
The main lesson from existing repositories is therefore twofold. First, some of the ingredients already exist. Second, none of these initiatives alone yet plays the role that ImageNet played for computer vision. 

\subsubsection{The JHTDB}
A particularly important reference point is the Johns Hopkins Turbulence Database (JHTDB) 
which is arguably one of the closest existing analogs to an open laboratory for turbulence data \citep{li2008public}. JHTDB provides remote access to multi-terabyte direct numerical simulation and large-eddy simulation datasets through web-service interfaces, allowing users to query velocity, pressure, gradients, and other quantities at arbitrary spatial and temporal locations without downloading the full database. The database includes homogeneous turbulence, wall-bounded turbulence, geophysical turbulence, and wind-farm simulations, with some datasets reaching very large resolutions, including world-record isotropic turbulence snapshots up to ($32768^3$) grid points and high-Reynolds-number channel-flow data.
The importance of JHTDB for AI applications is twofold. First, it provides access to high-quality, physically consistent DNS/LES data at scales that are otherwise inaccessible to most individual research groups.  Second, its query-based architecture anticipates one of the main needs of AI for turbulence: the possibility of extracting customized training samples, local space-time subdomains, particle trajectories, gradients, or filtered quantities from extremely large simulations without treating the database as a simple downloadable file archive. 
At the same time, JHTDB also illustrates the distinction between an open turbulence database and a full AI benchmark. From the perspective of generative AI, JHTDB is an exceptional data source, but it is not yet a benchmark ecosystem. It does not, by itself, define standardized machine-learning tasks, common training/validation/test splits, hidden test sets, baseline generative models, uncertainty-calibration protocols, or physics-aware leader-boards. 
 These issues are not minor technicalities: if training and test samples are too strongly correlated, or if the coarse-grained configuration destroys physically relevant small-scale information, an AI model may appear wrongly successful.
For fully developed turbulence, JHTDB has major strengths, the largest datasets are extremely valuable physically but difficult to use naively for AI: training on full 3D time-resolved data requires careful sampling, compression, parallel I/O, metadata handling, and reproducible pre-processing pipelines, see also the recent deployment of data  using zarr format \citep{yao2025analysis}.

\subsubsection{Blastnet}
BLASTNet (Bearable Large Accessible Scientific Training Network-of-Datasets) is an open-source ecosystem supporting machine-learning research in reacting and non-reacting fluid mechanics. It provides $\sim 5$ TB of simulation data, training and evaluation code, pretrained models, benchmarks, and tutorials for applications such as turbulence modelling, spatio-temporal prediction, super-resolution, and inverse problems. Its resources are distributed through GitHub and Kaggle to make large-scale scientific ML datasets more accessible and reusable \citep{blastnet}.  The benchmark analysis showed that predictive performance can scale with model size and computational cost, that architecture matters significantly, especially for smaller models, and that physics-based gradient losses can remain beneficial as model size increases. 
 BLASTNet therefore provides at least three important lessons. First, turbulence datasets should be made available in formats that are directly usable by ML practitioners. Second, repositories should include  also tutorials, baseline models, code, pre-trained weights, and benchmark tasks. Third, a useful benchmark should evaluate prediction error,  scaling behavior, computational cost, and the effect of physics-informed losses. 
At the same time, BLASTNet also has limitations if the target is a general benchmark for fully developed turbulence. Its emphasis is on compressible, reacting and non-reacting flow physics, which is extremely valuable but not equivalent to a complete coverage of turbulence. Moreover, although BLASTNet includes benchmark activities, it does not yet define a universal set of turbulence-generation tasks with hidden test sets, community leaderboards and physics-aware multi-objective metrics (Fig. \ref{fig:metrics}). BLASTNet is therefore best viewed  as a strong prototype showing how turbulence data can be made ML-ready and how benchmark tasks can be organized around large-scale three-dimensional flow fields.
 It remains focused on particular classes of compressible and reacting flows and on specific benchmark tasks, not  on basic turbulent extreme properties (see Big-4 in Figs. \ref{fig:bigfour1}-\ref{fig:bigfour4}). 

\subsubsection{Smart-TURB}
Among existing initiatives, the Smart-TURB portal \citep{smartturb2026} is also particularly relevant because it is explicitly conceived as a software infrastructure for turbulence and complex-flow data.
A central strength of Smart-TURB is its emphasis on statistically rich datasets (many uncorrelated samples). 
This requirement becomes particularly stringent for Lagrangian turbulence, for which Smart-TURB provides two especially valuable datasets. The first is TURB-Lagr \citep{biferale2023turblagr}, a database containing hundreds of thousands of three-dimensional particle trajectories obtained from DNS of homogeneous and isotropic turbulence. The database includes particle positions, velocities, and accelerations, together with velocity-gradient information sampled along the trajectories. TURB-Lagr was designed as an open testing ground for Lagrangian turbulence, data assimilation, and data-driven modeling, and is directly accessible through the Smart-TURB infrastructure.
Its relevance for generative AI goes well beyond its number of trajectories. Lagrangian signals combine strong temporal correlations, violent non-Gaussian acceleration fluctuations (see Fig. \ref{fig:bigfour3}), anomaloius scaling,  rare events, and non-Markovian dependence on the complete history experienced by a particle. Moreover, trajectories provide a naturally mesh-free representation, making TURB-Lagr particularly suitable for testing generative architectures that operate directly on time series, particle clouds, or irregularly sampled observations. TURB-Lagr has already demonstrated how a curated open data set can become an active platform for generative modeling. 
A complementary example is TURB-Smoke \citep{biferale2026turbsmoke}, which describes pollutants released from multiple point sources in a three-dimensional turbulent flow with a mean wind. The data set contains Lagrangian information on transported pollutant particles, together with the associated Eulerian flow and concentration fields. It was specifically designed to investigate source localization and inference from sparse, intermittent odor detections using heuristic, Bayesian, and machine-learning strategies. From a generative-AI perspective, TURB-Smoke extends the scope of TURB-Lagr from canonical single-particle dynamics to a physically and environmentally relevant inverse problem. 
Smart-TURB portal  also contains Eulerian databases, including TURB-Rot \citep{biferale2020turbrot}, a large collection of two- and three-dimensional snapshots of rotating turbulent flows. Such datasets broaden the range of physical regimes represented in the portal and may support Eulerian tasks such as field generation, reconstruction, and regime-conditioned sampling.\\
\subsection{Dataset hosting platforms and the NeurIPS model}
A useful reference for the turbulence community is provided by the NeurIPS Evaluations and  Datasets hosting guidelines, which require new datasets to remain available in the long term and recommend hosting them on dedicated machine-learning data platforms such as Harvard Dataverse, Kaggle, Hugging Face or OpenML, or on a custom hosting solution when the dataset requires it \citep{NeurIPS2026EvaluationsDatasets}. The NeurIPS guidelines also emphasize machine-readable metadata through the Croissant format, so that datasets can be inspected, verified, and assessed during review. 
However, for turbulence, the suitability of these platforms is strongly dependent on the size, structure, and intended use of the data. Harvard Dataverse is interesting for archival publication because it provides persistent citation, DOI-oriented data sharing, and long-term academic preservation. It is therefore well suited for curated benchmark subsets, metadata-rich datasets, reduced-order datasets, Lagrangian trajectory collections, or reference samples extracted from larger DNS/LES databases. It is less suitable if you require infrastructure for petabyte-scale or query-based turbulence databases, where remote access, sub-volume extraction, and HPC/cloud-native storage are required.

Kaggle is useful when the goal is to organize public competitions, distribute moderate-size data set and lower the entry barrier for the broader machine-learning community. Provides a familiar environment for participation in challenges, notebooks, public baselines, and comparison style. For turbulence, Kaggle could be effective for educational challenges, reduced benchmark tasks,  super-resolution involving small configuration patches. Its limitations are significant for fully developed turbulence: very large DNS fields, complex metadata, specialized physical diagnostics, parallel I/O and HPC-scale workflows do not naturally fit the standard Kaggle model.

Hugging Face is increasingly attractive for AI dissemination because it supports datasets, model checkpoints,  and community interaction in a single ecosystem. For generative AI in turbulence, this is particularly useful: a repository could distribute, beside data, also pretrained  models,  or patch-extraction scripts, together with inference demos. Its weakness is that the platform is primarily designed around machine-learning products, not on the full needs of large-scale scientific simulation.

OpenML is valuable because it was designed around reproducible machine-learning experimentation: datasets, tasks, runs, evaluation protocols, and metadata are treated as linked objects.  OpenML could be useful for   closure-modeling benchmarks, reduced-order features, or small-to-medium gridded datasets. Its current strengths are closer to the focus of conventional  ML than to massive multidimensional DNS fields, generative modeling of spatio-temporal data. 

No single general-purpose ML platform is sufficient for a complete turbulence benchmark ecosystem. A practical solution would be a hybrid architecture. The full-resolution data could remain in domain-specific repositories such as JHTDB, Smart-TURB, BLASTNet, UKTC or other HPC/cloud platforms, where chunked access, metadata, subvolume queries and large-scale storage are feasible. Curated benchmark subsets, metadata records, baseline models, and challenge could then be mirrored or indexed on the above cited platforms. In this way, turbulence data would remain scientifically faithful and technically scalable, while becoming discoverable, citable, and usable by the broader AI community.

Recent NeurIPS-related contributions show a growing interest in ML for turbulent and complex flows, from stochastic neural LES \citep{neuralIdealLES2023} and BLASTNet-based super-resolution \citep{blastnet} to airfoil datasets, CFD competitions and multiphysics benchmarks \citep{unifoil2025,ml4cfd2024,bubbleml2023}. 
However, most existing benchmarks remain limited with respect to the physics of fully developed three-dimensional turbulence (see the Big-4 Figures \ref{fig:bigfour1}-\ref{fig:bigfour4}). Many tasks are built around engineering objectives rather than knowledge-based turbulence-specific questions. 
The limitation is even clearer at the metrics level. Evaluation is often based on point-wise error, short-time roll-out accuracy, correlation coefficients, lift/drag prediction, or visually plausible reconstruction. {\bf These metrics are useful, but insufficient. } 
\subsection{Pitfalls to avoid}
\label{sec:pitfalls}
Summarizing, for AI applications, and especially for generative models, the central question is whether such repositories can support fair, reproducible and physically meaningful comparison between methods. Recent discussions on scientific machine-learning benchmarks stress that benchmark design is itself a scientific problem. For turbulence, this is particularly important.
A first requirement is the prevention of {\bf data leakage}. Turbulent datasets are highly correlated in space and time, so random splitting of snapshots or subdomains can lead to overly optimistic results. Training and test samples should therefore be separated by {\bf flow realization, time interval, spatial region, Reynolds number, forcing, geometry or boundary condition}, depending on the task. For example, a super-resolution benchmark should avoid evaluating a model on neighboring sub-volumes that are strongly correlated  to those seen during training. Similarly, a forecasting model should not be tested on time windows that are too close to training trajectories, and a specific  statement about dependency on the dimensionless time horizon must be offered. 
A second requirement is the inclusion of {\bf negative controls and heuristic baselines}. A benchmark should not compare only advanced neural architectures. It should include, depending on the application,  simple interpolation, spectral filtering, Gaussian random fields with matched spectra, POD/DMD reconstructions, stochastic eddy models, classical LES closures and other standard equations-based or phenomenological stochastic baselines. These controls are essential because a neural network may appear impressive only because the task is under-constrained or because the baseline is too weak (see the paradigmatic case discussed is Sec. \ref{sec:pdebaseline}). 
A third requirement is {\bf physics-aware evaluation}. Benchmarks should  rely on the already cited multi-scale turbulent diagnostics (Fig. \ref{fig:metrics}). For Generative AI additional tests are needed: sample diversity, mode collapse, uncertainty calibration, rare-event fidelity, conditional consistency with observations, and -above all- tests of memorization/generalization in the training phase (Sec. \ref{sec:mem_gen}).
Another requirement is the design of {\bf out-of-distribution tests}. These tests are essential to distinguish interpolation within a database from the discovery of transferable physical structure.
{\bf Experimental realism} is also important. Most current AI-ready turbulence datasets are based on numerical simulations, but many important applications of generative AI involve incomplete, noisy, biased or irregular experimental measurements. Benchmarks should therefore include PIV/PTV-like observations, sparse probes, finite-resolution sensors, missing regions, nonuniform sampling, measurement noise and uncertainty estimates. Alternatively, it should be possible to make a digital-twin of the experimental sources of aleatoric and systematic errors from the clean DNS data.  
{\bf Metadata completeness and standardized documentation}: AI-ready datasets should follow FAIR principles,
and should include enough scientific and technical metadata to make reuse meaningful (see also \cite{heidt2026fair} for an interesting perspective on how to extend FAIR principles beyond data management). For turbulence, this means documenting governing equations, non-dimensional parameters, Reynolds and Mach numbers, numerical scheme, grid resolution, timestep, forcing, boundary conditions, filtering, interpolation, non-dimensionalisation, resolution criteria, uncertainty estimates and known limitations. Borrowing from AI documentation practices, each dataset should be accompanied by a {\it flow card} summarizing provenance, intended use, pre-processing decisions, recommended tasks, limitations, and ethical or practical constraints. 
 {\bf Reproducible pre-processing} In turbulence, pre-processing is not neutral. Filtering, coarse-graining, interpolation, normalization, non-dimensionalization, patch extraction, and time sub-sampling can all change the physical content of the learning problem. A benchmark should therefore provide reference pre-processing scripts and require authors to report any deviations. For super-resolution, for example, the definition of the coarse field, spectral truncation, box filtering, finite-volume averaging or interpolation must be fixed, because each choice corresponds to a different physical inverse problem.
 {\bf Uncertainty quantification and calibration} As said, generative stochastic models are attractive partly because they can produce ensembles, not only single deterministic predictions. Benchmarks should therefore evaluate whether the predicted uncertainty is meaningful. 
 {\bf Computational transparency}. Leader-boards should report accuracy and computational cost: training data volume, number of parameters, training time, inference time, memory requirements, hardware used, and energy cost (when possible). This is important because a very large generative model may achieve better scores but be impractical for online control or coupling with numerical solvers, e.g. consider  the need to sample  a 1-billion hyperparameter network for each grid point evaluations of a CNN based sub-grid model, compared with a Smagorinsky local and interpretable model. Scientific ML benchmarking literature has emphasized that fair comparison requires attention to algorithms, architectures, and computing platforms, as well as a deep understanding of the trade-off between parsimony and expressivity and/or accuracy vs efficiency.  {\bf Model and code availability} A turbulence benchmark should encourage or require self-contained inference code, fixed evaluation scripts, versioned datasets, persistent identifiers, and reproducible environments. This would allow to evaluate models on hidden test data and would reduce ambiguity due to library versions, GPU hardware, undocumented post-processing or solver-specific conventions. 
{\bf Benchmark governance} A turbulence benchmark should not be controlled by a single group or remain static. It should be maintained by a community consortium, with rotating challenge organizers, versioned datasets, public evaluation rules, explicit citation credit for data producers, and mechanisms for retiring solved tasks. This is important because benchmarks can distort research priorities. 
Finally, a benchmark should include {\bf failure modes}. In turbulence, a model may fail in subtle ways: it may reproduce spectra but miss rare events produce visually realistic vortices but violate incompressibility; reconstruct low-order statistics but distort energy flux or miss time-irreversibility; perform well at one Reynolds number but blatantly fail at another; or generate plausible fields that are dynamically inconsistent. A mature benchmark should make such failures visible. The aim is  to identify the best-performing model and to understand where different AI approaches break-down and what this reveals about both the models and the physics of turbulence.

A negative lesson learned from ImageNet is that very large datasets can encode hidden biases, annotation errors, or privacy issues. ImageNet itself later faced scrutiny and remediation efforts, including privacy-related updates such as face blurring. For turbulence, the analogous issues would not be human privacy in most cases, but numerical bias, solver bias, Reynolds-number bias, idealized-geometry bias and overfitting to canonical flows.

\subsection{Community challenges}
It would be key for the commuity to organise and set-ups a few periodic challenges, to let the practitioners also from other fields to come and confront with the state-of-the-art. One could imagine an {\bf Annual turbulence generative AI challenge} eventually hosted with JFM, ERCOFTAC, APS DFD, CTR, NeurIPS/ICLR workshops, or similar venues and with multiple tracks:  reconstruction,  super-resolution,  forecasting,  closure, experimental data assimilation and a clear set of {\bf fixed baseline models} as, e.g., Classical interpolation,  POD/Galerkin,  Gaussian random fields with matched spectra, or data-driven approaches  CNN/UNet,  Fourier neural operators,  Diffusion models,  Flow matching,  Autoregressive transformers.
More importantly, challenges must be judged {\bf  with physics-aware scores}, not a single scalar score only, and multi-objective evaluation: accuracy, physical consistency, uncertainty, cost, robustness. 
Finally, a reproducibility package includes self-contained codes, standard inference scripts, compute budget reporting, and carbon/energy footprint reporting where relevant.
The goal should not be to award a universal winner, but to expose trade-offs: a model may be excellent for reconstruction but poor for temporal roll-out; robust on DNS but fragile on experimental data. \\
In conclusion,  a turbulence analog of ImageNet should not be a passive archive of large flow fields. It should be a living scientific instrument: a curated, versioned, and community-governed environment where data-driven models are tested against the physical complexity of turbulence.  If designed around physically meaningful tasks, shaped by and shared with the community,  could accelerate the transition of generative AI from impressive ad-hoc demonstrations to trustworthy tools for turbulence research.

%% file: chapters/Outlook.tex
\section{Outlook}
Generative AI offers a new probabilistic framework for turbulence. Some conclusions emerge from the applications, limitations, and benchmark requirements discussed in this Perspective.\\
Generative models are most compelling for non-unique inverse problems and stochastic closures. Their principal advantage is not the replacement of deterministic solvers in problems with a unique, well-resolved solution. It is the ability to represent distributions of physically admissible states when observations are incomplete or unresolved degrees of freedom influence the resolved dynamics. Data assimilation, super-resolution, stochastic subgrid modeling are therefore especially promising applications.
Realism concerns whether an ensemble reproduces the relevant invariant measure, multiscale correlations, physical constraints, and temporal evolution. For conditional problems, it also requires that the generated variability represent the uncertainty left unresolved by the observations rather than uncontrolled sampling noise.
High-order and rare-event validation is essential. Energy spectra and low-order moments are necessary but insufficient. Intermittency, anomalous scaling, gradient and acceleration statistics, scale-by-scale fluxes, coherent structures, and extreme events contain much of the physics that makes turbulence different from standard chaotic systems. A model that fails in these quantities may reproduce the appearance of turbulence without reproducing its dynamics.
The present evidence is stronger for reconstruction than for autonomous three-dimensional generation or long-time roll-outs. 
These remain important open challenges, rather than established capabilities.
The community infrastructure is now as important as the model architecture. Progress cannot be cumulatively assessed while studies use different datasets, splits, filters, baselines, and validation metrics. Open and versioned datasets, hidden test cases, standard preprocessing, physics-aware evaluation,  and documented failure modes are needed to determine which advances are robust. The appropriate objective should not be a single universal  target, but a federated benchmark ecosystem that exposes trade-offs among physical fidelity, uncertainty, generalization, and computational cost.
 Advancements in the field will require the development of generative modeling in close interaction with turbulence theory, numerical simulation, experiments, and community benchmark design.

%% file: chapters/Acknowledgements.tex
\section{Acknowledgments}
It is a pleasure for us to  acknowledge the major contribution  of Massimo Cencini to the initial conception of this Perspective, its overall structure, and the selection of its contents, as well as the continuous and stimulating exchanges we have shared on AI tools for Fluid Mechanics -and many others - over the past decade.
We are particularly grateful to Tianyi Li and Fabio Bonaccorso for their key role in the scientific and technical development of the data-driven research activities within our local group, and for the many discussions and collaborations that have directly informed this Perspective. We also warmly acknowledge the valuable interactions with the other components of our group:  Elisa Bellantoni, Roberto Benzi, Chiara Calascibetta, Maurizio Carbone,  Niccolo Cocciaglia, Elettra Consolini, Francesco Fossella, Andre Freitas, Fabio Guglietta, Giulio Cimini, Robin Heinonen, Lorenzo Piro, Jingran Qiu,  Mauro Sbragaglia, Martino Scarpolini and  Victor Valadao. We are equally grateful to the many external collaborators and friends who have helped to shape our understanding of the field: Hussein Aluie, Jeremie Bec, Giulio Biroli, Guido Boffetta, Alberto Carrassi, Antonio Celani, Luca Centurioni, Shiyi Chen, Sergio Chibbaro, Misha Chertkov, Patricio Clark di Leoni, Xander de Witt,  Greg Eyink, Rainer Grauer, Luca Guastoni, Kristian Gustafsson, Petros Koumoutsakos, Alessandra Sabina Lanotte, Stephane Mallat, Alexey Mailybaev,  Marc Mezard, Hridey Narula,  Prasad Perlekar, Federico Toschi, Massimo Vergassola, Ricardo Vinuesa, Angelo Vulpiani, Kiwon Um and Minping Wan.\\
This work was supported by the European Research Council (ERC) under the European Union’s Horizon 2020 research and innovation program Smart-TURB (Grant Agreement No. 882340); the HORIZON MSCA Doctoral Networks program under Grant Agreement No. 101072344, project AQTIVATE; and the Italian Ministry of University and Research (MUR) - FARE program (No. R2045J8XAW); the Italian Ministry of University and Research (MUR) - Fondo Italiano per la Scienza (FIS2) - 2023 Call, project DeepFL, CUP: E53C24003760001.